\documentclass[a4paper,fleqn,usenatbib]{mnras}

\usepackage[T1]{fontenc}
\usepackage{ae,aecompl}

\usepackage{graphicx}	
\usepackage{amsmath}	
\usepackage{amssymb}	

\usepackage{hyperref}
\usepackage{xspace}
\usepackage{rotating}
\usepackage{ulem}
\usepackage[flushleft]{threeparttable}

\newcommand{\hst}{\textit{HST}\xspace}

\newcommand{\masyr}{mas yr$^{-1}$\xspace}
\newcommand{\kms}{km s$^{-1}$\xspace}

\title[2D kinematics of massive stars near the GC]{2D kinematics of massive stars near the Galactic Center}

\author[M. Libralato et al.]{
  Mattia Libralato$^{1}$\thanks{E-mail: \href{mailto:libra@stsci.edu}{libra@stsci.edu}},
  Daniel J. Lennon$^{2,3}$,
  Andrea Bellini$^{1}$,
  Roeland van der Marel$^{1,4}$,\looseness=-4
  \newauthor
  ~Simon J. Clark$^{5}$,
  Francisco Najarro$^{6}$,
  Lee R. Patrick$^{7}$,
  Jay Anderson$^{1}$,
  Luigi R. Bedin$^{8}$,\looseness=-4
  \newauthor
  ~Paul A. Crowther$^{9}$,
  Selma E. de Mink$^{10,11}$,
  Christopher J. Evans$^{12}$,
  Imants Platais$^{13}$,\looseness=-4
  \newauthor
  ~Elena Sabbi$^{1}$,
  Sangmo Tony Sohn$^{1}$\\~\\
$^{1}$Space Telescope Science Institute 3700 San Martin Drive, Baltimore, MD 21218, USA \\
$^{2}$Instituto de Astrof\'isica de Canarias, E-38205 La Laguna, Tenerife, Spain \\
$^{3}$Universidad de La Laguna, Dpto. Astrof\'isica, E-38206 La Laguna, Tenerife, Spain \\
$^{4}$Center for Astrophysical Sciences, Department of Physics \& Astronomy, Johns Hopkins University, Baltimore, MD 21218, USA \\
$^{5}$School of physical sciences, The Open University, Walton Hall, Milton Keynes, MK7 6AA, United Kingdom \\
$^{6}$Centro de Astrobiología (CSIC/INTA), ctra. de Ajalvir km. 4, 28850 Torrejón de Ardoz, Madrid, Spain \\
$^{7}$Departamento de F\'{\i}sica, Ingenier\'{\i}a de Sistemas y Teor\'{\i}a de la Se\~nal, Universidad de Alicante, E-03690 San Vicente del Raspeig, \\ $\phantom{~}$Alicante, Spain \\
$^{8}$INAF-Osservatorio Astronomico di Padova, Vicolo dell'Osservatorio 5, Padova, I-35122, Italy \\
$^{9}$Department of Physics and Astronomy, Hicks Building, Hounsfield Road, University of Sheffield, Sheffield S3 7RH, United Kingdom \\
$^{10}$Center for Astrophysics, Harvard-Smithsonian, 60 Garden Street, Cambridge, MA 02138, USA \\
$^{11}$Anton Pannenkoek Institute for Astronomy, University of Amsterdam, 1090 GE Amsterdam, Netherlands \\
$^{12}$UK Astronomy Technology Centre, Royal Observatory Edinburgh, Blackford Hill, Edinburgh, EH9 3HJ, United Kingdom \\
$^{13}$Department of Physics and Astronomy, Johns Hopkins University, 3400 North Charles Street, Baltimore, MD 21218, USA
}

\date{Accepted 2020 October 18. Received 2020 September 28; in original form 2020 July 29}

\pubyear{2020}

\begin{document}
\label{firstpage}
\pagerange{\pageref{firstpage}--\pageref{lastpage}}
\maketitle

\everypar{\looseness=-4}

\begin{abstract}
  The presence of massive stars (MSs) in the region close to the
  Galactic Center (GC) poses several questions about their origin. The
  harsh environment of the GC favors specific formation scenarios,
  each of which should imprint characteristic kinematic features on
  the MSs.  We present a 2D kinematic analysis of MSs in a GC region
  surrounding Sgr A* based on high-precision proper motions obtained
  with the \textit{Hubble Space Telescope}. Thanks to a careful data
  reduction, well-measured bright stars in our proper-motion catalogs
  have errors better than 0.5 \masyr. We discuss the absolute motion
  of the MSs in the field and their motion relative to Sgr A*, the
  Arches and the Quintuplet. For the majority of the MSs, we rule out
  any distance further than 3--4 kpc from Sgr A* using only kinematic
  arguments. If their membership to the GC is confirmed, most of the
  isolated MSs are likely not associated with either the Arches or
  Quintuplet clusters or Sgr A*.  Only a few MSs have proper motions
  suggesting they are likely members of the Arches cluster, in
  agreement with previous spectroscopic results. Line-of-sight radial
  velocities and distances are required to shed further light on the
  origin of most of these massive objects. We also present an analysis
  of other fast-moving objects in the GC region, finding no clear
  excess of high-velocity escaping stars. We make our
  astro-photometric catalogs publicly available.
\end{abstract}

\begin{keywords}
  Galaxy: center -- Galaxy: open clusters and associations:
  individual: Arches -- Galaxy: open clusters and associations:
  individual: Quintuplet -- Stars: massive stars -- Proper motions
\end{keywords}

\section{Introduction}

At a distance of only 8 kpc \citep{2019A&A...625L..10G}, the Galactic
Center (GC) region can be studied in exquisite detail, more than any
other galactic center. The Central Cluster, the Arches and the
Quintuplet host rich massive-star (MS) populations, in addition to
which there are at least 50 isolated MSs \citep[see, e.g.,][and
references
therein]{1996ApJ...461..750C,2010ApJ...725..188M,2011MNRAS.417..114D,2015MNRAS.446..842D},
whose existence represents a conundrum for this environment
\citep[see, e.g.,][]{2010RvMP...82.3121G}.

Most of these MSs \citep{2011MNRAS.417..114D} lie clearly outside of
the two massive clusters and the central 20 pc region, dominated by
the gravitational potential of the central black hole \citep[see
Fig.~10 of][]{2010ApJ...725..188M}. The bulk of these stars is made up
of Wolf-Rayet stars, very luminous OB supergiants or Luminous Blue
Variables that were discovered by virtue of their strong emission
lines in the near-infrared (NIR), or strong X-ray emission
\citep[e.g.][]{2007ApJ...662..574M,2009ApJ...703...30M}. Hence, MSs
may only represent the tip of the iceberg since the weaker winds of
lower mass OB stars makes them more difficult to detect through the
analysis of their spectral lines.

The origin of isolated MSs is currently unknown, but the unique GC
environment favors some plausible formation scenarios. For example,
some of these stars might be remnants of disrupted clusters, runaway
stars (isolated OB-like objects with peculiar velocities greater than
30 \kms) ejected from the Arches or the Quintuplet or members of their
tidal tails resulting from the interaction of the clusters with Sgr A*
\citep[e.g.,][]{2014A&A...566A...6H}. Some of these isolated MSs might
instead result from the disruption of binaries via interaction with
the central black hole \citep[the Hills
mechanism][]{1988Natur.331..687H}. This interaction would be
responsible for the creation of hyper-velocity stars (HVSs). HVSs have
velocities of thousands of \kms, which are greater than the Galactic
escape velocity \citep{2015ARA&A..53...15B}, and would represent the
extreme outliers in the velocity distribution of isolated MSs.

\begin{figure*}
  \centering
  \includegraphics[width=\textwidth]{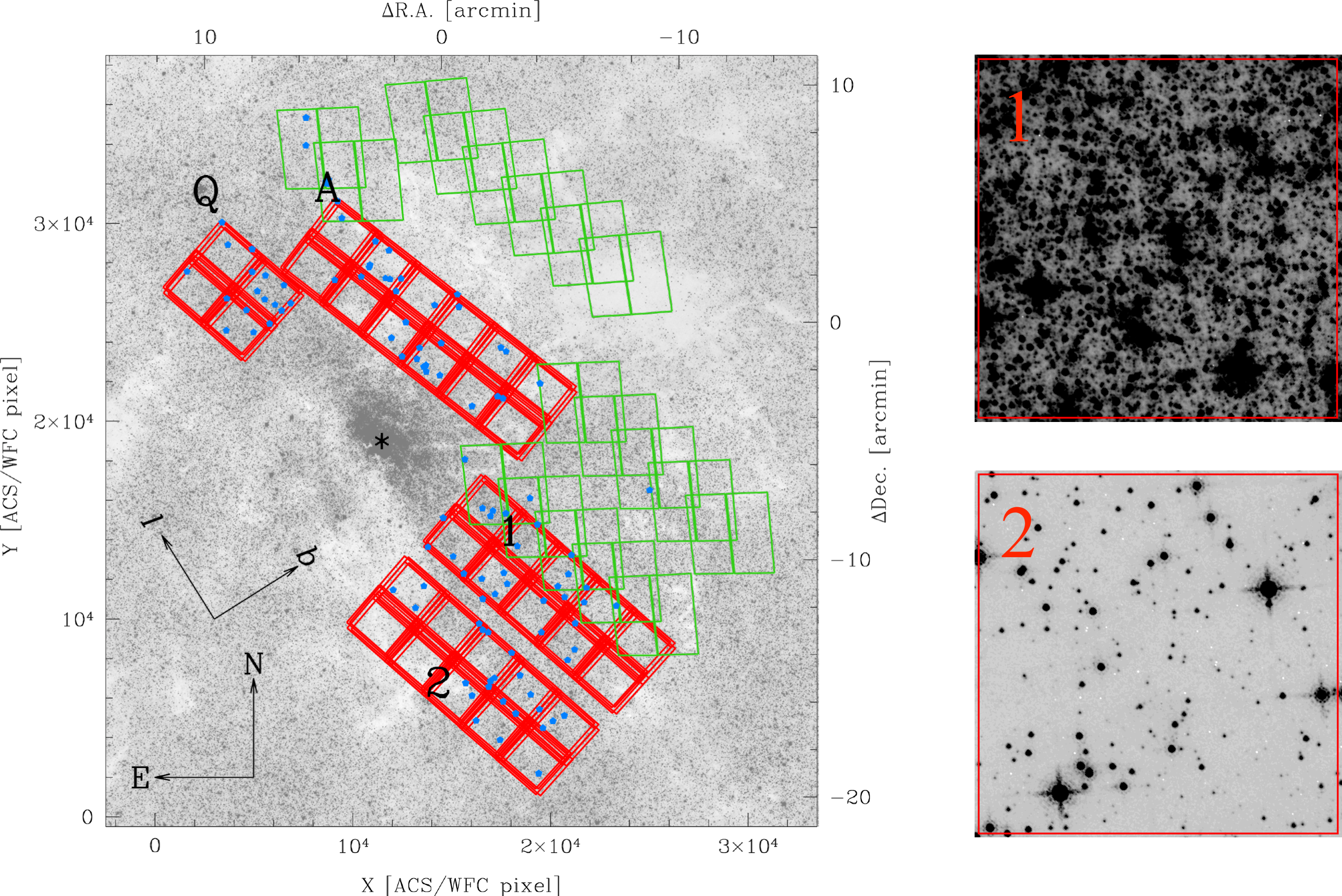}
  \caption{The FoV of the GO-12915+13771 data is shown in the left
    panel (pixel scale 50 mas pixel$^{-1}$). The red and green
    outlines depict the footprints of the WFC3/IR and ACS/WFC data,
    respectively, and are overlaid to a $K_{\rm S}$ mosaic from the
    ``VISTA Variable in the V\'ia L\'actea'' survey (via
    \href{http://archive.eso.org/scienceportal/home}{ESO Science
      Portal}). The black asterisk, ``Q'' and ``A'' mark the position
    of Sgr A*, the Quintuplet and the Arches, respectively. Blue
    symbols highlight the targets of our investigation for which we
    derive a PM measurement. The right panels show (with the same
    logarithmic scale) a zoom-in around a typical low- (1) and
    high-extinction (2) regions in the WFC3/IR stacked image.}
  \label{fig:figure1}
\end{figure*}
    
Some MSs have apparent lifetimes too short to justify their current
positions had they been ejected from the nearest clusters or
associations. It has been suggested that these stars might result from
a peculiar mechanism of star formation such as the one in which single
massive stars form from small molecular clouds
\citep{2004RMxAC..22..127O,2007MNRAS.380.1271P,2010ApJ...725.1886L}. This
idea assumes the MSs to be normal single stars and not the product of
more exotic systems \citep[e.g., binary mergers of less massive but
longer lived stars;][]{2014ApJ...782....7D,2019A&A...624A..66R}.

The proposed formation scenarios imprint distinct kinematic
fingerprints on the MSs, so that studying the motion of MSs can help
us understand how MSs have originated. Crowding and the complex
spectra of these massive stars make line-of-sight (LOS)
radial-velocity measurements challenging. High-precision proper
motions (PMs) can instead provide useful pieces of information about
their 2D motions in the plane of the sky. State-of-art PMs over large
field of views (FoVs) are nowadays provided by the Gaia Data Release 2
\citep[DR2,][]{2016A&A...595A...1G,2018A&A...616A...1G}
catalog. However, Gaia is ultimately blind toward the GC because of
extinction, and its depth extends to a few kpc at most in the GC
direction.

In this paper, we investigate the nature of MSs in the GC region by
measuring their PMs with \textit{Hubble Space Telescope} (\hst)
optical and IR data. We also search for fast-moving stars in the
region.

\section{Data sets and reduction}\label{datared}

The GC region was observed with the \hst's near-infrared (IR) channel
of the Wide-Field Camera 3 (WFC3; pixel scale $\sim$130 mas
pixel$^{-1}$) in October 2012 and August 2015 during programs GO-12915
and GO-13771 (both PI: Lennon). The FoV was covered by 36 WFC3/IR
pointings per epoch, each of which was observed with 4
images. Additional images were taken in parallel mode with the
Wide-Field Channel (WFC; pixel scale $\sim$50 mas pixel$^{-1}$) of the
Advanced Camera for Surveys (ACS). Parallel fields were covered by 18
ACS/WFC pointings of 1 image each. Table~\ref{tab:log} lists all the
observations. The FoV covered by the \hst data is shown in
Fig.~\ref{fig:figure1}.

\begin{table}
  \caption{List of observations used in this paper.}
  \centering
  \label{tab:log}
    \setlength\tabcolsep{3 pt}
  \begin{tabular}{cccccc}
    \hline
    \hline
    GO & Date & Instrument/Camera & Filter & $N$ $\times$ Exp. Time  \\
    \hline
    12915 & Oct. 2012 & WFC3/IR & F139M  & $108 \times 300$ s \\
          &           &         &        & $36 \times 250$ s \\
          &           & ACS/WFC & F850LP & $18 \times 101$ s \\
    13771 & Aug. 2015 & WFC3/IR & F139M  & $108 \times 300$ s \\
          &           &         &        & $36 \times 250$ s \\
          &           & ACS/WFC & F850LP & $18 \times 110$ s \\
    \hline
  \end{tabular}
\end{table}

We made use of \texttt{\_flt}-type exposures (images bias-subtracted
and flat-fielded but not resampled, produced by the official \hst
pipelines). ACS/WFC \texttt{\_flt} images were also corrected for
charge-transfer-efficiency (CTE) defects
\citep{2010PASP..122.1035A}\footnote{The WFC3/IR detector is not
  affected by CTE because of its different read-out mode with respect
  to that of charged-coupled devices.}. For each camera/filter, we
extracted positions and fluxes of isolated, bright sources in each
image by fitting spatially-variable point-spread functions (PSFs). The
PSFs for the ACS/WFC CCDs were tailored to each exposure starting from
the publicly-available \hst library
PSFs\footnote{\href{http://www.stsci.edu/~jayander/STDPSFs/}{http://www.stsci.edu/$\sim$jayander/STDPSFs/}.}
as described in \citet{2017ApJ...842....6B}. For WFC3/IR exposures,
the publicly-available library PSF models did not provide a
satisfactory fit to the stellar profiles even after perturbation. In
this case, we derived from scratch new, spatially-variable PSF models
using our data set, following the prescriptions given in
\citet{2016wfc..rept...12A}. These new models were then also perturbed
on an image by image basis. Stellar positions were corrected for
geometric distortion by means of the corrections available for \hst
WFC3/IR \citep{2016wfc..rept...12A} and ACS/WFC detectors
\citep{2006acs..rept....1A}\footnote{\href{http://www.stsci.edu/~jayander/STDGDCs/}{http://www.stsci.edu/$\sim$jayander/STDGDCs/}.}.

Our data reduction is based on two photometric reductions. The
first-pass photometry finds all detectable sources in a single wave of
finding and measures them without considering the presence of close-by
neighbors. This is a severe limitation in crowded regions like those
towards the GC (see the top-right panel in Fig.~\ref{fig:figure1}).
To overcome this issue, we performed a second-pass photometric
reduction by means of the software \texttt{KS2}, a sophisticated
\texttt{FORTRAN} routine currently used in the astro-photometric
analyses of stellar clusters
\citep[e.g.,][]{2016ApJS..222...11S,2017ApJ...842....7B,2018ApJ...853...86B,2018ApJ...861...99L,2019ApJ...873..109L,2018MNRAS.481.3382N}.

Starting from the outputs of the first-pass photometry, the
second-pass photometry makes use of all the images at once to find all
sources in the field. The combination of all the images enhances the
detectability of faint sources otherwise lost in the noise of single
exposures. Furthermore, the position and flux of each detected object
are measured after all its neighbors are subtracted from the image.

In order to combine all the images of a given epoch/camera/filter,
\texttt{KS2} requires the definition of a common astrometric and
photometric reference system. We initially cross-identified
well-measured, unsaturated, bright stars in our first-pass catalogs
with the Gaia DR2 catalog by means of six-parameter linear
transformations. The vast majority of Gaia stars within our FoV are
Disk stars, which we used to define our reference-frame system. In
total, we have about 3200 Gaia stars in our WFC3/IR field (with an
average of 77 stars per image) and 2500 stars in our ACS/WFC field
(with an average of 153 stars per image). The standard deviation of
the positional residuals between a single WFC3/IR catalog and the Gaia
catalog is, on average, 1.05 mas in the first epoch and 0.29 mas in
the second epoch. For the ACS/WFC data, standard deviations of the
positional residuals are 0.75 mas and 0.18 mas in the first and second
epoch, respectively. The differences between the values in the first-
and second-epoch data are mainly due to the Disk kinematics (GO-12915
data were obtained 2.7 yr before the Gaia DR2, while GO-13771 data
were obtained at about the same epoch of the Gaia DR2). Gaia positions
were projected onto a tangent plane centered at
${\rm (R.A.,Dec.)} = (266.368833, -28.920167)$ deg (an arbitrary point
in the center of our FoV) and transformed from degrees to pixels
adopting a pixel scale of 50 mas\,pixel$^{-1}$ (similar to that of the
ACS/WFC detector). The $X$ and $Y$ axes were oriented toward West and
North, respectively, and the reference tangent point was placed at
position $(14\,500,25\,000)$. Then, master-frame positions were
obtained by averaging the single-image positions once transformed into
this Gaia-based reference system. Our master frames allow us to obtain
an estimate of the positional precision we reached with our data. The
median 1D positional rms for bright stars in the WFC3/IR data is of
1.5 mas in both epochs, while for bright objects in the ACS/WFC data
with at least 2 measurements is 1.2 mas in the first epoch and 1.3 mas
in the second epoch.

The photometric registration of all the images into the same
photometric system is not as straightforward as the astrometric set up
due to the non-contiguous FoV (see left panel of
Fig.~\ref{fig:figure1}). Thus, we defined a common photometric system
as follows:
\begin{enumerate}
\item for each image, we measured the magnitude of bright, isolated
  objects on the corresponding WFC3/IR drizzle \texttt{\_drz} or
  ACS/WFC \texttt{\_drc} exposure using aperture photometry with a
  circular aperture with a five-pixel radius. We then transformed the
  PSF-based magnitudes to the aperture-based ones by applying the
  2.5$\sigma$-clipped median offset found between the two
  systems. This normalizes the magnitude of each source in each image
  to a 1-s exposure (note that the WFC3/IR images are in electrons per
  second, so the magnitude difference is close to 0) and makes the
  photometric calibration straightforward in the later stage;
\item we combined (where possible) multiple
  \texttt{\_drz}/\texttt{\_drc} exposures onto a common reference
  frame system. The magnitudes of the objects in the
  \texttt{\_drz}/\texttt{\_drc} master frame were computed by
  averaging the magnitudes of the stars in the single images once
  transformed onto the same reference system;
\item we cross-identified well-measured, bright stars in the
  zero-point-corrected first-pass catalogs defined in (i) with the
  \texttt{\_drz}/\texttt{\_drc} master frame defined in (ii), computed
  the residual zero-point differences and corrected the first-pass
  photometry by these zero-point values. This last step ensures that
  any uncorrected photometric zero-point residual between different
  pointings is minimized;
\item the resulting corrected first-pass stellar magnitudes were then
  averaged together and defined the magnitudes of our master
  frame. The zero-point differences between the corrected first-pass
  stellar magnitudes and the master-frame magnitudes are lower than
  0.01 mag on average and a few hundredth of a magnitude at most.
\end{enumerate}

Once a common astro-photometric system has been defined for each
epoch/camera/filter, we run \texttt{KS2}. \texttt{KS2} measures
stellar positions and fluxes using several different methods
\citep[see][for details]{2017ApJ...842....6B}. In this work, we make
use of the method based on PSF fitting of neighbor-subtracted stellar
images, which is optimized for PM analyses
\citep{2018ApJ...853...86B,2018ApJ...861...99L}.

Finally, we transformed out \texttt{KS2}-based photometry on to the
VEGA-mag flight system using the zero-points and infinite-aperture
corrections provided by the
STScI\footnote{\href{http://www.stsci.edu/hst/acs/analysis/zeropoints}{http://www.stsci.edu/hst/acs/analysis/zeropoints}
  for ACS/WFC and
  \href{http://www.stsci.edu/hst/instrumentation/wfc3/data-analysis/photometric-calibration/ir-photometric-calibration}{http://www.stsci.edu/hst/instrumentation/wfc3/
    \-data-analysis/photometric-calibration/ir-photometric-calibration}
  for WFC3/IR.}.

In general, the higher the number of images mapping the same region of
the sky, the more efficient is the \texttt{KS2} detection of faint
objects. However, \texttt{KS2} is designed to search and fit an object
within a radius of two pixels from the location of a local maximum on
the reference frame found by combining all images. Sources that moved
by more than two pixels in $\sim$2.8 yr (the temporal baseline between
the GO-12915 and GO-13771 programs) in our data set are most likely to
be missed by the \texttt{KS2} finding algorithms if we use all the
available data in a single run. Since the main goal of our project is
to find fast-moving objects, we run \texttt{KS2} for each epoch
separately.

As described in \citet{2016ApJS..222...11S} and
\citet{2017ApJ...842....6B}, \texttt{KS2} searches for and measures
objects that satisfy various criteria (isolation, signal above the sky
background, quality of the PSF fit, total number of peaks in multiple
images). The diagnostic parameters used to define these criteria do
not have the same values in all images, but they change with, for
example, the image quality (and the PSF), the value of the sky, the
level of crowding. We tested different combinations of these finding
criteria and chose the parameters that gave us a good compromise
between including real stars and excluding image artifacts or spurious
detections. Thus, if the local values of the diagnostic parameters are
very different from the global ones used to define \texttt{KS2}
searching criteria, \texttt{KS2} could have missed real stars or
included more spurious detections in the final list. In general, there
are no one-size-fits-all parameters for such a large FoV. Our final
photometric catalogs (one for each epoch) contain about 830\,000
objects.

\texttt{KS2} provides a series of diagnostic parameters that can be
used to select a sample of well-measured objects
\citep[e.g.,][]{2018ApJ...861...99L}. For the WFC3/IR catalogs, we
define as well-measured stars those fulfilling all the following
requirements: (a) their quality of PSF fit (\texttt{QFIT})
parameter\footnote{The \texttt{QFIT} parameter is a linear-correlation
  coefficient between the pixel values in the image and those of
  predicted by the PSF. A well-measured star has a \texttt{QFIT} close
  to unity.}  is larger than the 85-th percentile of the \texttt{QFIT}
value at any given magnitude, but we additionally kept all objects
with \texttt{QFIT} higher than 0.975 and rejected those with a
\texttt{QFIT} lower than 0.6, regardless of their percentile value,
(b) their magnitude rms is lower than the 85-th percentile at any
given magnitude (by analogy with the \texttt{QFIT}, sources with a
magnitude rms lower/higher than 0.1/0.5 mag are also kept/discarded),
(c) the ratio between the number of individual exposures actually used
to measure position/flux of a given star and the total number of
exposures in which the star was found is lower than 50\%, (d) their
fraction of neighbor flux within the fitting radius with respect to
the star flux before neighbor subtraction is less than 1, (e) the
absolute value of their shape parameter \texttt{RADXS}
\citep[excess/deficiency of flux just outside of the fitting radius
with respect to that expected from the PSF;
see][]{2008ApJ...678.1279B} is lower than the 85-th percentile at any
given magnitude (all sources with a \texttt{RADXS} lower than
$\pm$0.01 are considered as stellar-like objects, while detections
with a \texttt{RADXS} larger than $\pm$0.1 are significantly
narrower/broader than the PSF expectation, and hence discarded), (f)
their flux within the PSF fitting radius is at least 3$\sigma$ above
the local sky. For the ACS/WFC catalogs, crowding is not an issue and
bad-measured stars are clearly identified as outliers in the
distributions of the diagnostic parameters mentioned above. At odd
with the WFC3/IR case, to select well-measured stars we identified and
removed by eye the outliers in the distribution of each diagnostic
parameter. Finally, we combined the catalogs of the two runs of
\texttt{KS2} together and kept only objects that were measured in both
epochs.

\texttt{KS2} also outputs single exposure catalogs containing raw
position and flux of each detected source. These positions and fluxes
are superior over those obtained with the first-pass photometry stage,
since \texttt{KS2} measurements are deblended. Hereafter, we will
refer to these \texttt{KS2}-based catalogs simply as \textit{raw}
catalogs.

\subsection{Ancillary catalogs}

Most of the stars in our FoV are observed with only one filter in our
\hst data. In order to build color-magnitude diagrams (CMDs) for the
subsequent analyses, we linked our photometric catalogs to other
catalogs from previous or ongoing surveys of the GC:
\begin{itemize}
  \item the catalog of \citet{2011MNRAS.417..114D}, made as part of
    the Paschen $\alpha$ survey of the GC with the \hst Near-Infrared
    Camera and Multi-Object Spectrometer
    \citep[NICMOS,][]{2010MNRAS.402..895W}, provides F187N- and
    F190N-filter photometry;
  \item the ``GALACTICNUCLEUS'' $JHK_{\rm S}$ catalog of
    \citet{2019A&A...631A..20N}, obtained with HAWK-I@VLT speckle data;
  \item the PSF-based catalog of \citet{2019A&A...629A...1S}, made with
    the ``VISTA Variable in the V\'ia L\'actea'' (VVV) data.
\end{itemize}
Depth and completeness of these catalogs are heterogeneous.

The HAWK-I@VLT instrument has a higher angular resolution than that of
the VIRCAM@VISTA imager (106 mas pixel$^{-1}$ versus 339 mas
pixel$^{-1}$), which is an advantage in crowded environments such as
the region close to the GC. For this reason, we chose to rely mainly
on the GALACTICNUCLEUS $JHK_{\rm S}$ catalog. However, while this
catalog completely covers our WFC3/IR data, it only partially overlaps
with our ACS/WFC data. As such, we used the VVV data to study the
ACS/WFC region. We found magnitude zero-point differences between the
GALACTICNUCLEUS and the VVV photometry. We corrected the VVV
photometry to match that of the GALACTICNUCLEUS catalog by linearly
interpolating the median magnitude difference between the two catalogs
in different 0.5-mag bins. The zero-point corrections between the VVV
and GALACTICNUCLEUS photometry are of the order of 0.1--0.2 mag in $J$
filter and 0.4--0.6 mag in $K_{\rm S}$ filter.

As for the GALACTICNUCLEUS catalog, the Paschen $\alpha$ data cover
the entire WFC3/IR area and only part of the ACS/WFC region. The
Paschen $\alpha$ catalog also contains a list of MSs in the GC region.

\section{Proper motions}\label{pm}

State-of-the-art techniques currently used to study the internal
kinematics of globular clusters with \hst
\citep[e.g.,][]{2014ApJ...797..115B} are based on the selection of a
reference population of stars. Cluster stars are typically chosen as a
reference because, to first order, they are moving toward the same
direction on the sky and have a tight distribution in the vector-point
diagram (VPD). However, this method is not easy to apply to our
project. The different stellar populations in the field have complex
PM distributions in the VPD as a result of the Galactic kinematics and
the reflex motion of the Sun. The uneven dust distribution toward the
GC (e.g., the right panels of Fig.~\ref{fig:figure1}) adds further
complications. For example, highly-reddened regions only leave nearby
objects visible, while regions with a less severe reddening contain a
mix of stars at different distances out to beyond the GC. The combined
effect of kinematics and dust distribution could bias our PMs if we
select reference samples with different bulk PMs across the field of
view. For this reason, PMs are computed in a different fashion, by
taking advantage of the Gaia DR2 catalog \citep[for a similar
technique, see also][]{2018MNRAS.481.5339B}. PMs are computed
independently for the two sets WFC3/IR and ACS/WFC data, so as to
minimize mismatches between images taken using significantly different
bandpasses, but at the unavoidable expense of having fewer images at
our disposal in the overlapping regions between the two data sets. In
Appendix~\ref{appendix:hst}, we provide comparisons of independent PM
measurements in the overlapping regions as a sanity check.

For each of the two data sets, we defined an absolute reference system
using the Gaia DR2 catalog as follows. First, we projected the Gaia
DR2 catalog as described in Sect.~\ref{datared}. We used Eq.~(2) of
\citet{2018A&A...616A..12G} to define an orthographic projection of
the Gaia DR2 PMs into Equatorial Coordinates. Then, we used Gaia-DR2
PMs to predict the position of each source in the Gaia catalog at the
average epochs of the GO-12915 (2012.8) and GO-13771 (2015.6) data
sets, respectively. We removed from the Gaia sample all objects that
(i) are fainter than $G=19$, (ii) have a PM error in either coordinate
larger than 0.6 \masyr, or (iii) do not fulfill the requirement
described by Eq.~(C.1) of \citet{2018A&A...616A...2L}.

For each star in the catalog, we used six-parameter, global linear
transformations to transform its position on to the Gaia-DR2 reference
system of the corresponding epoch (hereafter, we refer to these
positions as transformed positions). In addition, we also produced a
set of transformed positions by applying a local adjustment in order
to mitigate the possible presence of small, uncorrected systematic
residuals in, e.g., the \hst PSF models and the geometric-distortion
solution. This adjustment \citep[the so-called ``boresight''
correction described in][]{2010ApJ...710.1032A} is defined as the
average of the positional residuals between the transformed \hst
positions and the Gaia-DR2 positions of the closest $N$ stars to the
target. These local corrections are based on the closest $N = 24$ (for
the WFC3/IR data) or $N = 25$ (for the ACS/WFC data) sources to each
object\footnote{The number of reference stars used to calculate these
  local corrections was empirically chosen as the largest value that
  allowed to obtain a PM measurement for all stars across the FoV,
  even in regions covered by only one image per epoch.}. The median
distance of the furthest reference star in a WFC3/IR image is about
750 WFC3/IR pixels ($\sim$90 arcsec), while in an ACS/WFC image is
about 2200 ACS/WFC pixels ($\sim$110 arcsec).

\begin{figure}
  \centering
  \includegraphics[width=\columnwidth,keepaspectratio]{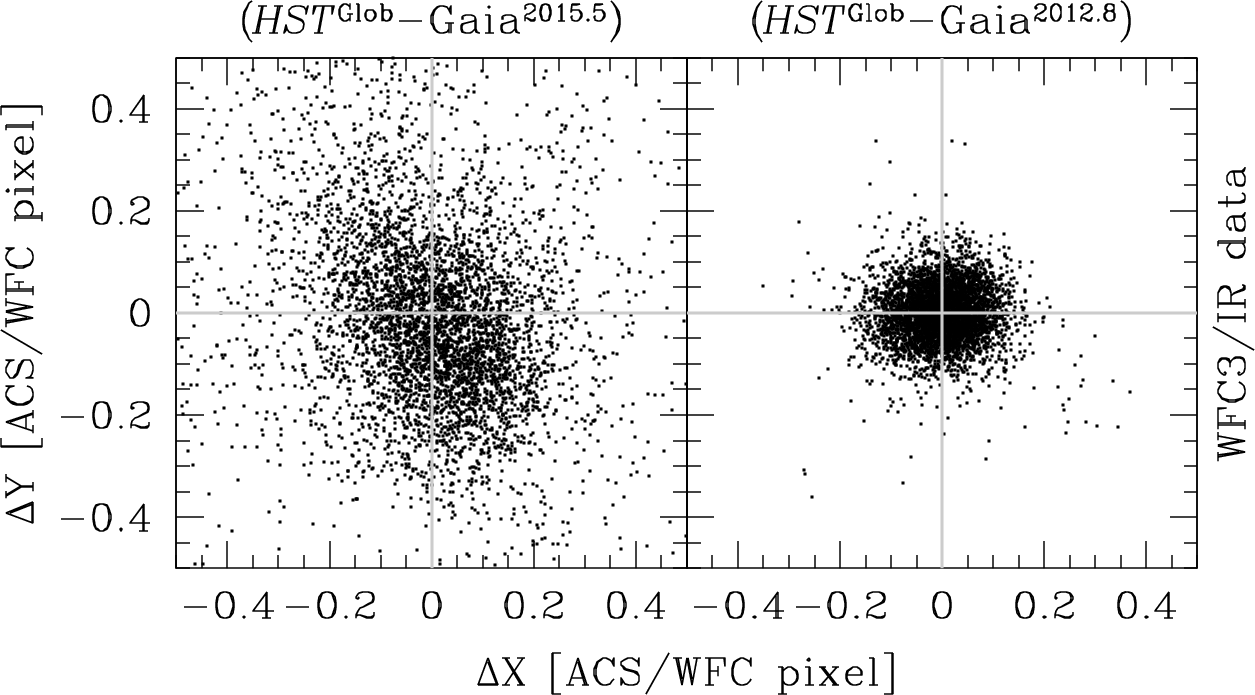}
  \caption{Example of positional residuals between the 2012.8 WFC3/IR
    positions ($HST^{\rm Glob}$) and the Gaia-DR2 positions at epoch
    2015.5 (left panel) and at epoch 2012.8 (right
    panel). Six-parameter, global linear transformations were used to
    transform the stellar positions between the reference
    systems. Displacements are in units of ACS/WFC pixels.}
  \label{fig:figure2}
\end{figure}

\begin{figure}
  \centering
  \includegraphics[width=\columnwidth,keepaspectratio]{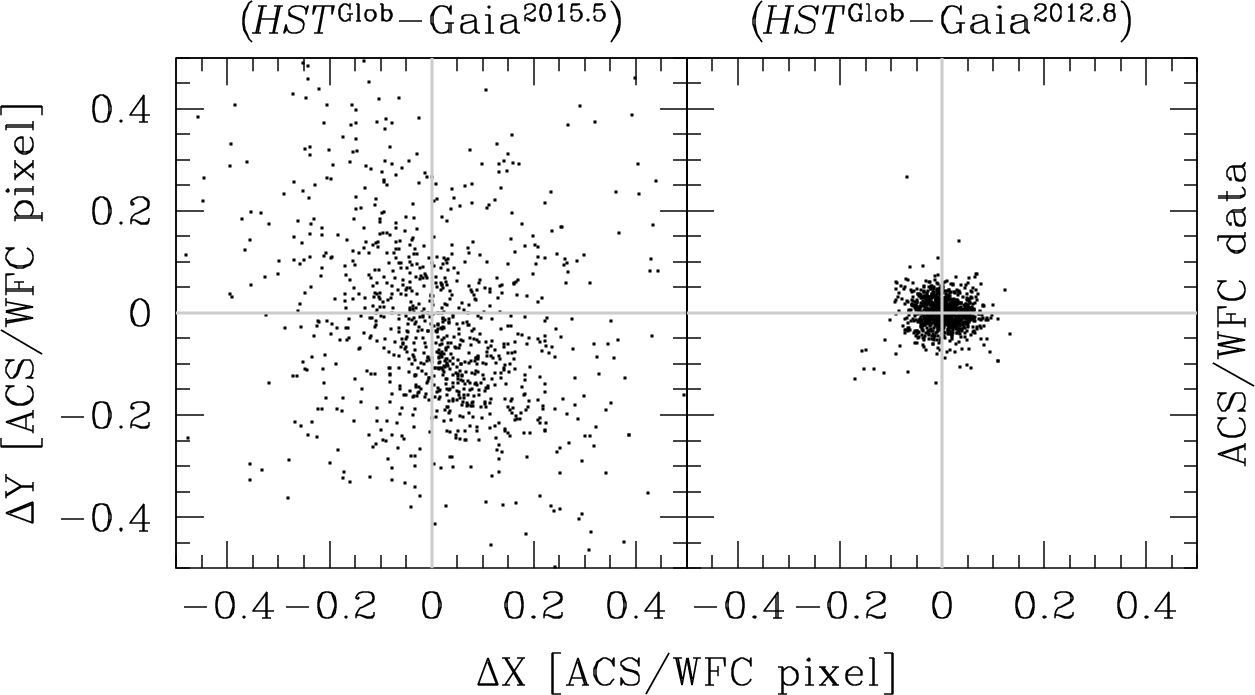}
  \caption{As in Fig.~\ref{fig:figure2}, but for the ACS/WFC data.}
  \label{fig:figure3}
\end{figure}

Figures~\ref{fig:figure2} and \ref{fig:figure3} show the impact of the
stellar PMs on the distribution of positional residuals between \hst
and Gaia. In the left panels, we show the positional residuals between
the 2012.8 \hst positions and the Gaia-DR2 positions at the reference
epoch of the Gaia-DR2 catalog (2015.5). In the right panels, we show
the positional residuals of the same stars after Gaia-DR2 positions
are moved at the epoch of \hst observations. A tight distribution of
the positional residuals is a proxy of accurate transformation between
the frames. The distributions of the positional residuals in the right
panels clearly show how important it is to take into account stellar
motions even at this initial cross-matching stages. The comparison of
the positional residuals between the 2015.6 \hst positions and the
Gaia-DR2 positions with and without moving the Gaia-DR2 positions at
the epoch of \hst observations shows negligible differences, as
expected.

\begin{figure*}
  \centering
  \includegraphics[width=0.8\textwidth,keepaspectratio]{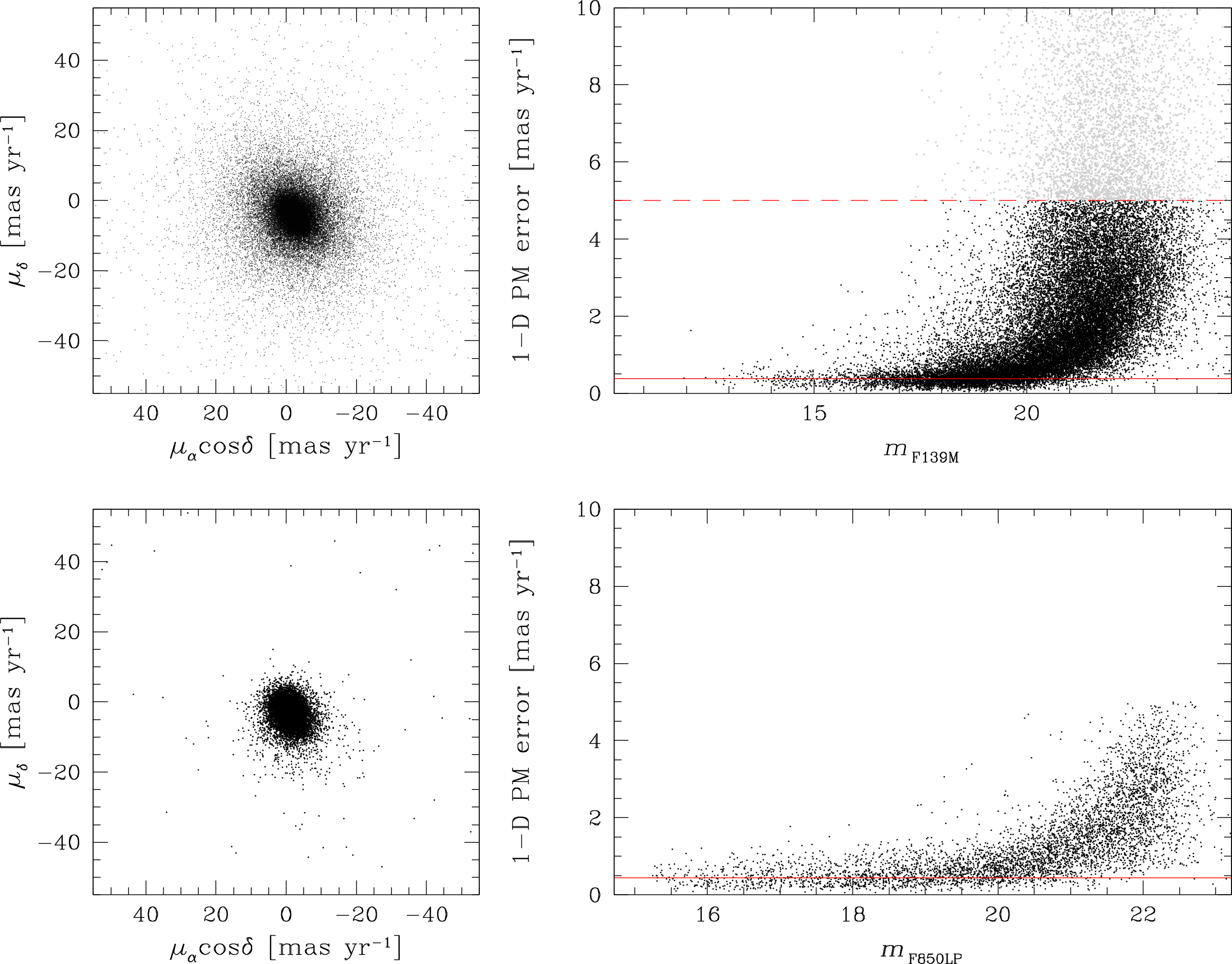}
  \caption{(Top panels): the VPD (left) and the 1-D PM error as a
    function of the $m_{\rm F139M}$ magnitude (right) for the
    WFC3/IR-based PMs. In the right panel, the red, solid horizontal
    line is set at the median value of the PM errors for bright stars
    (0.38 \masyr). Grey points are stars with a PM error larger than a
    threshold set at 5 \masyr (red, dashed horizontal line; see the
    text for details). These stars are not considered in our
    analysis. Only 5\% of the points are shown in each panel for
    clarity. (Bottom panel): the VPD and the 1-D PM error versus
    $m_{\rm F850LP}$ magnitude for the ACS/WFC-based PMs. The median
    PM error for bright stars is 0.45 \masyr (red, solid horizontal
    line).}
  \label{fig:figure4}
\end{figure*}

Global transformations proved to be superior (tighter distribution of
positional residuals) for the WFC3/IR data, while the opposite applies
to ACS/WFC data. This is probably related to the available number of
stars in common between Gaia and \hst catalogs. In general, there are
plenty of common stars in the ACS images that can be used to derive
the boresight correction, which makes the local adjustment
effective. On the other hand, there are not many stars that are
unsaturated and well measured in both Gaia and WFC3 data: much larger
searching radii are typically needed for the WFC3 data to find the
closest 24 stars for the boresight correction, resulting in negligible
improvements over the global transformations alone. For this reason,
we chose to use the globally-transformed positions for the
WFC3/IR-based PMs and the locally-corrected positions for the
ACS/WFC-based PMs. Finally, it should be highlighted that the
positional residuals have a larger dispersion along the $X$ axis. The
$X$ and $Y$ axes are aligned and oriented as $-$R.A. and $+$Dec. by
construction. In this GC field, Gaia positional and PM errors are
larger along the R.A. than the Dec. direction because of the scanning
pattern of the satellite, and are the probable cause of this feature.

The final single-epoch position of a star is computed as the robust
average of the transformed positions of all images at the epoch. An
outlier-rejection step was performed with a jackknife resampling
technique similarly to what is described in
\citet{2014ApJ...797..115B}.

After these steps, we had two catalogs, one for each epoch of our
data. The positions of the objects in these two catalogs are in the
same absolute reference system defined by the Gaia DR2 catalog, only
at different epochs. Therefore, our absolute PMs can be simply defined
as the difference between the positions in the two epochs, divided by
the average temporal baseline. The PM errors were obtained by adding
in quadrature the positional errors in each epoch and dividing by the
temporal baseline.

\begin{figure}
  \includegraphics[height=12cm,keepaspectratio]{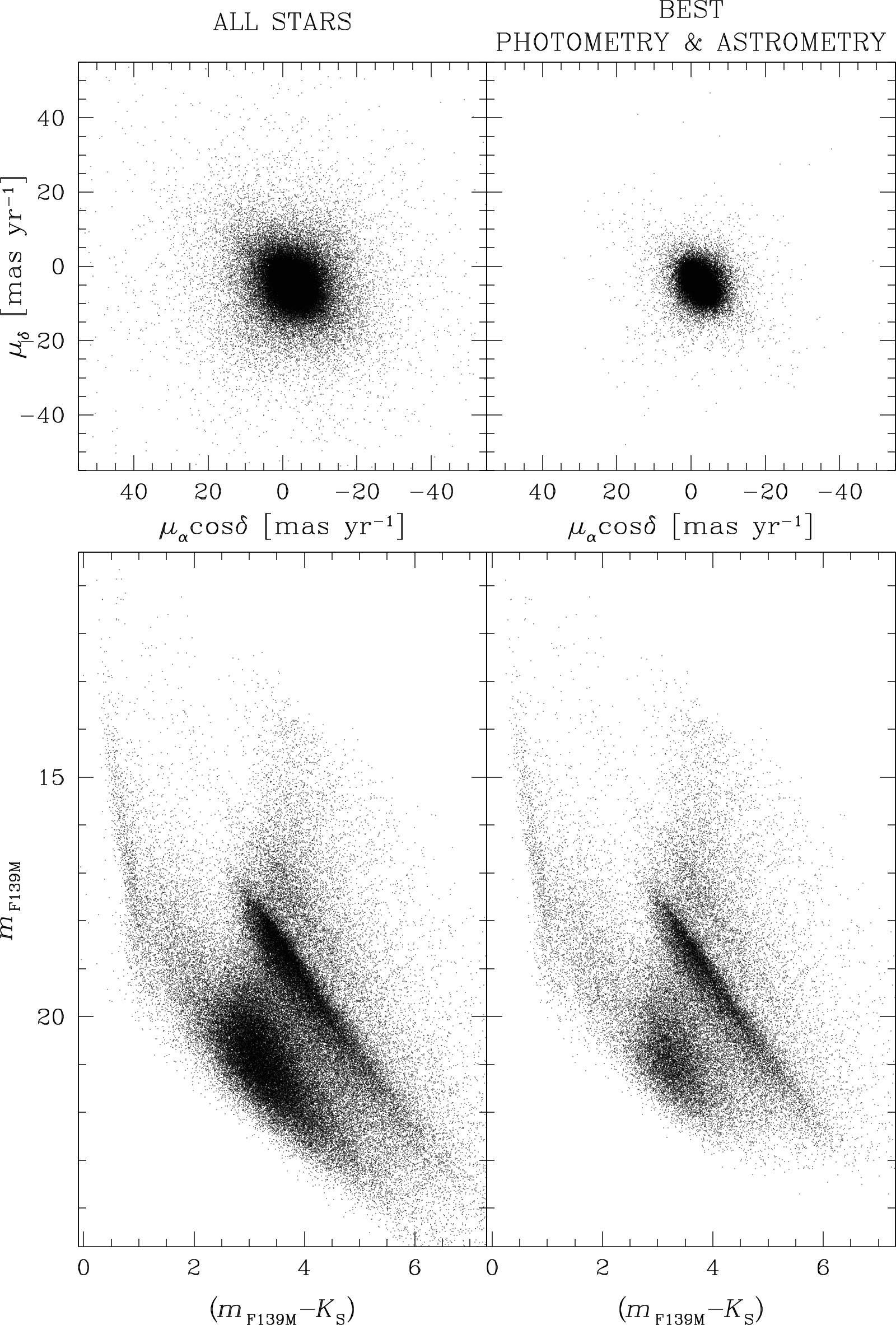}
  \caption{In the left panels, we show the VPD (top) and the
    $m_{\rm F139M}$ versus ($m_{\rm F139M}-K_{\rm S}$) CMD (bottom)
    with all stars with a WFC3/IR-based PM and a $K_{\rm S}$ magnitude
    measurement. In the right panels, we show only stars that passed
    all photometric and astrometric selection criteria. Only 25\% of
    the stars are shown for clarity.}
  \label{fig:figure5}
\end{figure}

\begin{figure}
  \includegraphics[height=12cm,keepaspectratio]{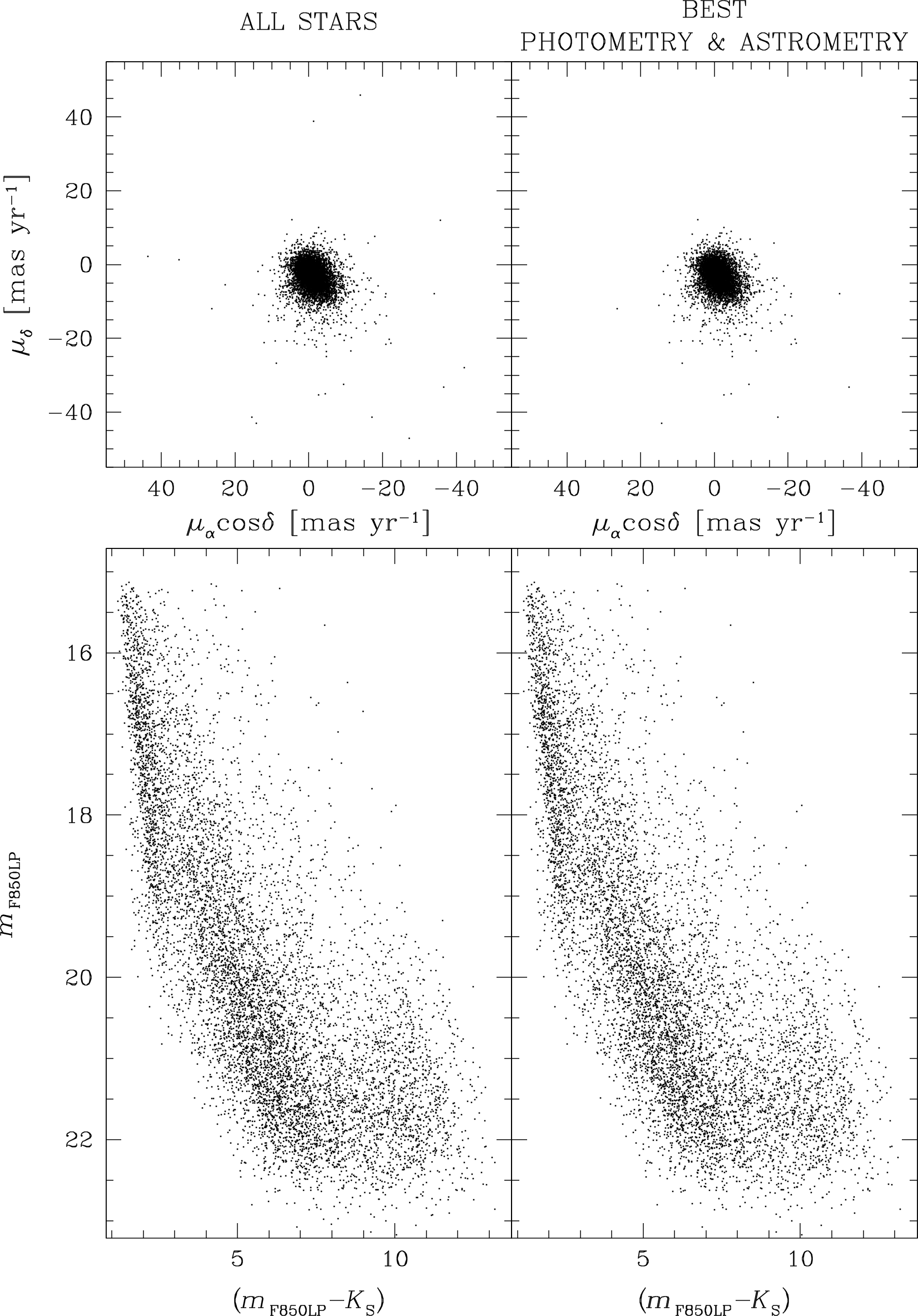}
  \caption{Similar to Fig.~\ref{fig:figure5}, but for the ACS/WFC
    data.}
  \label{fig:figure6}
\end{figure}

The main features and caveats of our PM catalog are the following:
\begin{itemize}
\item our PMs are absolute, not relative;
\item any systematic error in the Gaia DR2 catalog is also present in
  our catalog;
\item crowding and detector cosmetics facilitate mismatches between
  the catalogs, especially at faint magnitudes where the
  signal-to-noise ratio is low. Fast-moving objects should be
  carefully checked. In our analysis, we exclude all objects that
  moved by more than 5 pixels in 2.86 yr, i.e., have a PM greater than
  about 70 mas yr$^{-1}$. At the distance of Sgr A*, this value
  corresponds to $\sim$2600 \kms. We chose this threshold as a
  compromise between removing mismatches in our catalog and keeping
  potential HVSs;
\item the median PM error of bright stars in the WFC3/IR-based PM
  catalog is 0.38 \masyr, while in the ACS/WFC-based PM catalog is
  0.45 \masyr (see Fig.~\ref{fig:figure4});
\item stars measured in only one exposure in an epoch do not have
  positional errors for that epoch. For them, the catalog lists a flag
  value of 99.99 \masyr for their PM error. These stars were included
  in the analyses of MSs and fast-moving objects.
\item the positional error of stars measured in only two exposures in
  an epoch is defined as the absolute difference between the two
  values divided by $\sqrt{2}$. Therefore, their corresponding PM
  errors are likely underestimated.
\end{itemize}

Figure~\ref{fig:figure4} shows VPD (left panels) and PM error as a
function of magnitude (right panels) for stars in the WFC3/IR (top)
and ACS/WFC (bottom) PM catalogs. The red, solid, horizontal lines in
the right panels are set at the median PM error of bright stars. In
the following analysis, we considered as well-measured stars those
with a PM error lower than the 85-th percentile of the error
distribution at any given magnitude for the WFC3/IR-based PMs and
lower than a threshold drew by hand for the ACS/WFC-based PMs. In the
WFC3/IR PM catalog, we also kept all objects with a PM error lower
than 0.7 \masyr (i.e., about twice the median PM error of bright, well
measured stars) and discarded those with a PM error larger than 5
\masyr (red, dashed horizontal line in the top-right panel of
Fig.~\ref{fig:figure4}) in addition to the percentile-based selection
on the PM error. No upper limit was set to the PMs derived from the
ACS/WFC data because of the higher astrometric quality.

Appendix~\ref{appendix:gaia} presents an in-depth comparison between
our \hst and Gaia-DR2 PMs. While there is general agreement, we find
position-dependent systematics across the FoV, in particular in the
outer regions near the edge. Stars brighter than $G \sim 18$ do not
present significant color- and magnitude-dependent systematic
errors. The lower astrometric quality of the Gaia-DR2 PMs for fainter
stars is likely the main cause of the systematic trends visible in
Figs.~\ref{fig:figure32} and \ref{fig:figure33}.

The comparison between the WFC3/IR- and the ACS/WFC-based PMs is shown
in Appendix~\ref{appendix:hst}. There is a difference between the
ACS/WFC and the WFC3/IR PMs along the R.A. direction. However, this
systematic difference disappears if we consider relatively isolated
stars in the WFC3/IR data. Crowding is therefore an important
contributor of systematic errors, particularly for faint stars. Our PM
catalogs are made publicly available
(Appendix~\ref{appendix:catalog}). All the aforementioned caveats and
the quality checks in Appendixes~\ref{appendix:gaia} and
\ref{appendix:hst} should be considered when using these PM catalogs.

The left panels of Fig.~\ref{fig:figure5} show the VPD and the
$m_{\rm F139M}$ versus ($m_{\rm F139M}-K_{\rm S}$) CMD for all stars
in common between our WFC3/IR and the GALACTICNUCLEUS catalogs, while
those on the right present similar plots just for well-measured
stars. It is worth noticing the X-shaped trails of larger-PM stars in
the VPD, which is aligned with the X/Y axes of the detector (see the
orientation of the WFC3/IR data in Figure~\ref{fig:figure1}). We found
that the X-shaped trails are more noticeable the fainter the
stars. Recently, \citet[][and references therein]{2018PASP..130f5004P}
report and analyze the so-called ``brighter-fatter'' effect in the
WFC3/IR data, which causes charge redistribution among the pixels. The
investigation of this effect is out of the scope of our paper, but the
magnitude dependency and the X/Y orientation of the effect we see in
the VPDs seem to suggest a relation with the brighter-fatter effect.

VPDs and $m_{\rm F850LP}$ versus ($m_{\rm F850LP}-K_{\rm S}$) CMDs for
stars in common between our ACS/WFC and the VVV catalogs are presented
in Fig.~\ref{fig:figure6}. Optical data are basically blind to most of
the Bulge population, and the vast majority of the stars measured in
the ACS/WFC data are Disk stars.

\section{A kinematic view of the GC region}\label{science}

In this section, we provide a brief description of the astronomical
scene in our field. In the following, we consider only well-measured
stars that passed all photometric and astrometric quality selections,
unless declared otherwise. In this and the following Sections, all
figures related to WFC3/IR-based PMs show only a fraction of the stars
in the field, for clarity.

\subsection{IR view}\label{nir}

\begin{figure}
  \centering
  \includegraphics[width=\columnwidth]{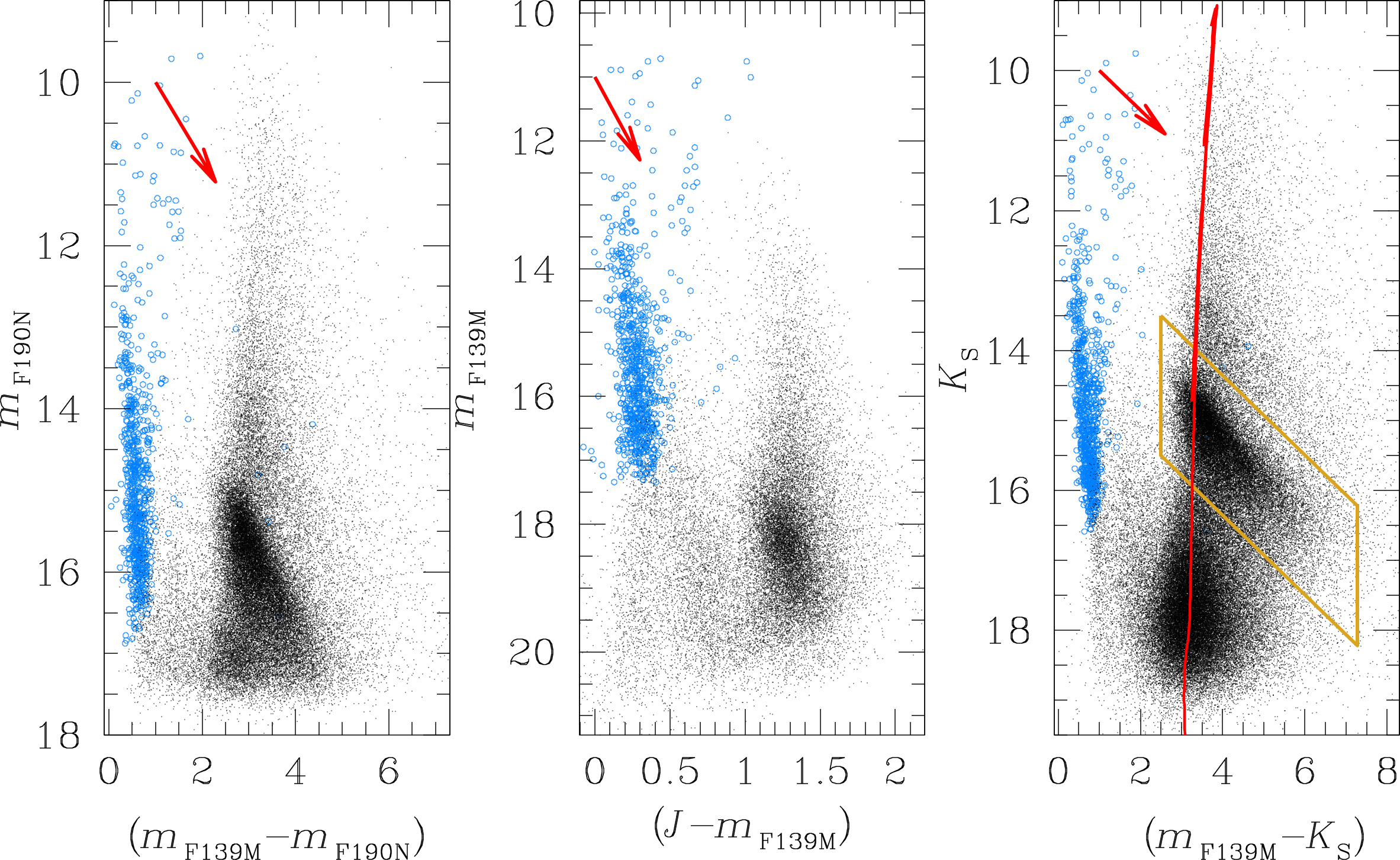}
  \caption{CMDs based on the WFC3/IR data. (Left panel):
    $m_{\rm F190N}$ versus ($m_{\rm F139M}-m_{\rm F190N}$)
    CMD. (Middle panel): $m_{\rm F139M}$ versus ($J-m_{\rm F139M}$)
    CMD. (Right panel): $K_{\rm S}$ versus ($m_{\rm F139M}-K_{\rm S}$)
    CMD. Blue open circles depict stars in common with the Gaia DR2
    catalog. In each CMD, the direction of the reddening vector is
    indicated by a red arrow. The reddening vector is computed by
    assuming an arbitrary extinction $A_{\rm F139M}$ in each plot and
    the extinction index $\alpha=2.3$ of
    \citet{2018A&A...620A..83N}. A PARSEC isochrone for 10-Gyr-old
    stars with $[{\rm
      Fe/H}]$$=$$-$0.16 at 8.2 kpc from the Sun representing the Bulge
    population is fit by eye as reference in the rightmost panel. The
    gold rectangle marks the location of the Bulge Red Clump.}
  \label{fig:figure7}
\end{figure}

Various populations can be distinguished along the line of sight
towards the GC. We use the CMDs shown in Fig.~\ref{fig:figure7} to
disentangle them. The CMDs immediately reveal the presence of a narrow
bluer sequence (hereafter, the blue sequence) and a broader redder
sequence (hereafter, the red sequence).

Stars in common with the Gaia-DR2 catalog are shown as blue open
circles. These objects are mainly located along the blue sequence and
have a median parallax of $\sim$0.5 mas ($\sim$2 kpc), i.e., they are
Disk stars. The majority of the stars in the CMDs are instead part of
the red sequence. The broadening of the red sequence is a proxy of the
severe extinction that affects these stars and suggests that these
stars are Bulge/Bar objects within a few kpc from the GC. As a
reference, we plot in the rightmost panel a 10-Gyr-old PARSEC
isochrone \citep{2012MNRAS.427..127B} with $[{\rm
  Fe/H}]$$=$$-$0.16 \citep[e.g.,][]{2017AJ....154..239R} and with
distance of 8.2 kpc from the Sun, to represent an old stellar
population in the Bulge of the Galaxy. Finally, sources between the
blue and the red sequences are an admixture of Disk stars further than
$\sim$2 kpc from the Sun and Bulge/Bar objects with a less-severe
extinction (plus the scatter due to the photometric errors).

\begin{figure}
  \centering
  \includegraphics[width=\columnwidth,keepaspectratio]{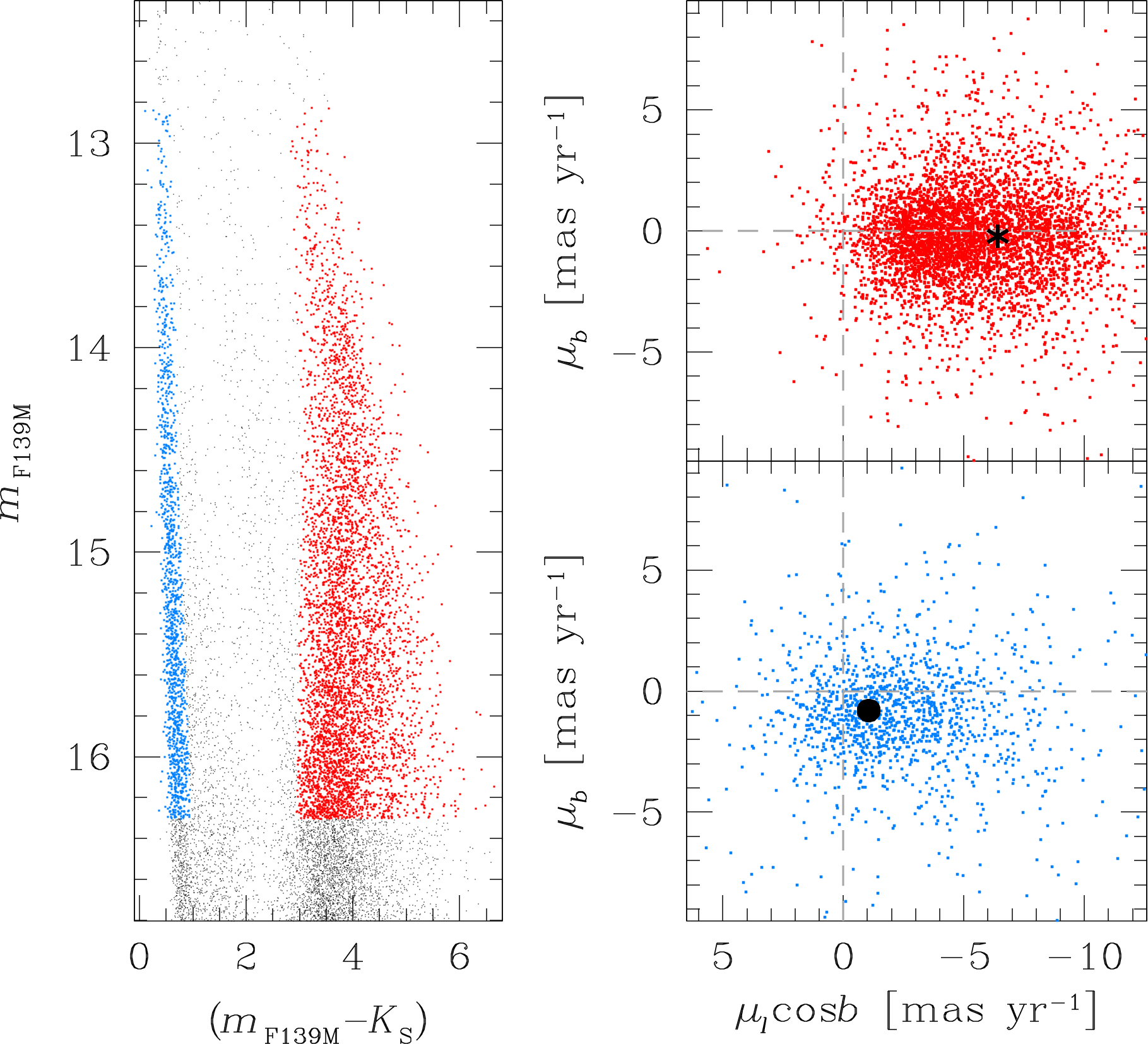}
  \caption{The left panel presents the $m_{\rm F139M}$ versus
    ($m_{\rm F139M}-K_{\rm S}$) CMD of the stars in our WFC3/IR-based
    PM catalog. Disk stars are shown as blue points, while Bulge/Bar
    objects are depicted in red. The corresponding VPDs of the PMs in
    Galactic coordinates are shown in the right panels. The grey,
    dashed lines are set at $\mu_l \cos b$$=$0 \masyr and
    $\mu_b$$=$0 \masyr. The black dot in the bottom-right VPD
    represents the expected PM of Disk stars at 2 kpc from the Sun,
    while the star on the top-right VPD is the PM of Sgr A* (see
    Table~\ref{tab:ref}).}
  \label{fig:figure8}
\end{figure}

We can reach the same conclusions with astrometric arguments.  In the
left panel of Fig.~\ref{fig:figure8}, we selected a sample of bright
stars along the Disk (in blue) and Bulge/Bar (in red) sequences in the
CMD. The VPDs in Galactic coordinates\footnote{In this paper, Galactic
  PM errors are computed following a Monte Carlo approach similar to
  that described in \citet{2020MNRAS.497.4733L}. For each star, we
  used 10\,000 samples of its Equatorial PM. These samples were drawn
  by a Gaussian distribution with average and $\sigma$ equal to the
  absolute PM and errors of the star. Then, we converted the PMs of
  these samples from the Equatorial to the Galactic reference
  system. Finally, we defined as the Galactic PM error of the star the
  standard deviation of the obtained distributions.} in the right
panel of Fig.~\ref{fig:figure8} demonstrate that our PMs are accurate
enough to detect the different motions of the two populations.

We know from the Gaia-DR2 catalog that these Disk stars are, on
average, at a distance of a few kpc from the Sun. According to the
recent work of \citet{2019ApJ...885..131R}, the rotation velocity of
the Galaxy at the distance of these stars is $\sim$237 \kms. By
assuming from the same paper the rotation velocity for the Sun of
$\sim$247 \kms, this means that the Disk stars in the blue sequence of
the CMD should have an apparent motion along the Galactic longitude
$l$ of about $-10$ \kms, i.e., $-1.05$ \masyr. The Sun also has a
motion of $+$7.6 \kms perpendicular to the plane of the Galaxy. If we
assume that our Disk stars have a negligible vertical motion, the
expected apparent motion along the Galactic latitude $b$ of the Disk
stars at 2 kpc should be of $-7.6$ \kms, i.e., $-0.8$ \masyr. In the
bottom-right VPD in Fig.~\ref{fig:figure8}, we plot as a black dot the
expected apparent PM of Disk stars. The good agreement between the
expected motion of Disk stars in our FoV and the PMs of blue-sequence
stars supports the idea that these objects are within a few kpc from
the Sun.

\begin{figure}
  \centering
  \includegraphics[width=\columnwidth,keepaspectratio]{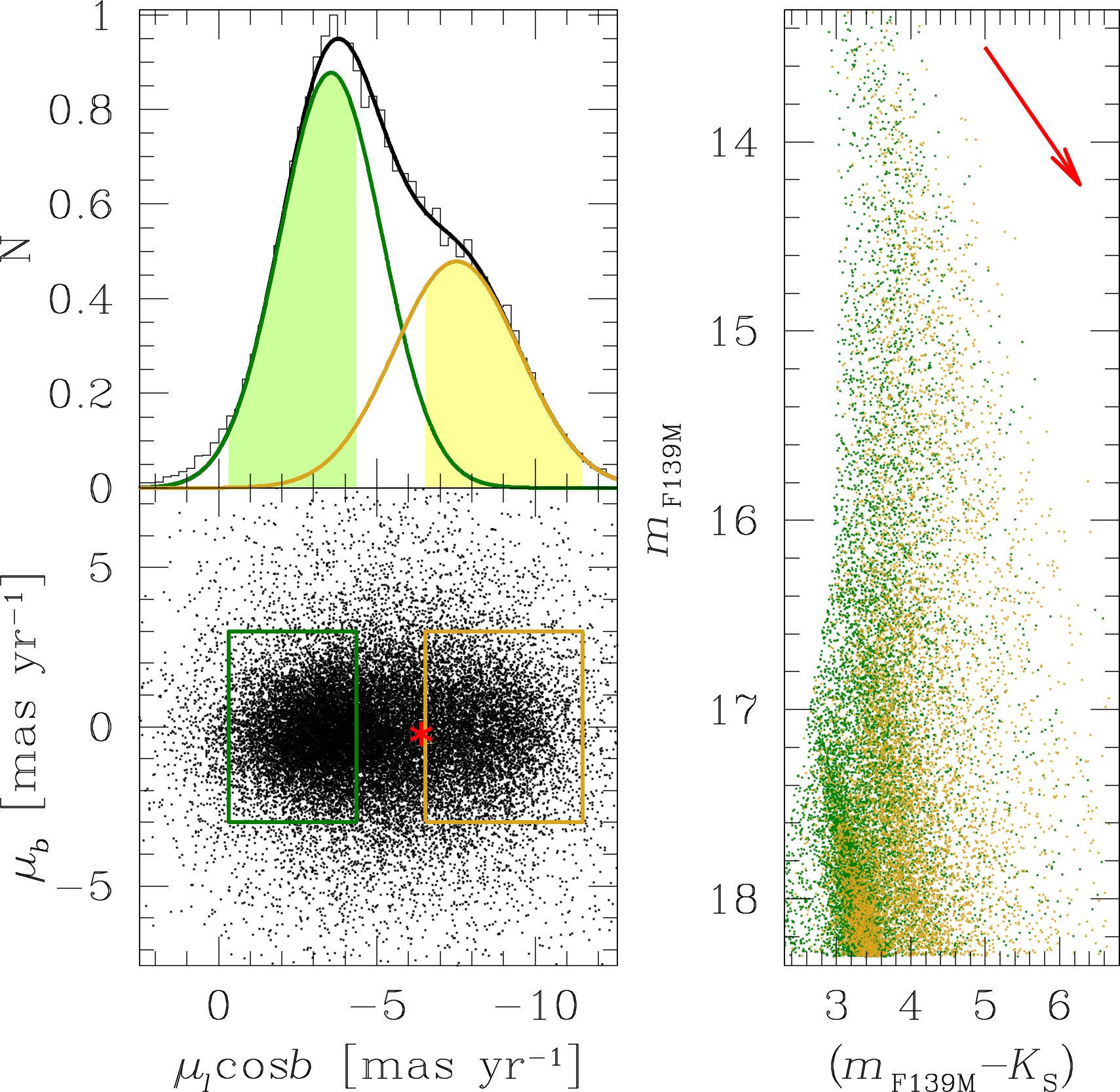}
  \caption{(Top-left panel): histogram of the $\mu_l \cos b$ PMs of
    bright, well-measured Bulge/Bar stars. The dual Gaussian fit to
    the histogram is shown with a black line, while the individual
    Gaussian components are shown in green and yellow. We used these
    Gaussian functions to select two samples of stars for each group
    (shaded pale green and yellow regions). To avoid contamination
    between the two groups, we selected only stars from 0.5$\sigma$ to
    2$\sigma$ from the peaks of the Gaussian functions. (Bottom-left
    panel): VPD of the absolute PMs in Galactic coordinates. The red
    asterisk marks the PM of Sgr A*. (Right panel): $m_{\rm F139M}$
    versus ($m_{\rm F139M}-K_{\rm S}$) CMD. Only stars enclosed in the
    green and yellow rectangles are shown in the CMD and are
    color-coded accordingly. The red arrow represents the reddening
    vector.}
  \label{fig:figure9}
\end{figure}

\begin{figure}
  \centering
  \includegraphics[width=\columnwidth,keepaspectratio]{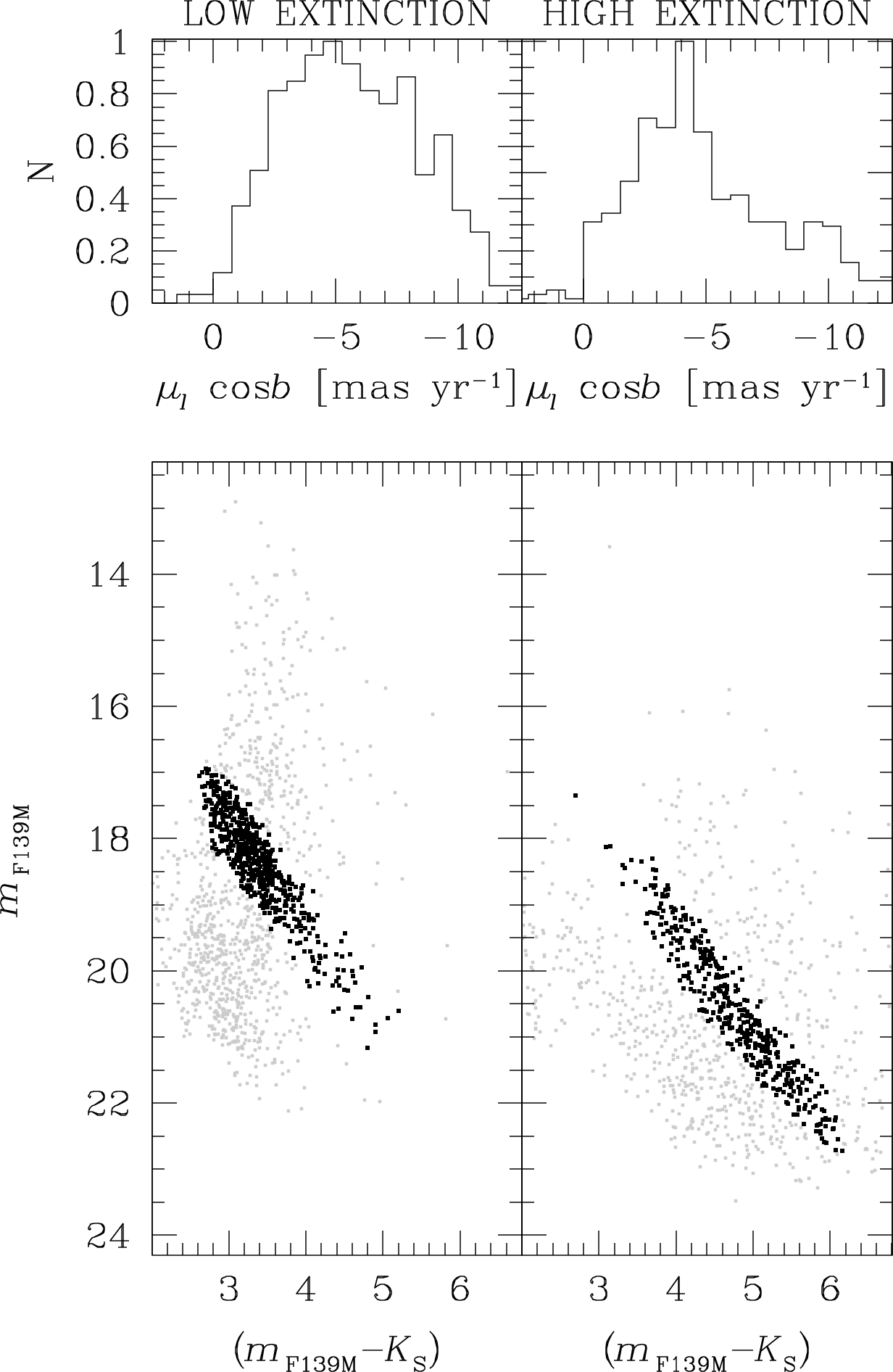}
  \caption{Top panels present the histograms of $\mu_l \cos b$ for
    red-clump stars in low- (left) and high-extinction (right)
    regions. The corresponding $m_{\rm F139M}$ versus
    ($m_{\rm F139M}-K_{\rm S}$) CMDs for stars in the two regions are
    shown in the bottom panels. Black points are red-clump stars,
    while the other grey points represent all other Bulge/Bar stars in
    these regions.}
  \label{fig:figure10}
\end{figure}

Stars belonging to the red sequence have a PM distribution different
from those of the Disk stars (Fig.~\ref{fig:figure8}). The black star
in the top-right VPD marks the absolute PM of Sgr A* from
\citet{2020ApJ...892...39R}. Again, red-sequence stars have PMs
similar to that of Sgr A*, thus suggesting they are part of the same
GC region. Interestingly, the VPD of Bulge/Bar stars highlights the
presence of two different groups of stars, whose distributions are
mainly elongated along the $l$ direction.

To better understand the nature of the two stellar groups in the GC
region, we extended the sample to fainter stars for a better
statistics, made a histogram (with bin width of 0.25 \masyr) of the
PMs along the $l \cos b$ direction, and fit the histogram with a dual
Gaussian function (top-left panel of Fig.~\ref{fig:figure9}). We found
a tighter group (green contour) centered at $\mu_l \cos
b$$\sim$$-3.6$ \masyr, and a broader group (yellow contour) centered
at $\mu_l \cos
b$$\sim$$-7.5$ \masyr. The mean PM of the former group is (just) a few
\masyr ($\sim$100 \kms at the distance of Sgr A*) larger than that of
Sgr A* and might contain stars in front of the GC. The latter group
has a PM similar to that of Sgr A*, and it might represent stars at
the distance of Sgr A*. The $m_{\rm F139M}$ versus
($m_{\rm F139M}-K_{\rm S}$) CMD of these two groups of stars (right
panel) shows that stars with a PM similar to that of Sgr A* are
fainter, suggesting they are further from the Sun than the other
group. Yellow points are also redder than green dots, suggesting that
the uneven dust distribution in the field could also cause the
magnitude-color difference shown in the CMD. However, the direction of
the reddening vector does not seem consistent with the direction of
the color/magnitude offset between yellow and green points, which
means that extinction alone cannot explain the observed offset in the
CMD.

To verify that distance from the GC is the main factor in explaining
the kinematic and photometric differences between the two groups of
stars, we performed a simple test. We selected five regions from the
stacked image with high stellar densities, i.e., with low extinction
(as in panel 1 of Fig.~\ref{fig:figure1}), and five heavily-absorbed
regions (as those in panel 2 of Fig.~\ref{fig:figure1}). Regions with
high extinction should contain mainly stars closer to us, while
regions with low extinction should present a mix of stars at different
distances, up to the GC and possibly beyond. In
Fig.~\ref{fig:figure10}, we plot the CMDs for the stars in each
region. Then we selected red-clump stars in both CMDs. In the top
panels, we show the corresponding histograms (bin width of 0.25
\masyr) of the PMs along the $l \cos b$ direction. The
``high-extinction'' histogram shows a clear drop of stars with
$\mu_l \cos b \lesssim -6$ \masyr with respect to the
``low-extinction'' histogram, as expected if objects with
$\mu_l \cos b \lesssim -6$ \masyr are those closer to (in front or
behind) the GC.

\subsection{Optical view}\label{optical}

\begin{figure}
  \centering
  \includegraphics[width=\columnwidth]{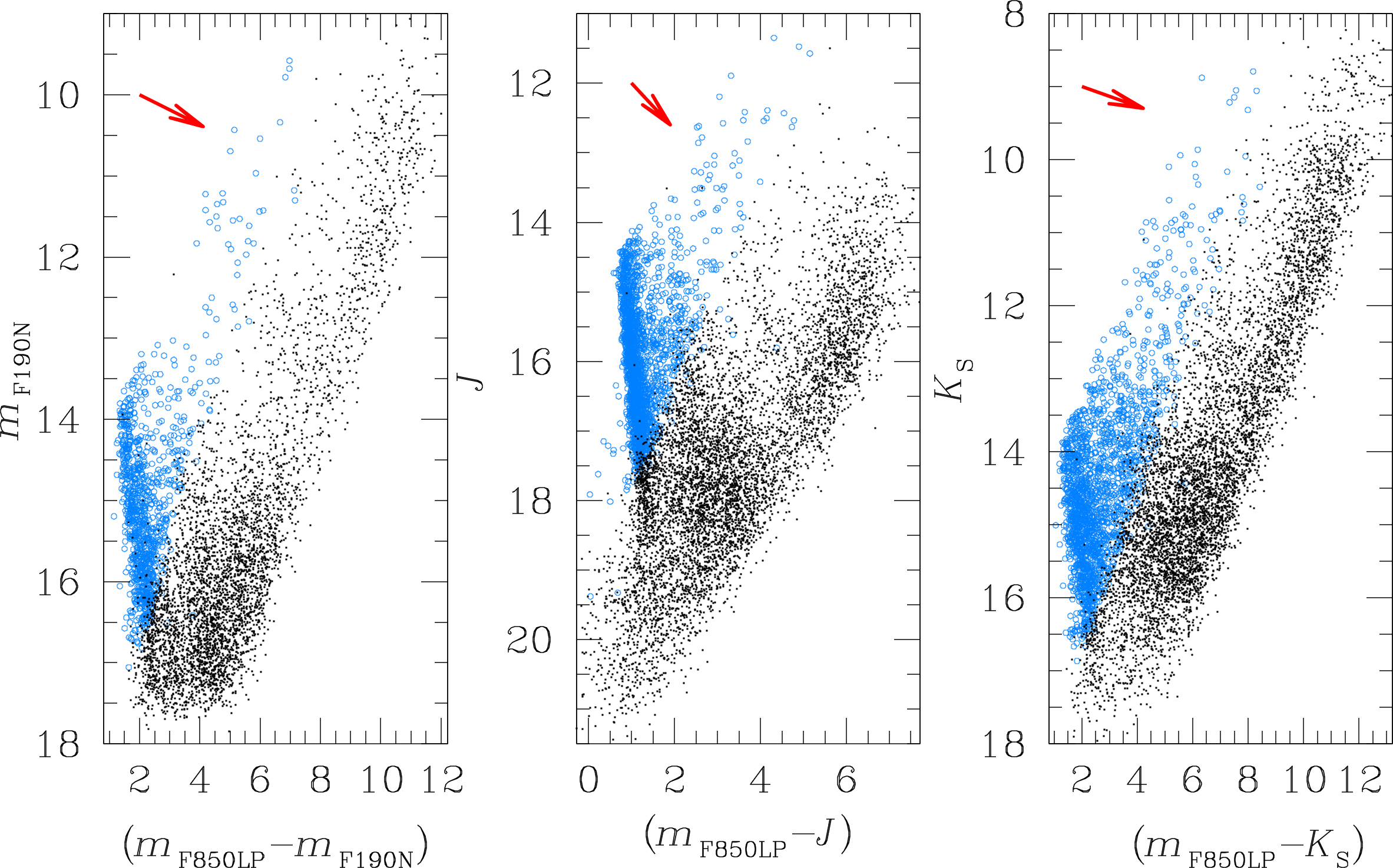}
  \caption{CMDs based on the ACS/WFC data. (Left panel):
    $m_{\rm F190N}$ versus ($m_{\rm F850LP}-m_{\rm F190N}$)
    CMD. (Middle panel): $J$ versus ($m_{\rm F850LP}-J$) CMD. (Right
    panel): $K_{\rm S}$ versus ($m_{\rm F850LP}-K_{\rm S}$) CMD. As in
    Fig.~\ref{fig:figure7}, stars in common with the Gaia-DR2 catalog
    are shown as blue open circles. The red arrows represent the
    direction of the reddening vector and it was computed by assuming
    an arbitrary extinction $A_{\rm F850LP} = 2.5$ in each plot and
    the extinction index $\alpha=2.3$ of \citet{2018A&A...620A..83N}.}
  \label{fig:figure11}
\end{figure}

Figure~\ref{fig:figure11} presents an overview of CMDs made with the
ACS/WFC data. Most of the stars are close-by Disk stars, but some
bright Bulge/Bar objects seem still present. However, the
interpretation of the different features visible in these CMDs is not
straightforward in these color combinations with the F850LP filter.

\begin{figure}
  \centering
  \includegraphics[width=\columnwidth,keepaspectratio]{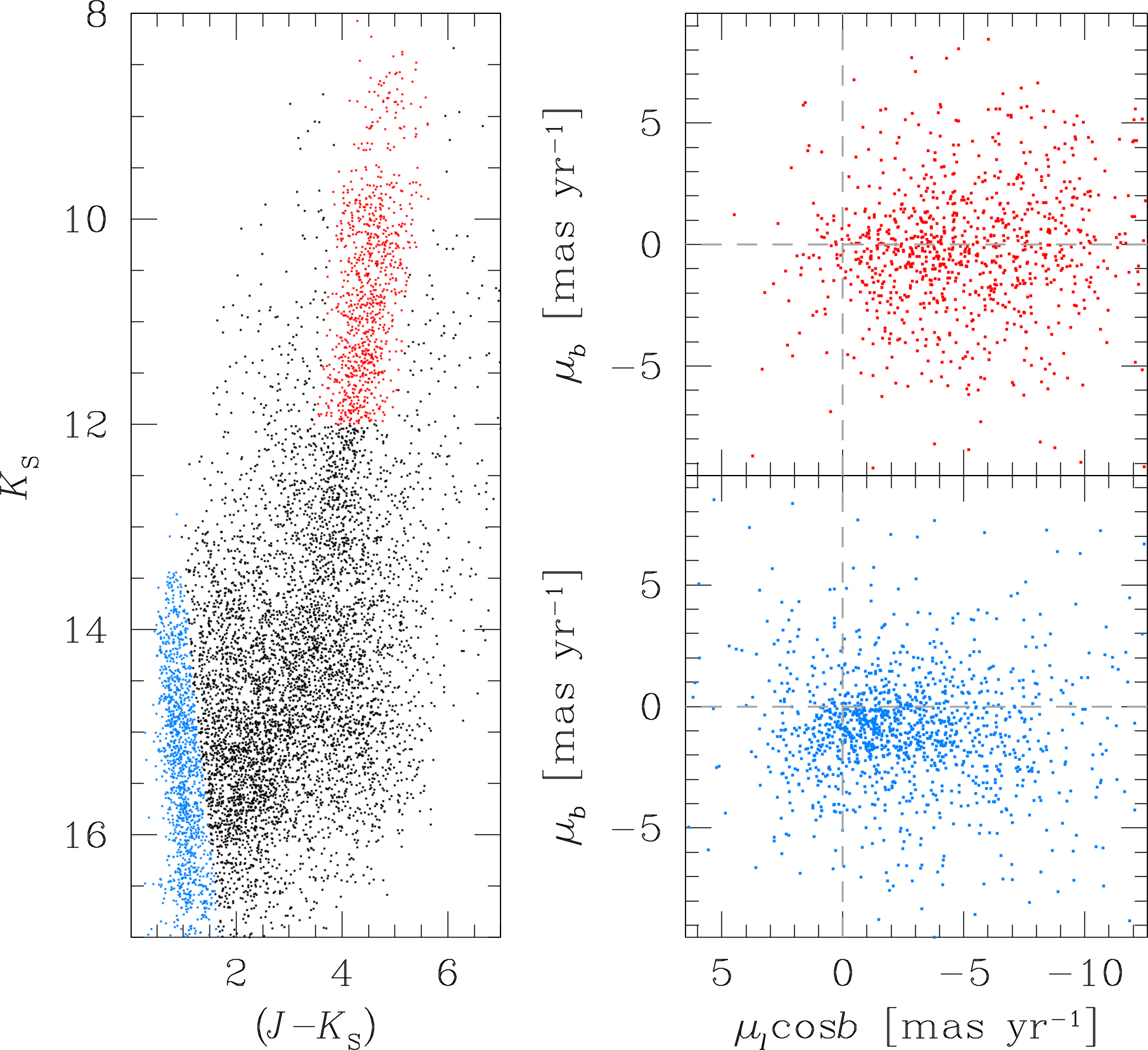}
  \caption{The $K_{\rm S}$ versus $(J-K_{\rm S})$ CMD of the stars in
    the ACS/WFC data for which we have a PM measurement is shown in
    the left panel. Blue and red points highlight a sample of Disk
    objects and a group of Bulge/Bar stars. The right panels present
    the VPDs of the PMs in Galactic coordinates for the Bulge/Bar (top
    panel) and Disk (bottom panel) stars.}
  \label{fig:figure12}
\end{figure}

In the left panel of Fig.~\ref{fig:figure12}, we plot the $K_{\rm S}$
versus $(J-K_{\rm S})$ CMD for the stars in the ACS/WFC data for which
we have PM measurements. We selected a sample of stars in the Disk
sequence and one in the brightest end of the redder sequence of the
CMD. The right panels of Fig.~\ref{fig:figure12} present the VPDs of
the PMs of the blue and red stars. As in Fig.~\ref{fig:figure8}, their
PMs show the different kinematics of the these stars.

\begin{figure}
  \centering
  \includegraphics[width=\columnwidth,keepaspectratio]{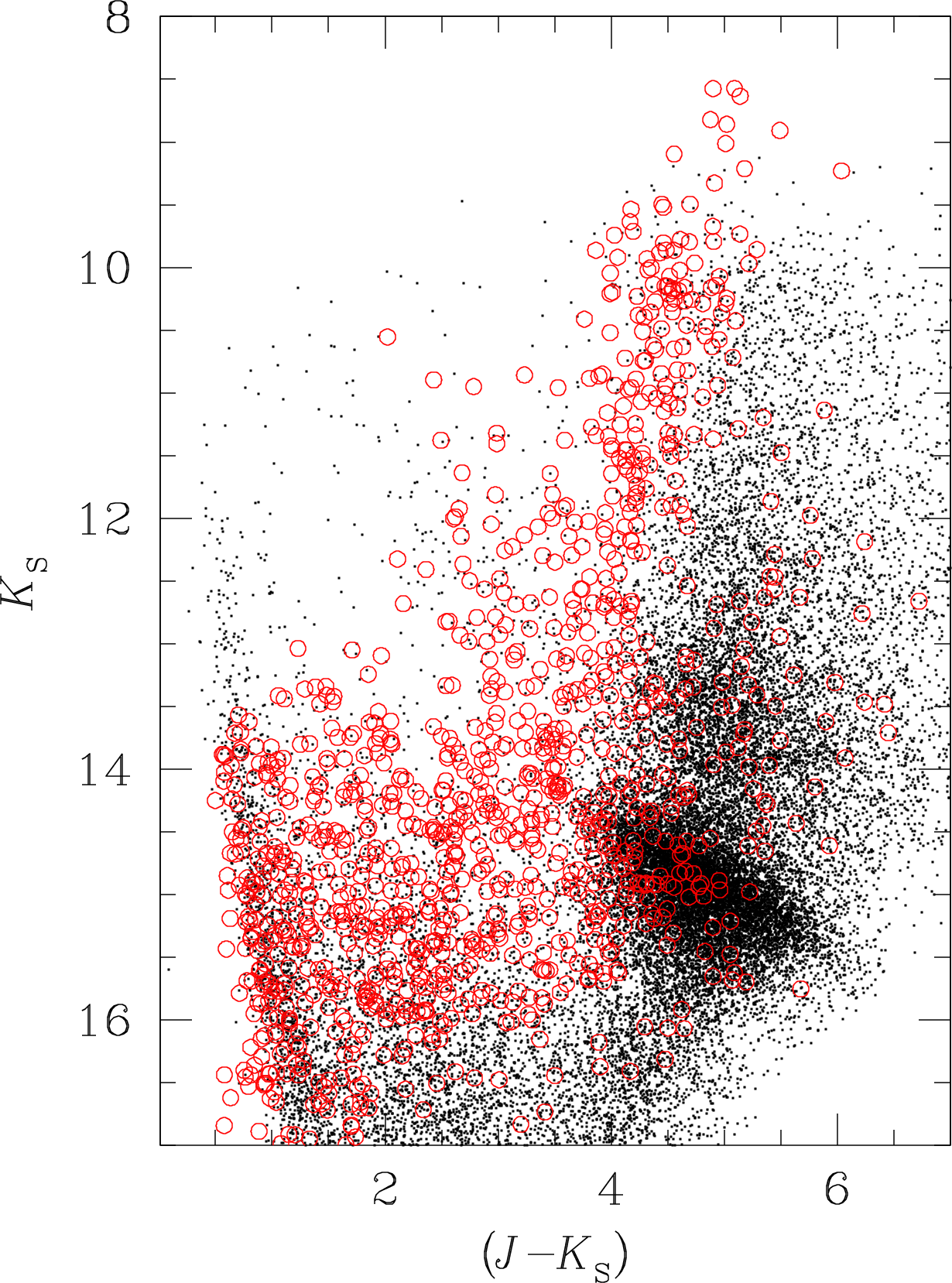}
  \caption{The $K_{\rm S}$ versus $(J-K_{\rm S})$ CMD of the stars in
    the WFC3/IR data (black points) and in the ACS/WFC data (red, open
    circles).}
  \label{fig:figure13}
\end{figure}

Bulge/Bar stars in the GC region are faint in F850LP filter and the
only stars in common between the ACS/WFC data and the VVV catalog are
probably bright stars in front of the GC, closer to the Sun than those
visible in the WFC3/IR FoV. Figure~\ref{fig:figure13} shows the
$K_{\rm S}$ versus $(J-K_{\rm S})$ CMD of stars in the WFC3/IR FoV
(black points) and in the ACS/WFC data (red, open circles). Stars with
$(J-K_{\rm S}) \lesssim 4$ are present in both WFC3/IR and ACS/WFC
data. Stars redder than $(J-K_{\rm S}) \sim 4$ in common between the
WFC3/IR and ACS/WFC catalogs are only those in the bright blue side of
Bulge/Bar sequence, as expected if Bulge/Bar stars in ACS/WFC field
are closer and less reddened than those in the WFC3/IR data.

\begin{table*}
  \caption{Position and PMs of Sgr A*, the Arches and the Quintuplet.}
  \centering
  \label{tab:ref}
  \begin{threeparttable}
    \begin{tabular}{ccccc}
      \hline
      \hline
      Object & R.A. & Dec. & ($\mu_\alpha \cos\delta$,$\mu_\delta$) & ($\mu_l \cos b$,$\mu_b$) \\
             & deg & deg & \masyr & \masyr \\
      \hline
      Sgr A* & 266.4168371 & $-29.00781056$ & $(-3.156\ \pm\ 0.006,-5.585\ \pm\ 0.010)$ & $(-6.411\ \pm\ 0.008,-0.219\ \pm\ 0.007)$ \\
      Arches & 266.4604 & $-28.8244$ & $(-1.45\ \pm\ 0.23,-2.68\ \pm\ 0.14)$ & $(-3.05\ \pm\ 0.17,-0.16\ \pm\ 0.20)$ \\
      Quintuplet & 266.5578 & $-28.8300$ & $(-1.19\ \pm\ 0.09,-2.66\ \pm \ 0.18)$ & $(-2.89\ \pm\ 0.16,-0.38\ \pm \ 0.12)$ \\
      \hline
    \end{tabular}  
    \begin{tablenotes}
    \item \textbf{Notes.} (i) Position and PMs of Sgr A* are from
      \citet{2020ApJ...892...39R}. (ii) Positions of the Arches and
      the Quintuplet clusters are from the Simbad database, while PMs
      are from \citet{2020MNRAS.497.4733L}.
    \end{tablenotes}
  \end{threeparttable}
\end{table*}

\subsubsection{The Arches cluster}\label{arches}

Recently, \citet{2020MNRAS.497.4733L} linked the PM catalog of
\citet{2015A&A...578A...4S} to an absolute system by means of the
Gaia-DR2 catalog, and measured the absolute PM of the Arches as:
\begin{equation}
  \begin{gathered}
    (\mu_\alpha \cos\delta,\mu_\delta)^{\rm Arches\,-\,L20} \\
    = \\
    (-1.45\ \pm\ 0.23,-2.68\ \pm\ 0.14) \textrm{ \masyr.}
  \end{gathered}
  \label{eq:arches1}
\end{equation}

The Arches cluster is present in the North-East region of the FoV and
it is covered by 2 ACS/WFC images per epoch. As an independent check
of our PMs, we computed the absolute PM of the Arches using our
\hst-based PMs and compared this new estimate with that of
\citet{2020MNRAS.497.4733L}.

Most of Arches' members are not detected by \texttt{KS2} because they
are either too faint or did not pass the detection criteria of
\texttt{KS2} (see discussion in Sect.~\ref{datared}).
\citet{2018A&A...617A..65C} identified MSs likely members of the
Arches cluster according to their LOS velocities. Some of these MSs
are also present in our catalog (see later discussion in
Sect.~\ref{MS}). The average PM of these MSs measured in two images
per epoch, which represents an estimate of the absolute PM of the
Arches, is:
\begin{equation}
  \begin{gathered}
    (\mu_\alpha \cos\delta,\mu_\delta)^{\rm Arches\,-\,This\,work\,(1)} \\
    = \\
    (-1.24\ \pm\ 0.78,-2.44\ \pm\ 0.68) \textrm{ \masyr.}
  \end{gathered}
  \label{eq:archesdirect}
\end{equation}
This value is in agreement with the estimate of
\citet{2020MNRAS.497.4733L} at the 1$\sigma$ level, although our PM
errors are larger because of the faintness (i.e., less-precise PMs) of
our reference objects.

We also took advantage of Disk stars in our ACS/WFC PM catalog and
converted the PMs of \citet{2015A&A...578A...4S} from relative to
absolute values following the prescription of
\citet{2020MNRAS.497.4733L}. Our second independent estimate of the
absolute PM of the Arches is:
\begin{equation}
  \begin{gathered}
    (\mu_\alpha \cos\delta,\mu_\delta)^{\rm Arches\,-\,This\,work\,(2)} \\
    = \\
    (-1.34\ \pm\ 0.41,-2.11\ \pm\ 0.20) \textrm{ \masyr.}
  \end{gathered}
  \label{eq:arches2}
\end{equation}
The agreement with the estimate of \citet{2020MNRAS.497.4733L} is at
the $\sim$2$\sigma$ level. The average PM error of the Gaia stars used
in the work of \citet{2020MNRAS.497.4733L} for the
relative-to-absolute conversion is about 0.43 \masyr, while that of
the stars in our PM catalog used for the same task is 0.76
\masyr. Therefore, we chose to keep the original PM estimate of
\citet{2020MNRAS.497.4733L} as the absolute PM of the Arches in the
rest of the paper.

\begin{figure*}
  \centering
  \includegraphics[height=11.0cm]{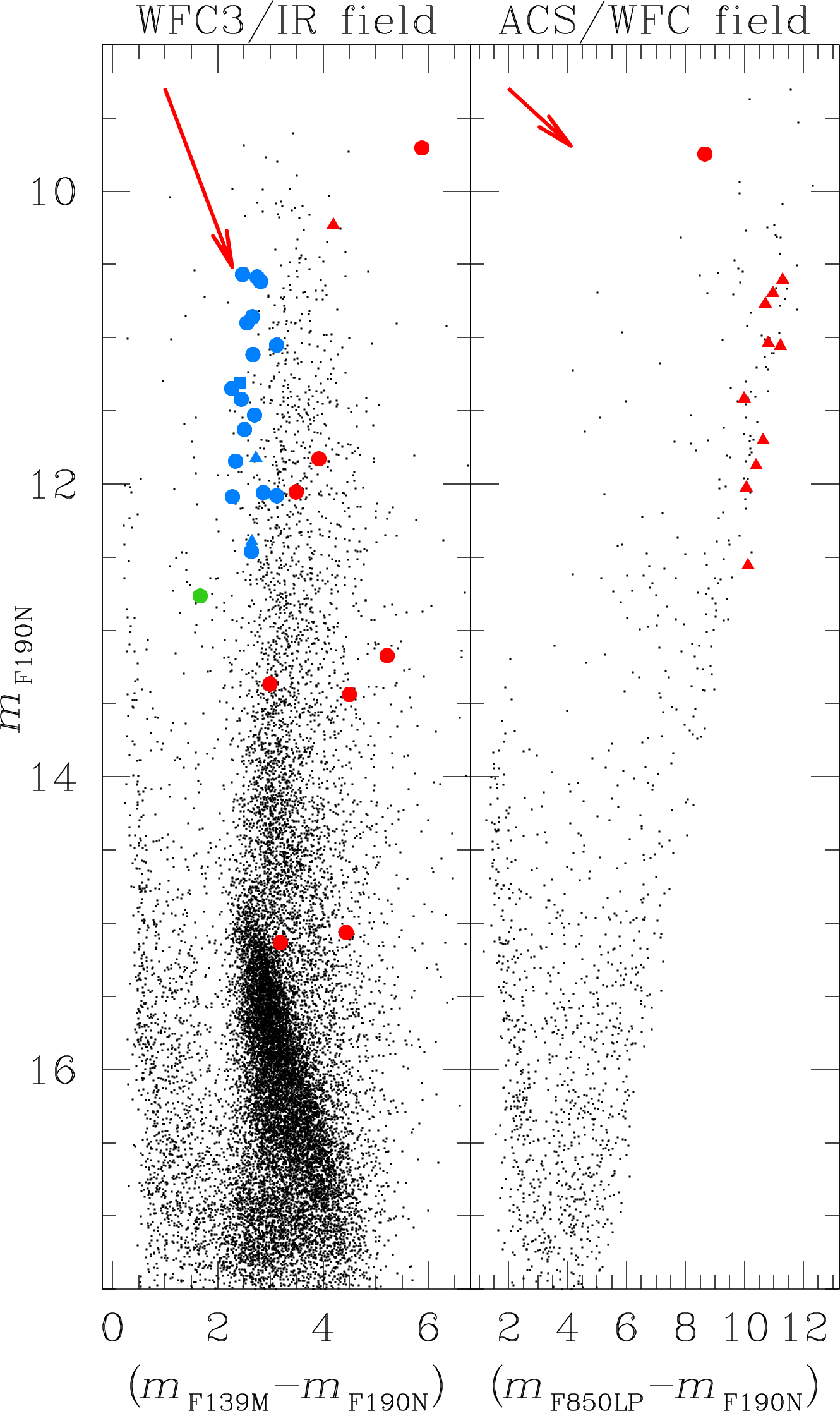}
  \includegraphics[height=11.0cm]{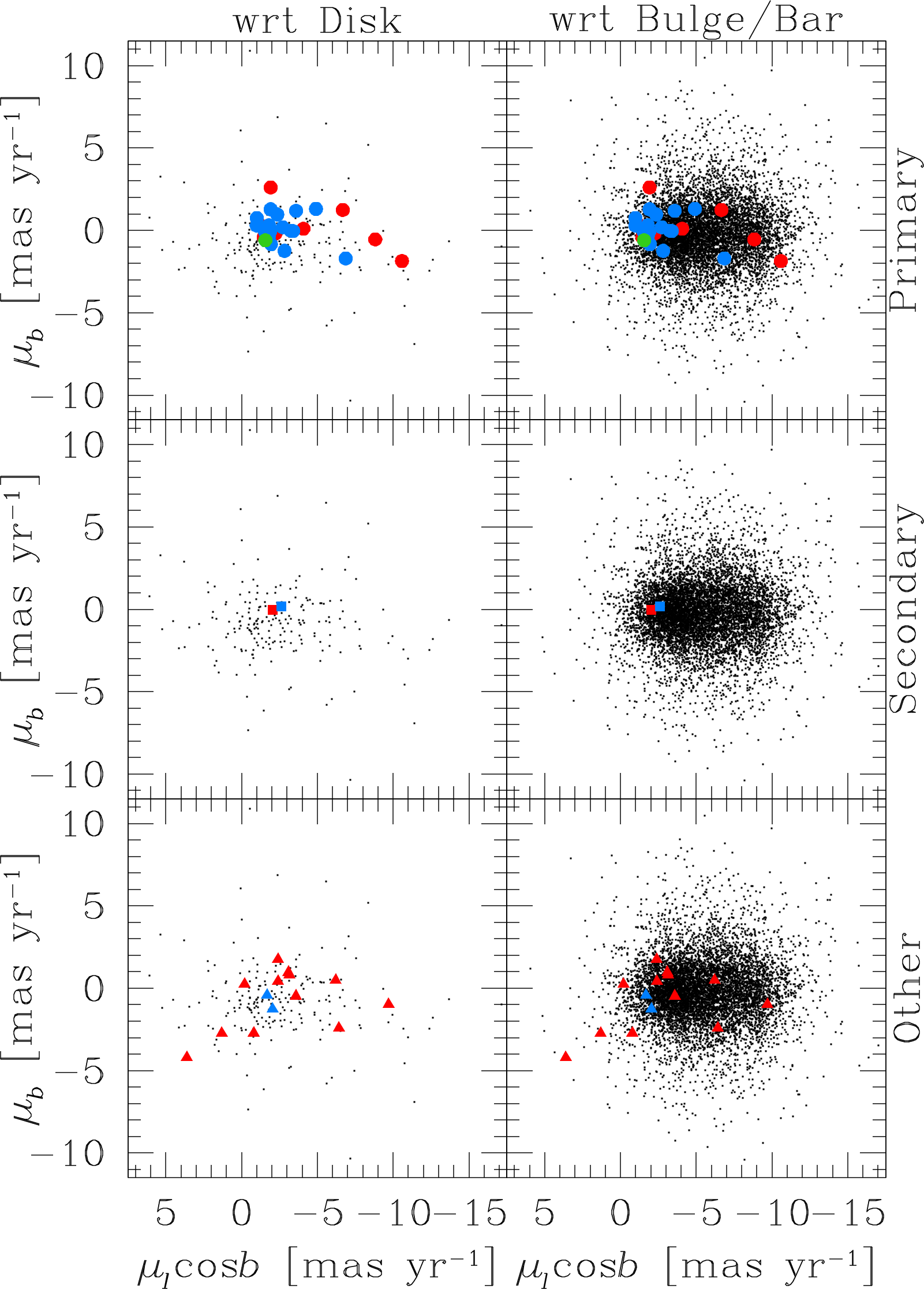}
  \caption{In the left and middle-left panels, we plot the
    $m_{\rm F190N}$ versus ($m_{\rm F139M}-m_{\rm F190N}$) CMD for the
    stars in the WFC3/IR field and the $m_{\rm F850LP}$ versus
    ($m_{\rm F850LP}-m_{\rm F190N}$) CMD for the stars in the ACS/WFC
    field, respectively. We selected three samples of objects in the
    CMDs: green points refer to stars bluer than the Bulge/Bar
    sequence, while blue (red) points are the bluest (reddest) MSs in
    the Bulge/Bar sequence. Circles, squares and triangles represent
    confirmed MSs in the Primary, Secondary and Other samples,
    respectively. The middle-right and right panels show the VPDs of
    the PMs of the MSs in the Primary (top), Secondary (middle) and
    Other (bottom) lists compared to the samples of Disk and Bulge/Bar
    stars defined in Sect.~\ref{nir}, respectively.}
  \label{fig:figure1415}
\end{figure*}

\section{Kinematics of massive stars}\label{MS}

Most MSs in our field were identified by \citet{2011MNRAS.417..114D}
as part of the NICMOS survey of
\citet{2010MNRAS.402..895W}. \citet{2011MNRAS.417..114D} found several
Paschen $\alpha$ emitting candidates, most of which are evolved
massive stars with strong winds. The authors released a list of 152
confirmed Paschen $\alpha$ emitters (hereafter Primary list) and a
list of 189 potential candidates for which it was not possible to
perform a clear classification (Secondary list).

\begin{figure}
  \centering
  \includegraphics[width=0.875\columnwidth]{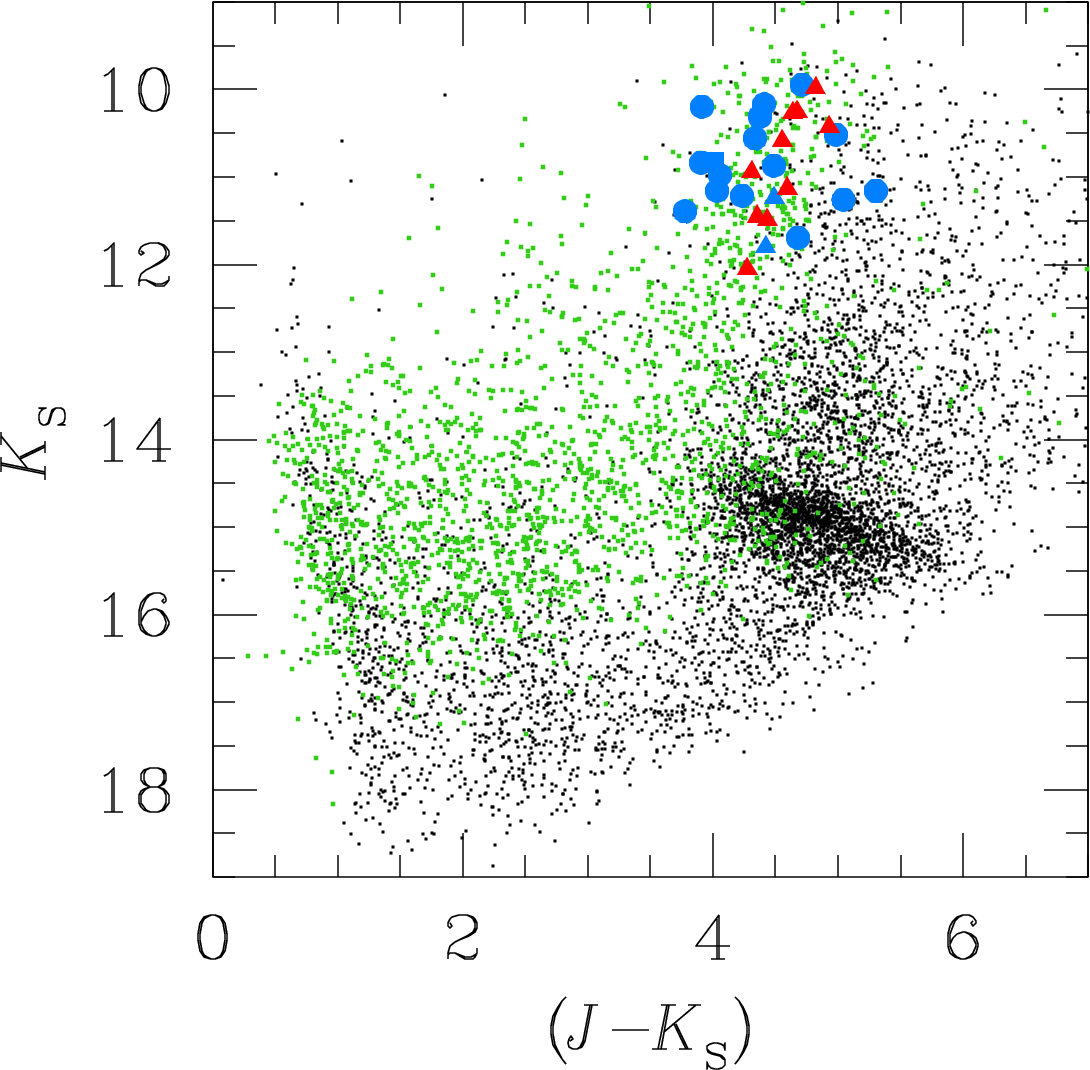}
  \caption{$K_{\rm S}$ versus $(J-K_{\rm S})$ CMD. Black points are
    stars in the WFC3/IR field, while green points represent stars in
    the ACS/WFC field. Blue points are the MSs in the blue side of the
    Bulge/Bar sequence as shown in Fig.~\ref{fig:figure1415}. The MSs
    that are likely members of the Arches according to
    \citet{2018A&A...617A..65C} are shown as red triangles.}
  \label{fig:figure16}
\end{figure}

We also included MSs from Clark et al. (submitted). Recently, Clark et
al. (submitted) undertook a NIR spectroscopic survey of the Arches,
Quintuplet and diffuse MS population in the GC region using SINFONI
and KMOS on the ESO Very Large Telescope (VLT). The list of their
targets comprises two subsets. The first subset is made by MSs that
were previously classified as such from low S/N and/or spectral
resolution observations
\citep{1999ApJ...510..747C,2006ApJ...638..183M,2007ApJ...662..574M,2010ApJ...725..188M,2010ApJ...710..706M,2013AJ....146..109D,2015MNRAS.446..842D,2019ApJ...872..103G}. These
stars have the following characteristics: (i) have a Paschen $\alpha$
excess indicative of a powerful stellar wind
\citep{2011MNRAS.417..114D}, (ii) are associated with an X-ray source
indicative of a colliding wind binary \citep{2009ApJ...703...30M}, or
(iii) their mid-IR properties are indicative of a highly luminous,
potentially dust post-main-sequence star
\citep{2019ApJ...872..103G}. The second subset comprises the remaining
candidates characterized by pronounced Paschen $\alpha$ emission from
the catalog of \citet{2011MNRAS.417..114D}, but they have not been
spectroscopically classified yet. Other 17 MSs were returned from this
cohort, with the remaining $\sim$30 stars comprising cool, low mass
interlopers along the line of sight. In addition to these lists, we
also considered MSs in the Arches cluster from
\citet{2018A&A...617A..65C}. We refer to the collection of MSs in our
field not included in the Primary and Secondary lists of Paschen
$\alpha$ emitters of \citet{2011MNRAS.417..114D} simply as Other list.

The spectral-type characterization made by Clark et al. (submitted)
allow us to define three samples of objects: confirmed MSs, candidate
MSs and non-massive objects. In total, we measured the PM of 43
confirmed MSs, 64 candidate MSs and five non-massive objects. The list
of these objects is presented in Table~\ref{tab:mslist}. In the
following, we discuss in detail the kinematics of the confirmed MSs,
while we provide a shorter analysis for the candidate MSs and the
non-massive stars. Not all these objects, especially those at the
faint-end of our catalogs, passed all the astro-photometric quality
selections described in Sects.~\ref{datared} and \ref{pm}, but we
study their kinematics anyway. In Table~\ref{tab:ref}, we report
positions and PMs of Sgr A* \citep[from][]{2020ApJ...892...39R}, the
Arches and the Quintuplet \citep[from][]{2020MNRAS.497.4733L} as a
reference.

\subsection{Confirmed MSs}\label{confirmed}

\subsubsection{Population structure and Distribution}\label{confirmed_pop}

The location of confirmed MSs in the CMD (left and middle-left panels
of Fig.~\ref{fig:figure1415}) suggests that almost all MSs have a
Bulge/Bar origin, with only one possible interloper (green
point). Blue and red points mark MSs in the blue and red side of the
Bulge/Bar sequence in the CMD, respectively. This selection is not
possible in the ACS/WFC field and we simply plot probable Bulge/Bar
MSs redder than the arbitrarily color
$(m_{\rm F850LP}-m_{\rm F190N}) = 6.2$ as red points. Middle-right and
right panels of Fig.~\ref{fig:figure1415} present a comparison between
the PMs of the MSs in the Primary, Secondary and Other lists and the
PMs of stars in the Disk (middle-right panels) and Bulge/Bar (right
panels) samples described in Sect.~\ref{nir}. We considered Disk and
Bulge/Bar objects in the same magnitude interval of the MSs
($m_{\rm F190N} \lesssim 16$). The green point (star \# 18668 in the
Primary list) has a PM similar to that of the Disk or Bulge/Bar
stars. However, its position in different CMDs makes it challenging to
infer its membership.

\begin{figure*}
  \centering
  \includegraphics[width=0.8\textwidth]{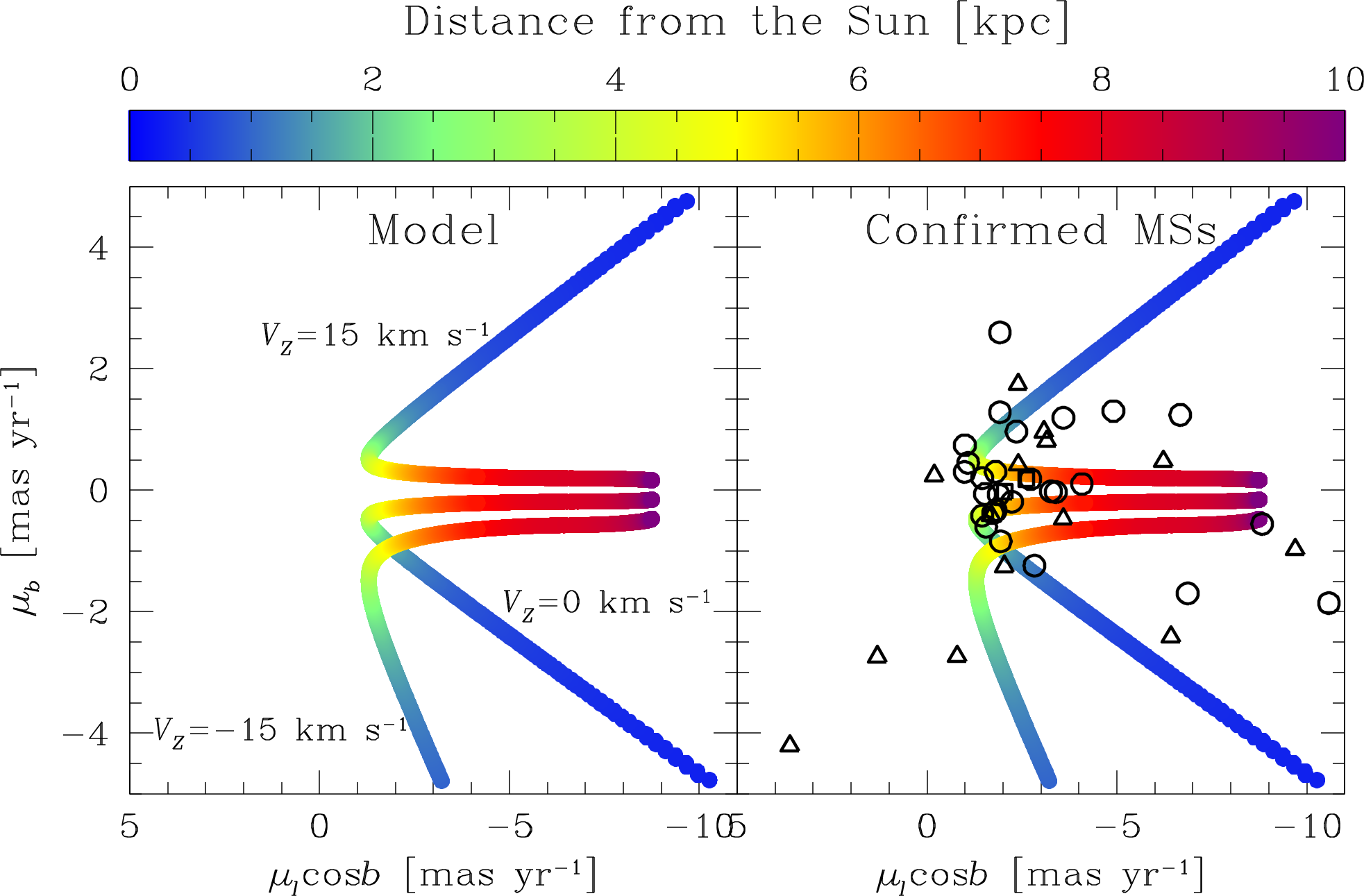}
  \caption{In each panel, we plot the VPDs of the Galactic PMs for a
    set of simulated Disk stars in the Galactic plane (see the text
    for details), color-coded according to the distance from the
    Sun. Black points in the right VPD show the PMs of the confirmed
    MSs. Open circles, squares and triangles refer to objects in the
    Primary, Secondary and Other lists, respectively.}
  \label{fig:figure17}
\end{figure*}

At a first glance, MSs have kinematic properties different from those
of Disk stars within a few kpc from the Sun. The comparison between
the PM distributions of the MSs and Bulge/Bar objects is not
straightforward. Bulge/Bar objects along the line of sight are located
in a wide range of distances from Sgr A*, either in front or behind
it, which reflect in different PM distributions along the $l \cos b$
direction. On the other hand, the PM distributions of Bulge/Bar
objects at various distances from Sgr A* along the $b$ direction in
our VPD seem rather similar. We compared the $\mu_b$ PM distribution
of MSs with those of Bulge/Bar objects. We selected only stars
brighter than $m_{\rm F190N} = 13$ and with a PM error lower than 1
\masyr, and measured their velocity dispersions $\sigma_b$. We find
that Bulge/Bar stars have $\sigma_{\mu_b} = (2.06 \pm 0.02)$ \masyr,
while MSs have $\sigma_{\mu_b} = (0.95 \pm 0.17)$ \masyr. The velocity
dispersion of the MSs shown in blue in Fig.~\ref{fig:figure1415} is
$\sigma_{\mu_b} = (0.79 \pm 0.17)$ \masyr. These values suggest that
most MSs are a distinct population from the rest of the Bulge/Bar
stars chosen for the comparison. Specifically, their lower
$\sigma_{\mu_b}$ (which corresponds to $\sim$30 \kms at the distance
of the GC) indicates that they are a near-planar population that is
not as vertically extended as the Galactic Bulge. The $\sigma_{\mu_b}$
of selected Bulge/Bar stars differs from what we would expect for
stars very close to the GC ($\sim$3 \masyr). Because of the direction
of the reddening vector and the magnitude cut at $m_{\rm F190N} = 13$,
we are likely selecting objects in front of Sgr A* (see
Sect.~\ref{nir} and CMDs in Fig.~\ref{fig:figure10}) that have a
different velocity dispersion than the Bulge population
\citep[e.g.,][]{2008ApJ...684.1110C}.

\begin{figure}
  \centering
  \includegraphics[width=\columnwidth]{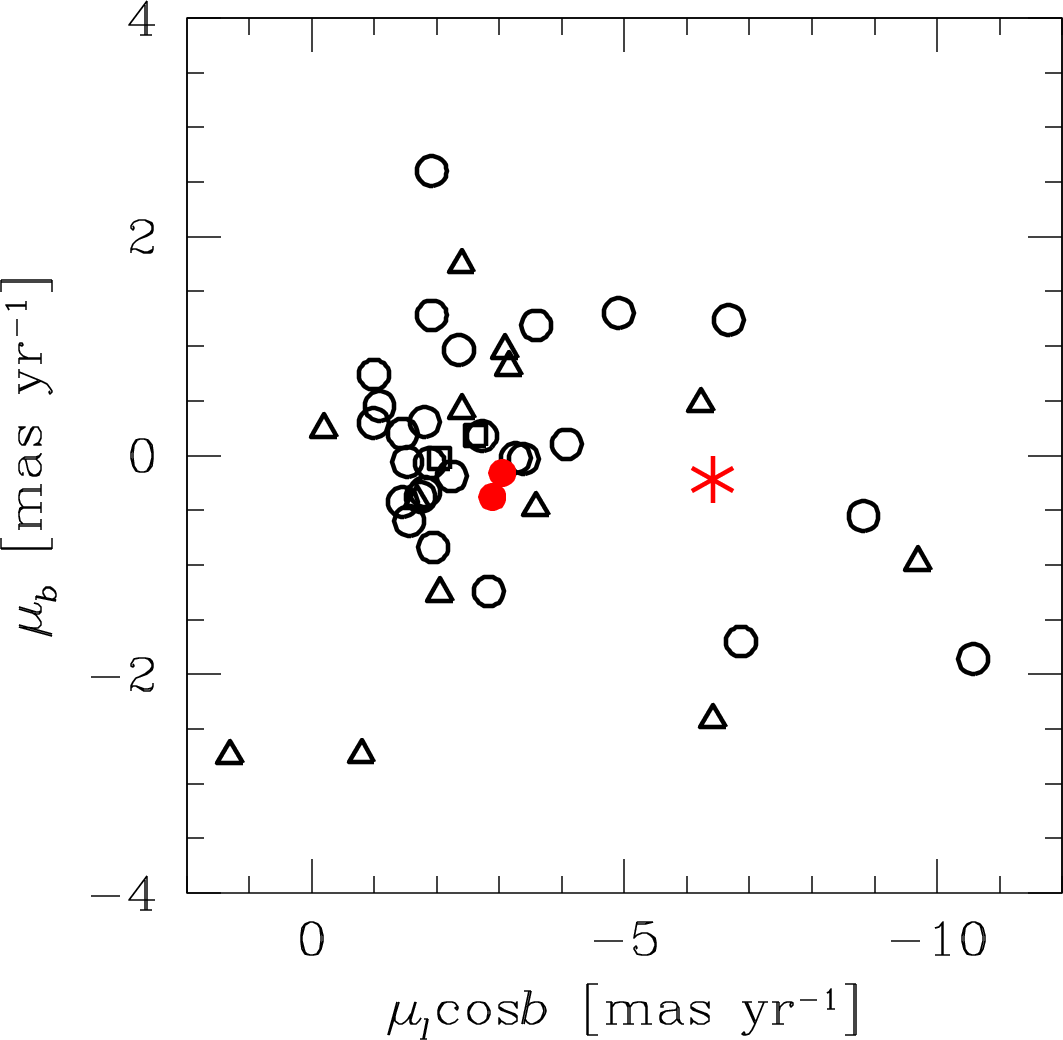}
  \caption{VPDs of the PMs of the confirmed MSs (black points). The
    red star marks the PM of Sgr A*, while the two red points indicate
    the PMs of the Arches and the Quintuplet clusters. As in
    Fig.~\ref{fig:figure1415}, circles, squares and triangles
    represent MSs in the Primary, Secondary and Other samples,
    respectively.}
  \label{fig:figure18}
\end{figure}

The MSs in blue in Fig.~\ref{fig:figure1415} seem part of a defined
sequence in the CMD, and are clustered together in the VPD.
Figure~\ref{fig:figure16} shows the $K_{\rm S}$ versus $(J-K_{\rm S})$
CMD for the stars in the WFC3/IR field (black points). Green points
represent stars in the ACS/WFC field. As discussed in
Sect.~\ref{optical}, stars in the ACS/WFC catalog in common with the
ancillary catalogs are Bulge/Bar objects closer to the Sun than those
in the WFC3/IR catalog. Again, blue points are the MSs in the blue
side of the Bulge/Bar sequence. The color width of these MSs in the
CMD is similar to that of the Bulge/Bar stars shown in green with
$(J-K_{\rm S}) \gtrsim 3.5$. This piece of evidence suggests that the
MSs shown in blue in Fig.~\ref{fig:figure1415} could be in front of
the GC, closer to the Sun than all other confirmed MSs, and could be
part of the general ``field'' population. Red points in
Fig.~\ref{fig:figure16} represent MSs likely members of the Arches
from \citet{2018A&A...617A..65C}. The similar CMD locations for these
two groups of star could instead indicate that isolated and Arches'
MSs are at the same distance. The different colors of MSs and
Bulge/Bar objects shown in black in Fig.~\ref{fig:figure16} could
reflect the different intrinsic nature of the two groups of stars
(young, massive objects versus old, low-mass stars). If the Arches is
in the GC, isolated MSs shown in blue in Fig.~\ref{fig:figure1415} are
in the GC as well, although they are not part of a cluster but general
field objects. Finally, the relative position of Arches and Bulge/Bar
stars in our CMD is qualitatively in agreement with those in the CMD
shown in \citet{2019ApJ...870...44H}.

The MSs in our field could be born in star-forming regions in the
Galaxy spiral arms and be just projected toward the GC, although it
seems unlikely since they lie on the Bulge/Bar sequence on the
CMDs. We investigated this hypothesis as follows. We simulated 1000
stars with the same Equatorial coordinates, equal to the average
(R.A.,Dec.) of the MSs in our field, but different distances from the
Sun (0--10 kpc with steps of 0.01 kpc). The Galactocentric motion
$(V_X,V_Y,V_Z)$ of each object was assigned by means of the Galactic
rotation curve measured by \citet{2019ApJ...885..131R}. Specifically,
we used their best model ``A5'' and the code in their Appendix~B
\citep[rescaled by the distance of Sgr A* of 8.178 kpc
of][]{2019A&A...625L..10G} to compute the circular rotation as a
function of Galactocentric distance. We also simulated a peculiar
motion for each source by introducing a velocity component along the
radial direction from the GC (between $-15$ and 15 \kms). We
decomposed these radial and circular velocities of each star in the
$(V_X,V_Y)$ velocities. $V_Z$ was varied between $-15$ and 15 \kms,
again to simulate a peculiar motion. Finally, we converted
$(V_X,V_Y,V_Z)$ in to Galactic PMs and compared them to the MS PMs. We
neglected the influence of the Galactic Bar at small Galactocentric
distances since we are mainly interested in the connection between MSs
and the Galactic spiral arms. The result is shown in the left panel of
Fig.~\ref{fig:figure17}. Simulated points are color-coded according to
their distance from the Sun (we plot only the extreme cases of radial
Galactocentric motion of $-15$ and 15 \kms and $V_Z = -15$ \kms or
$V_Z = 15$ \kms for clarity).

The comparison between the PMs of the confirmed MSs (right panel) and
our simulations shows that most MSs in the VPD have PMs that are not
consistent with those of the Disk stars within 4 kpc from the Sun. If
MSs are distant Disk stars and the discrepancy between observations
and simulations is mainly due to unaccounted for peculiar motions or
to the influence of the Bar, Fig.~\ref{fig:figure17} seems to imply
that most MSs are located in the innermost $\sim$3--4 kpc of the
Galaxy. As stated in \citet{2019ApJ...885..131R}, star-forming regions
in the 3-kpc arm are likely associated to the Bar of the Galaxy rather
than being a true Spiral arm, while maser sources in the 4-kpc
(``Norma'') arm have large peculiar motion because they are near the
end of the Bar. These pieces of evidence support a Bulge/Bar origin
for most confirmed MSs. However, a distance of 3--4 kpc from the GC is
unlikely for the following reasons. First, these stars are very bright
and should have been visible in the Gaia catalog if located at a
distance of 3--4 kpc from the GC. Second, some isolated MSs have been
associated to radio, X-ray and/or mid-IR features known to be in the
GC \citep[e.g.,][]{1996ApJ...461..750C}. Finally, the extinction for
these objects is larger than $A_V \gtrsim 20$ mag
\citep{2009ApJ...703...30M}, which would require an unusual, ad-hoc
high extinction for spiral-arm objects at 3--4 kpc from Sgr A*.
Therefore, the 3--4 kpc has to be considered an upper limit for the
distance of MSs from Sgr A*.

Unfortunately, as we still lack distances, we cannot a priori discard
origins within or outside the GC region. In the following, we assume
these MSs to be close to the GC and investigate if any of them are
associated with either Sgr A*, the Arches or the Quintuplet.

\begin{table*}
  \caption{Overview of the confirmed MSs selected as escaping
    candidates.}
  \centering
  \label{tab:confirm}
  \begin{threeparttable}
    \setlength\tabcolsep{3 pt}
    \begin{tabular}{ccccccccc}
      \hline
      \hline
      ID & $\Delta$PM & $\theta$ & Flight time & Possible origin & List & Spectral Type\\
         & \masyr & deg & Myr & & & & \\
      \hline
      17180 & $4.89 \pm 0.37$ & $-8.6 \pm 0.1$ & $0.10 \pm 0.01$ & Sgr A* & Primary & WC9 \\
      20255 & $4.73 \pm 0.28$ & $11.1 \pm 0.1$ & $0.11 \pm 0.01$ & Sgr A* & Primary & B1-2 Ia$+$/WNLh \\
      20612 & $3.15 \pm 0.49$ & $-4.7 \pm 0.2$ & $0.19 \pm 0.01$ & Sgr A* & Primary & O6-7 Ia$+$ \\
      \hline
      2061 & $6.71 \pm 0.62$ & $0.6 \pm 0.1$ & $0.22 \pm 0.01$ & Arches & Other & WCLd \\
      1398622 & $1.01 \pm 0.69$ & $2.0 \pm 0.4$ & $0.10 \pm 0.04$ & Arches & Secondary & B1-3 Ia$+$ \\
      \hline
      15593 & $4.19 \pm 0.86$ & $4.2 \pm 0.3$ & $0.08 \pm 0.01$ & Quintuplet & Primary & O6-7 Ia+ \\
      \hline
    \end{tabular}  
    \begin{tablenotes}
    \item \textbf{Notes.} (i) The $\Delta$PM in column (2) is relative
      to the origin source (i.e., the absolute PM of the origin source
      was subtracted from the PM of the star). (ii) The PM position
      angle $\theta$ and its error are defined as the median and the
      error to the median values, respectively, of the 1000
      realizations of $\theta$ obtained as described in the
      text. (iii) WC $=$ Carbon-type Wolf-Rayet; WN $=$ Nitrogen-type
      Wolf-Rayet; O $=$ O supergiant.
    \end{tablenotes}
  \end{threeparttable}
\end{table*}

If the confirmed MSs are in the GC region, the tight distribution in
the VPD (Fig.~\ref{fig:figure18}) for most of them suggests that these
MSs could have originated from the same molecular cloud orbiting
around Sgr A*. The difference between the bulk PM of most MSs and the
Arches and the Quintuplet clusters is likely an indication that MSs
and the clusters are experiencing a different Galactic potential and
are at different distances from Sgr A*, with the MSs probably further
from Sgr A* than the clusters. However, the PM difference between
isolated MSs and the clusters could simply be due to different
peculiar motions, and all these objects could be at the same distance
from Sgr A*. Only a small group of objects have a broad PM
distribution in the VPD and might represent the closest sample of MSs
to the GC. A third component of the motion (LOS radial velocities) for
the MSs is required to better constrain the location of their
formation.

\begin{figure}
  \centering
  \includegraphics[width=\columnwidth]{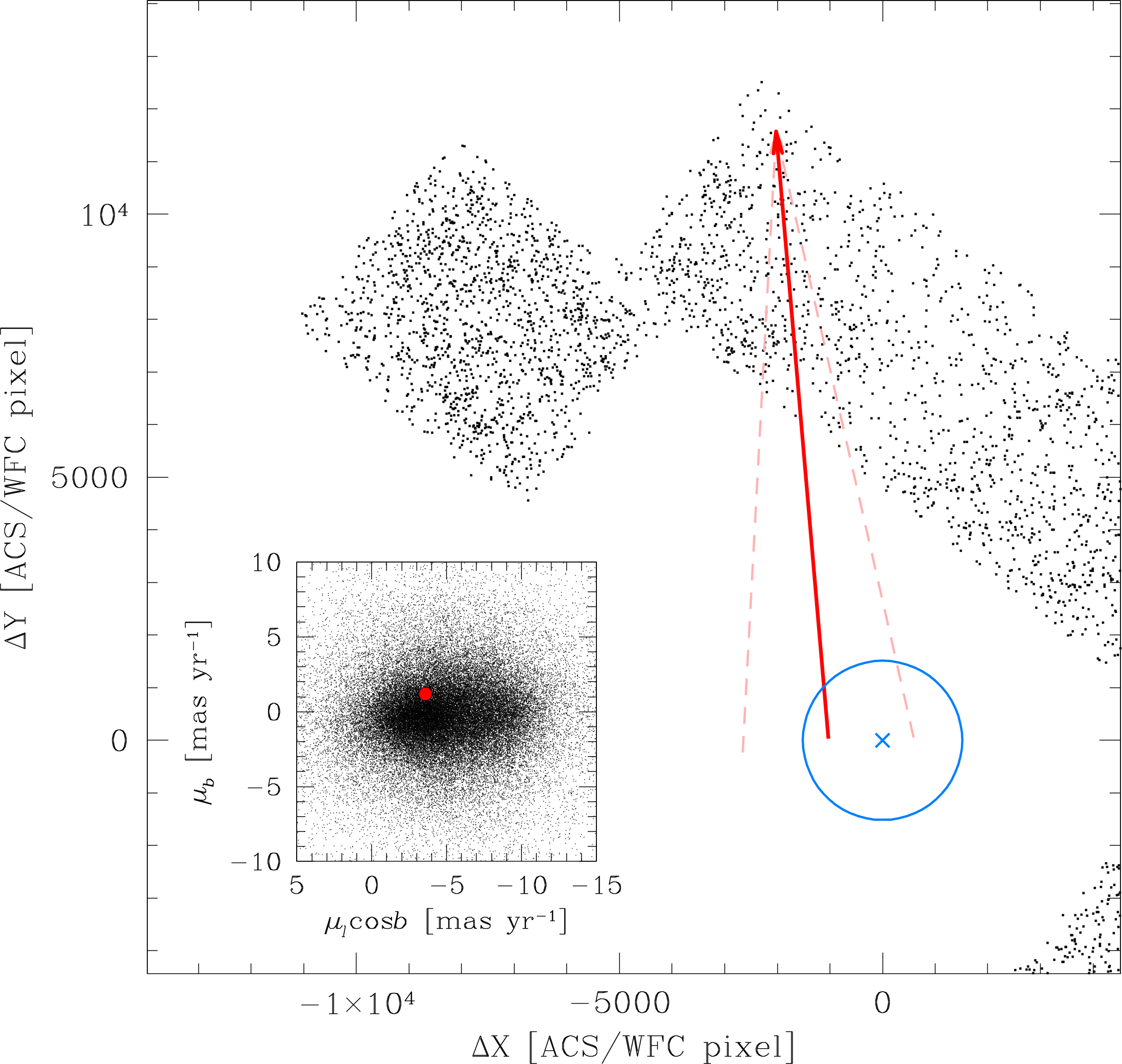}
  \caption{PM of the MS \# 20612 relative to Sgr A*. The red solid
    arrow starts at the time of the closest approach (see
    Table~\ref{tab:confirm}) and ends at the current position of the
    star. Pink dashed arrows define the confidence region. The blue
    circle has a radius of $\sim$1.26 arcmin and is centered on the
    position of Sgr A* (blue cross). The absolute PM of the source is
    shown in the VPD in the inset. Only 25\% of the bright sources are
    shown. North is up, East is to the left.}
  \label{fig:figure19}
\end{figure}

\subsubsection{Runaway star candidates}\label{confirmed_run}

Some MSs could indeed have been born in a cluster or close to Sgr A*,
and then ejected. Even if we do not know the distance of the MSs (nor
those of the Arches and the Quintuplet), we can still select
candidates whose PMs suggest they are moving radially away from these
possible birth places. First, we defined a reference system in which
the possible origin source (Sgr A*, the Arches or the Quintuplet) is
at rest and at pixel (0,0). Then, we computed the PM position angle
$\theta$ between the direction of the PM vector and the direction of
the star to the source. This angle is defined to be 0 deg if the star
is moving radially from the source, and $\pm$180 deg if it is moving
radially towards the source. We defined as candidates all objects that
verify the following conditions:
\begin{itemize}
\item (i) $|\theta| < 10$ deg;
\item (ii) the closest distance to the origin backward in time, based
  on the relative PM vector ($\pm$1$\sigma$), is equal or smaller than
  a limit radius. For Sgr A*, we defined the limit radius as its
  influence radius of $\sim$3 pc \citep{2018A&A...609A..27S}, i.e.,
  1.26 arcmin at the distance of 8.178 kpc, while for the Arches and
  the Quintuplet, we used their angular size provided by the Simbad
  database\footnote{\citet{2000A&AS..143....9W} and
    \href{http://simbad.u-strasbg.fr/simbad/}{http://simbad.u-strasbg.fr/simbad/}
    .};
\item (iii) the closest approach happened within the age of the origin
  source. For Sgr A*, we assumed an age of 6 Myr \citep[as that of the
  Central Star Cluster;][]{2007A&A...468..233M}. For the Arches and
  the Quintuplet, we considered ages of 2.5 Myr
  \citep{2008A&A...478..219M} and 4 Myr \citep{2010A&A...524A..82L},
  respectively.
\end{itemize}
We repeated the computation 1000 times, each time adding a random
noise to the PM, and verified if conditions (i), (ii) and (iii) were
met. The noise added to each star was randomly picked from a Gaussian
distribution with $\sigma$ equal to the PM error of the star. Stars
measured in only one image per epoch do not have a PM error. For them,
we assigned the median PM error of close-by stars at the same
magnitude level. If a candidate passed all three conditions (i), (ii)
and (iii) at least once, we inspected the results and discarded the
candidate if its PM errors are too large (i.e., wide range of
$\theta$). An example of the results for the star \# 20612 is shown in
Fig.~\ref{fig:figure19}.

\begin{figure*}
  \centering
  \includegraphics[width=\textwidth]{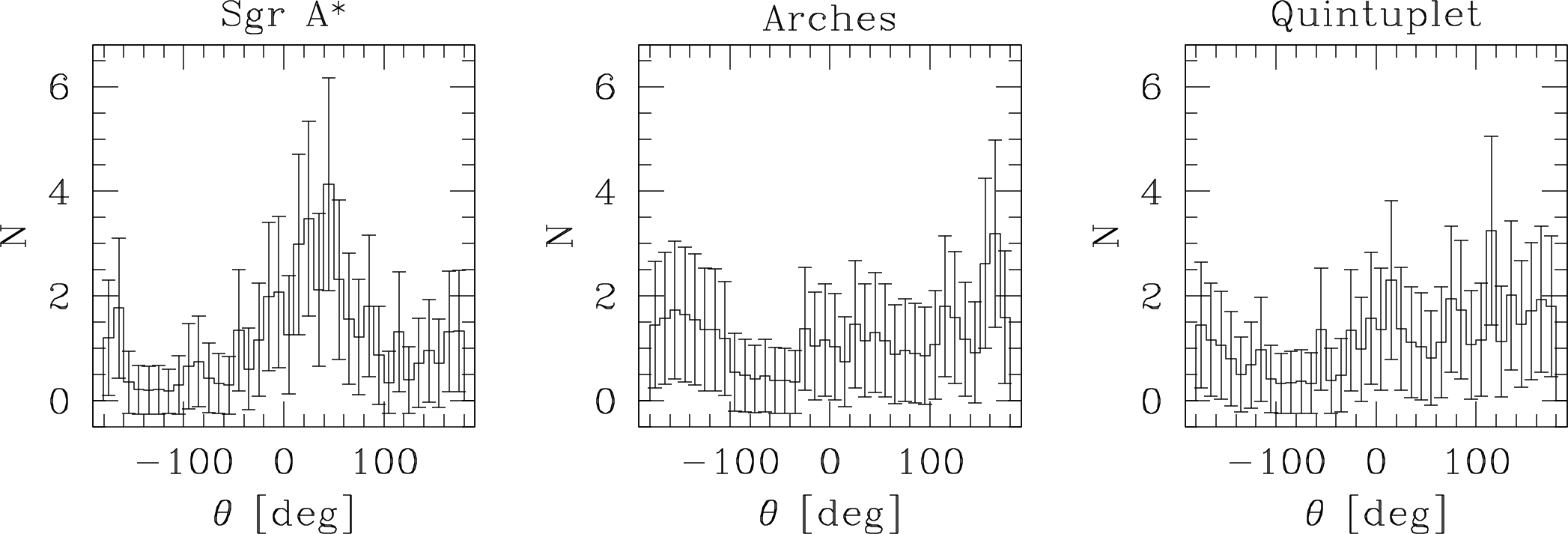}
  \caption{Histograms of the PM position angle $\theta$ for confirmed
    MSs. Error bars are defined as $\sqrt{N}$. From left to right, we
    show the histograms of $\theta$ computed with respect to Sgr A*,
    the Arches and the Quintuplet.}
  \label{fig:figure21}
\end{figure*}

In the reference system where Sgr A* is at rest and is placed at the
origin of the coordinates, MSs escaped from the Sgr A* region are
moving as the main stream of stars in the field (parallel to the
Galactic plane, toward positive Galactic longitude $l$, see VPD in the
inset in Fig.~\ref{fig:figure19}) and, according to Sect.~\ref{nir},
are likely in front of the GC. The Arches and the Quintuplet clusters
are located at the edge of the FoV and have Eastward motions larger
than most of the stars in the field. In the reference system where the
Arches or the Quintuplet are at rest, most stars in our field are
moving towards West and have a motion similar to what we expect for a
star escaping from the clusters. These conditions (position and
kinematics of the clusters) make the few MS escapers we found not that
special, thus weakening the escaping nature of our potential MS
escapers. Furthermore, the absolute PMs of MSs suggest they might not
be at the same distance of the clusters. All these features are
probably an indication that these objects are not genuine escapers and
their motion is just the result of perspective effect. The unknown
distance of MSs is the limiting factor that prevents us from reaching
definitive conclusions. Table~\ref{tab:confirm} summarizes the
parameters of these stars. We report the flight time of these objects,
defined as the distance from the current position to the point of
closest approach divided by the relative PM of the MS in the reference
frame where the origin source (Sgr A*, the Arches or the Quintuplet)
is at rest. The quoted errors on the flight times are only internal
and do not include any source of systematic errors.

The MSs that appear to have come from Sgr A* are not HVSs, thus
excluding the ejection via Hill mechanism. Various mechanisms can
eject stars over different timescales. An ejection resulting from the
disruption of a binary by a supernova explosion is plausible if the
age of the star-forming region is longer than the shortest lifetime of
a star \citep[$\sim$3 Myr; e.g.,][]{2017A&A...601A..29Z}. The
$\sim$2.5-Myr age of the Arches disfavors supernova-related events for
the candidate escapers. A three-body interaction can expel the
least-massive star in the system and leave behind a binary object made
by two stars with a mass of at least 100 $M_\odot$ \citep[the typical
mass of the MSs in this sample; see][]{2015MNRAS.446..842D}, larger
than that of the ejected object. Mass segregation in the two clusters
\citep[e.g.,][]{2019ApJ...870...44H,2019ApJ...877...37R} makes massive
stars to preferentially sink towards the cluster center, favoring
their dynamical interaction. Finding massive-star binary systems in
the core of the Arches and the Quintuple would not contradict a
three-body ejection mechanism as explanation for massive
escapers. Nevertheless, the flight time of all these escapers is
suspiciously small, thus suggesting a fortunate alignment of the PM
vectors.

Star \# 1398622 in the ACS/WFC catalog\footnote{Star \# 1398622 is
  measured in one image per epoch.} in the Secondary list is
classified as an ejected candidate from the Arches. Its absolute PM is
in agreement with the bulk motion of the Arches within 1$\sigma$,
suggesting that it could be a member of the cluster. The tidal radius
of the Arches is about 1.6 pc \citep{2013A&A...556A..26H}, which
corresponds to $\sim$40 arcsec at the distance of Sgr A*. Even if the
cluster is located 2 kpc in front of the GC (and its tidal radius
becomes about 53 arcsec), star \# 1398622 would still be outside the
tidal radius. For this reason, this star is an interesting target to
follow up with LOS-radial-velocity measurements.

Figure~\ref{fig:figure21} presents the histograms (bin width of 10
deg) of the PM position angle $\theta$ of confirmed MSs with respect
to Sgr A* (left panel), the Arches (middle panel) or the Quintuplet
(right panel). Each histogram shown in the Figure is the average of
1\,000 histograms, each of which was obtained by measuring $\theta$
after adding a random noise to the PM errors as described above. By
adding the random noise, we randomly blurred/sharpened and shifted the
distributions in the histograms. The final average histogram is less
dependent on the bin width and the starting point \citep[see,
e.g.,][]{2019ApJ...873..109L}. MSs do not show any significant peak at
$\theta \sim 0$ deg in the histograms. There is a hint at the
$\sim$2.8$\sigma$ level of a peak at $\theta \sim 30$ deg in the
histogram for Sgr A*. Given the relative position and PM of the
confirmed MSs with respect to Sgr A*, we still favor the idea that the
sample of potential escapers previously discovered is mainly the
result of a perspective effect.

\subsubsection{Arches' members}\label{confirmed_arches}

\begin{figure}
  \centering
  \includegraphics[width=\columnwidth]{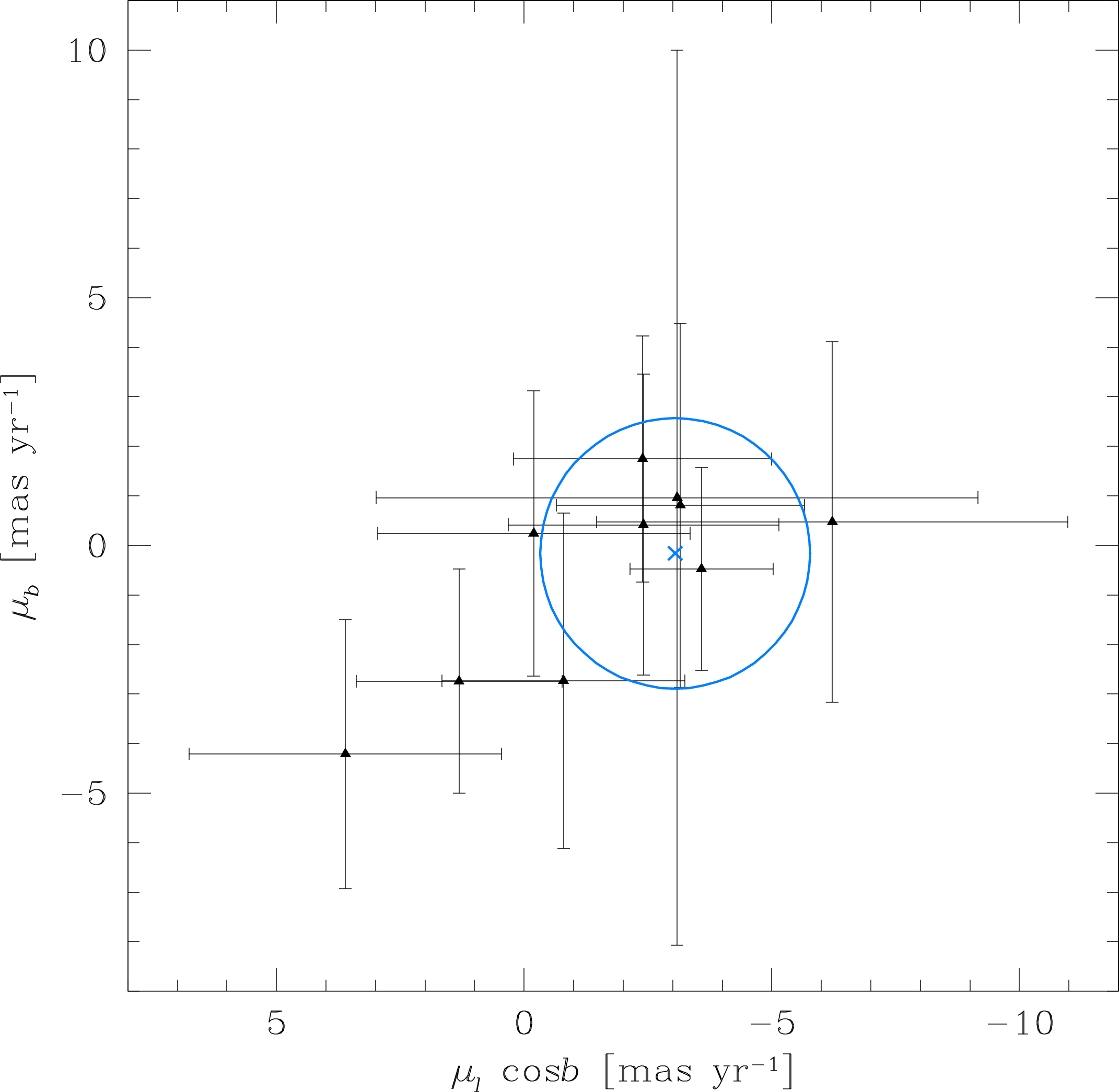}
  \caption{VPD of the absolute PMs of confirmed MSs in the ACS/WFC
    field close to the Arches (black triangles with error bars). The
    blue cross marks the absolute PM of the Arches cluster. The blue
    ellipse is centered on the cross and has semi-axes equal to the
    median PM errors of stars at the magnitude level of the ten
    massive objects.}
  \label{fig:figure20}
\end{figure}

Ten MSs in the ACS/WFC field close to the Arches are included in the
list of Arches' MSs of \citet{2018A&A...617A..65C}. The VPD of the PMs
of these stars is shown in Fig.~\ref{fig:figure20}. The blue cross
marks the PM of the Arches and the ellipse has semi-axes equal to the
median PM errors of stars at the magnitude level of the ten massive
objects in that region of the FoV. Eight out of ten stars are likely
members at the 1$\sigma$ level, and all objects have PMs consistent
with that of the Arches at the 2$\sigma$ level. These MSs are listed
in Table~\ref{tab:arches}.

\begin{table}
  \caption{List of confirmed massive stars likely members of
    the Arches cluster.}
 \centering
 \label{tab:arches}
 \begin{threeparttable}
   \begin{tabular}{ccc}
     \hline
     \hline
     ID & List  & Spectral Type \\
        & & \\
     \hline
     1412291 & Other & WN8-9h \\
     1412292 & Other & WN8-9h \\
     1412295 & Other & WN8-9h \\
     1412379 & Other & O6-6.5 Ia \\
     1412381 & Other & O6-7 Ia$+$ \\
     1412382 & Other & WN7-8h \\
     1412383 & Other & O4-5 Ia$+$ \\
     1412384 & Other & WN8-9h \\
     1412386 & Other & O4-5 Ia \\
     1412470 & Other & WN8-9h \\
     \hline
   \end{tabular}
   \begin{tablenotes}
   \item \textbf{Notes.} (i) PMs of the stars \# 1412379, 1412383, 1412386,
     1412470 were measured with one image per epoch. (ii) All stars
     are included in the list of Arches' massive stars of
     \citet{2018A&A...617A..65C}.
   \end{tablenotes}
 \end{threeparttable}
\end{table}

\subsubsection{Cluster tidal tails}\label{confirmed_cluster}

The confirmed MSs in our field could be former members of the Arches
or the Quintuplet in a tidal tail of the
clusters. \citet{2019ApJ...870...44H} and \citet{2019ApJ...877...37R}
analyzed in detail the Arches and the Quintuplet clusters. Neither
studies found hints of tidal-tail structures out to 3 pc from the
clusters' centers, and the authors observed (weak for the Quintuplet
and strong for the Arches) evidence of mass segregation. Both pieces
of information make unlikely to find massive stars in a tidal
tail. However, the GC region is very extreme, and a few massive
objects could have been ejected at early stages of the clusters'
evolution.

\begin{figure*}
 \centering
 \includegraphics[width=\textwidth]{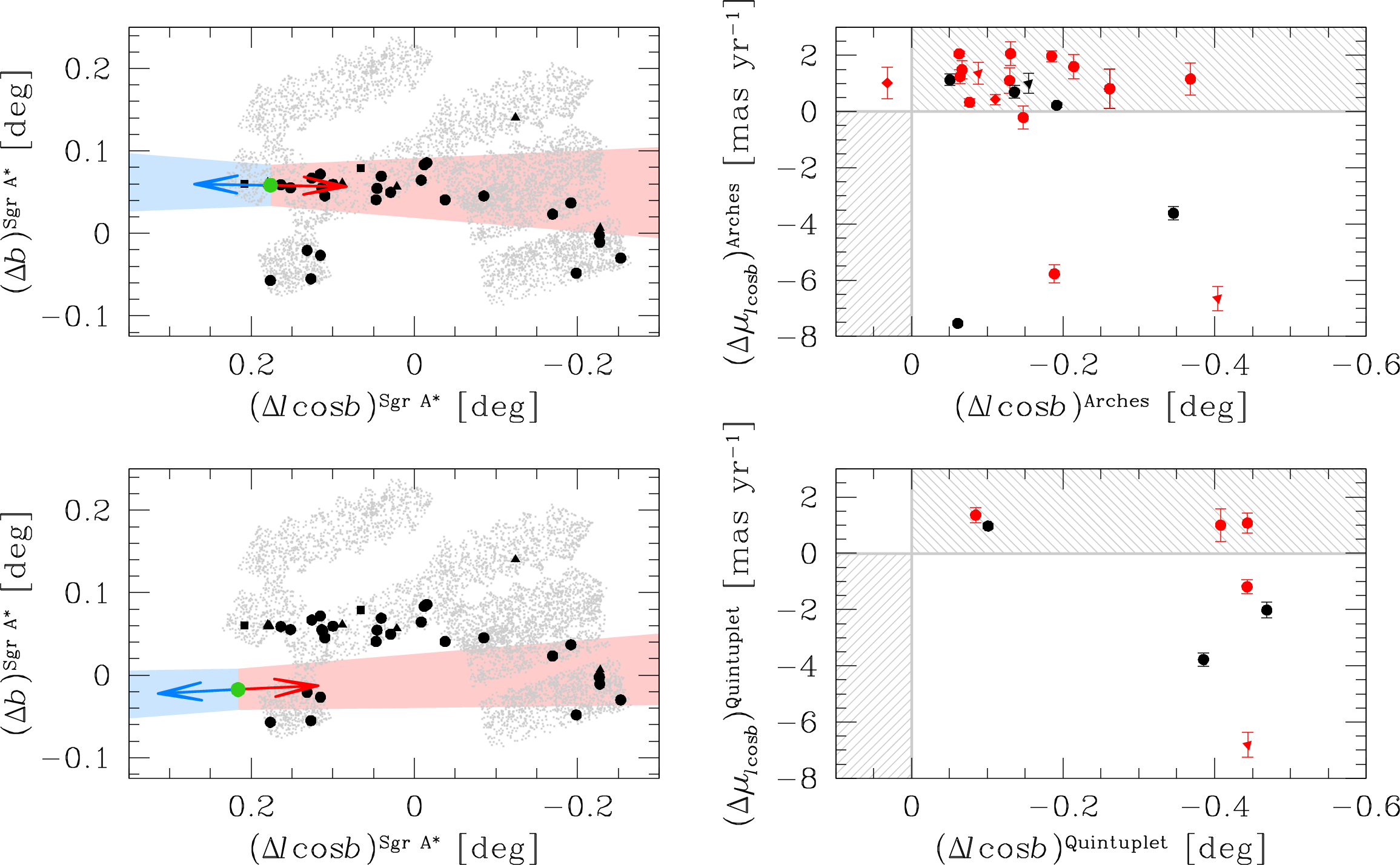}
 \caption{(Top panels): in the left panel, we show the FoV covered by
   our \hst observations in the reference frame where Sgr A* is at
   rest. We highlighted in black confirmed MSs (circles, squares and
   triangles represent MSs in the Primary, Secondary and Other
   samples, respectively). The current Arches position is shown as a
   green point. All other bright stars are shown in gray. The
   future/past motion of the Arches in $10^5$ yr is shown as a
   blue/red arrow. Cyan and pink regions mark the expected locations
   of the leading and trailing tails, respectively, of the cluster
   given the Arches' PM relative to Sgr A* (and its error). In the
   right panel, we show the PM of the MSs relative to the Arches as a
   function of position relative to the Arches along the $l$
   direction. Only MSs within the cyan or pink regions outside three
   times the cluster's angular size (from Simbad) are
   considered. Stars in the gray areas are likely not part of the
   tidal tail (see text fot details). Red points are MSs that have
   $|\Delta\mu_b^{\rm Arches}| \le 1$ \masyr. All other points are
   shown in black. (Bottom panels): as above, but for the Quintuplet
   cluster.}
 \label{fig:figure22}
\end{figure*}

We defined the direction of motion of the clusters with respect to Sgr
A* in the plane of the sky by means of the PMs listed in
Table~\ref{tab:ref}. Red and blue arrows in the left panels in
Fig.~\ref{fig:figure22} represent the expected motion of the two
clusters (the Arches on top and the Quintuplet on bottom, shown as
green dots) over $10^5$ yr backwards and forwards in time,
respectively. Cyan and pink regions highlight the location of the
leading and trailing tails given the PMs of the clusters relative to
Sgr A* (we also took into account for the PM errors). Black points
mark the locations of the confirmed MSs. MSs within the shaded regions
are potential members of a tidal tail. The right panels of
Fig.~\ref{fig:figure22} present the PMs of the MSs (only those within
the cyan/pink regions further than three times the cluster's angular
size from Simbad) relative to the Arches/Quintuplet as a function of
the position relative to the cluster along the $l$ direction. Any
tidally-lost star must have $\mu_b$ similar to that of the Arches or
the Quintuplet. Red points are MSs that have $\mu_b$ within 1 \masyr
from that of the corresponding cluster, while black points are all
other MSs. Stars in a leading/trailing tails should have a
positive/negative PM relative to the cluster, respectively.
Interestingly, position and motion of star \# 1398622 in our ACS/WFC
catalog are consistent with being part of the leading tail of the
Arches. Again, all six kinematic parameters of the MSs and of the
clusters are required to completely understand the location of their
origin.

\begin{figure*}
  \centering
  \includegraphics[height=11.5cm]{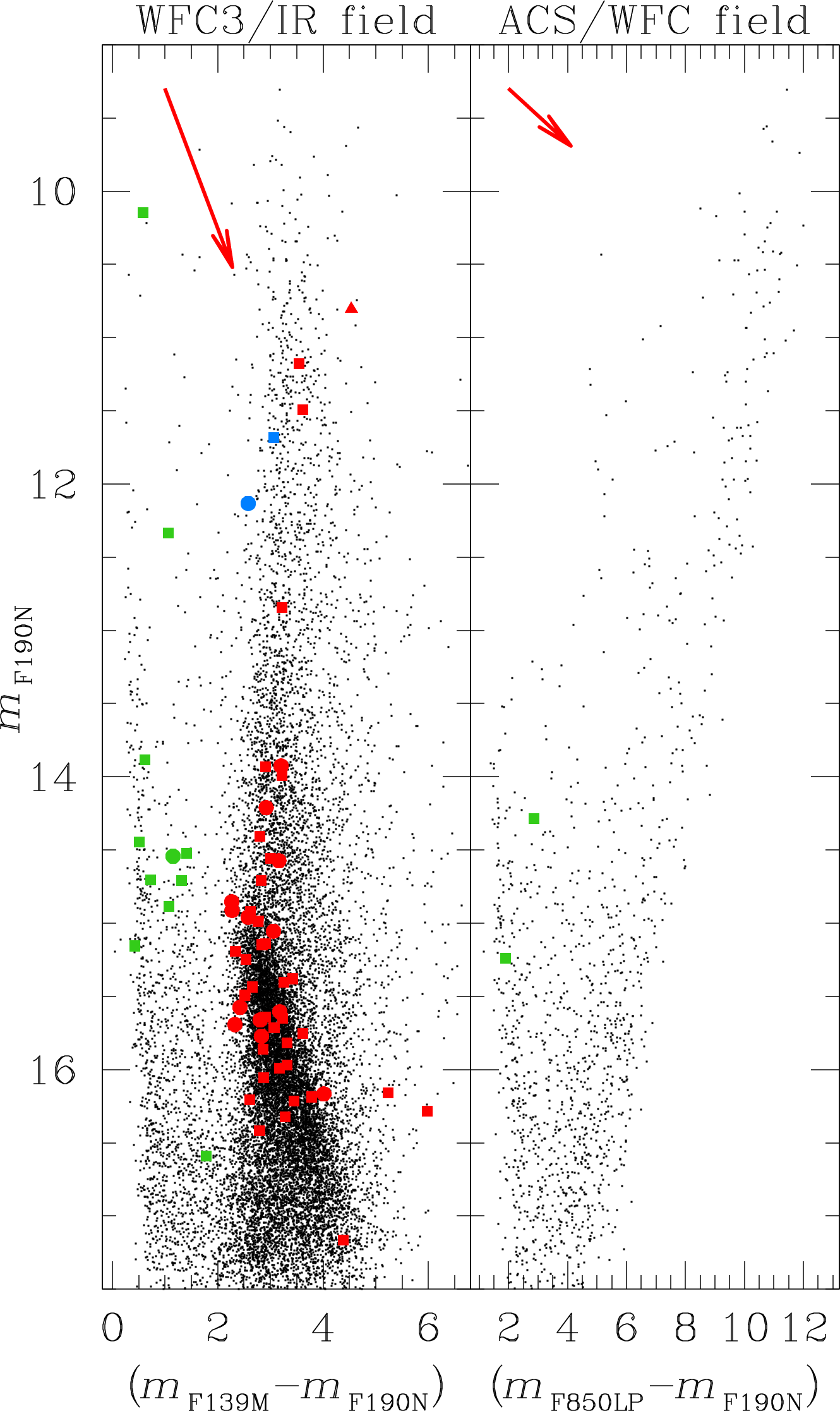}
  \includegraphics[height=11.5cm]{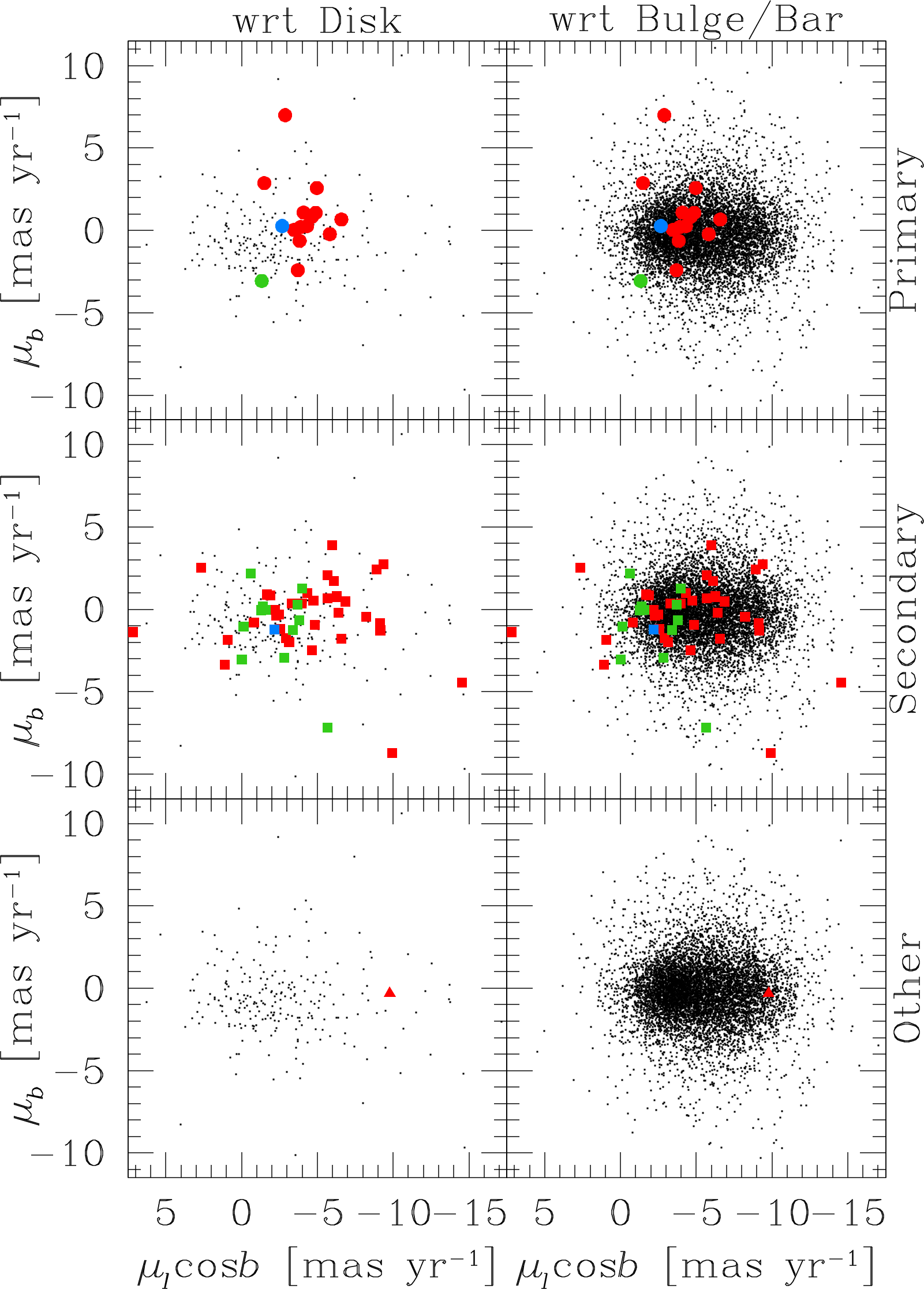}
  \caption{Similar to Fig.~\ref{fig:figure1415} but for candidate MSs.}
  \label{fig:figure2324}
\end{figure*}

\subsubsection{Candidate MSs}\label{candidates}

Figure~\ref{fig:figure2324} shows the CMDs and the VPDs for the
candidate MSs in our lists. Color and shape codings are the same as in
Sect.~\ref{confirmed} and Fig.~\ref{fig:figure1415}. Green points are
likely interlopers in the Disk. Five objects have a parallax
measurement in the Gaia-DR2, suggesting they are located within 2 kpc
from the Sun. The location of the remaining stars in CMDs based on
different color combination indicates that the remaining stars are
likely Disk objects as well. Therefore, we excluded all these 14 stars
for which CMDs or the Gaia-DR2 parallaxes clearly rule out a Bulge/Bar
connection. CMD locations and PMs of the remaining MS candidates
suggest that these objects are likely in the Bulge/Bar.

\begin{figure}
  \centering
  \includegraphics[width=0.9\columnwidth]{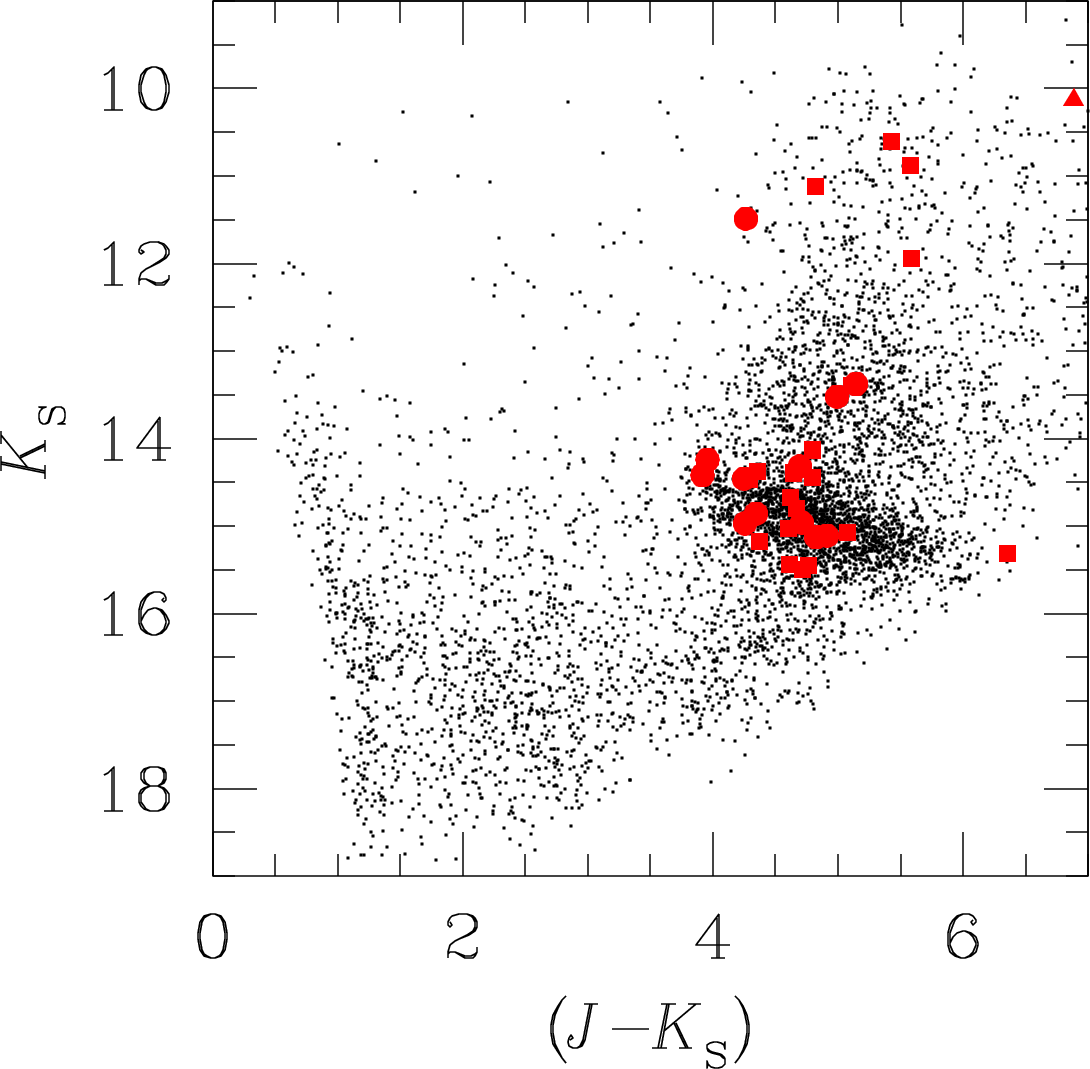}
  \caption{$K_{\rm S}$ versus $(J-K_{\rm S})$ CMD for stars in the
    WFC3/IR field (black points). Candidate MSs are shown as red
    points.}
  \label{fig:figure25}
\end{figure}

Figure~\ref{fig:figure25} presents the $K_{\rm S}$ versus
$(J-K_{\rm S})$ CMD of the stars in the WFC3/IR field. Stars
highlighted in red are the candidate MSs likely located in the
Bulge/Bar. Most MSs are in the Bulge Red Clump. We measured the
velocity dispersion $\sigma_b$ of these candidate MSs and, as a
reference, of Bulge/Bar objects in the Bulge Red Clump. We find that
candidate MSs have $\sigma_{\mu_b} = (2.18 \pm 0.41)$ \masyr, while
Bulge/Bar stars have a $\sigma_{\mu_b} = (2.88 \pm 0.01)$ \masyr, thus
suggesting that the two groups of objects could be at the same
distance from Sgr A* at the $\sim$1.7$\sigma$ level. These MS
candidates are likely a more genuine Bulge/Bar ``field'' population
than the confirmed MSs discussed in Sect.~\ref{confirmed_pop}.

\begin{figure*}
  \centering
  \includegraphics[width=\textwidth]{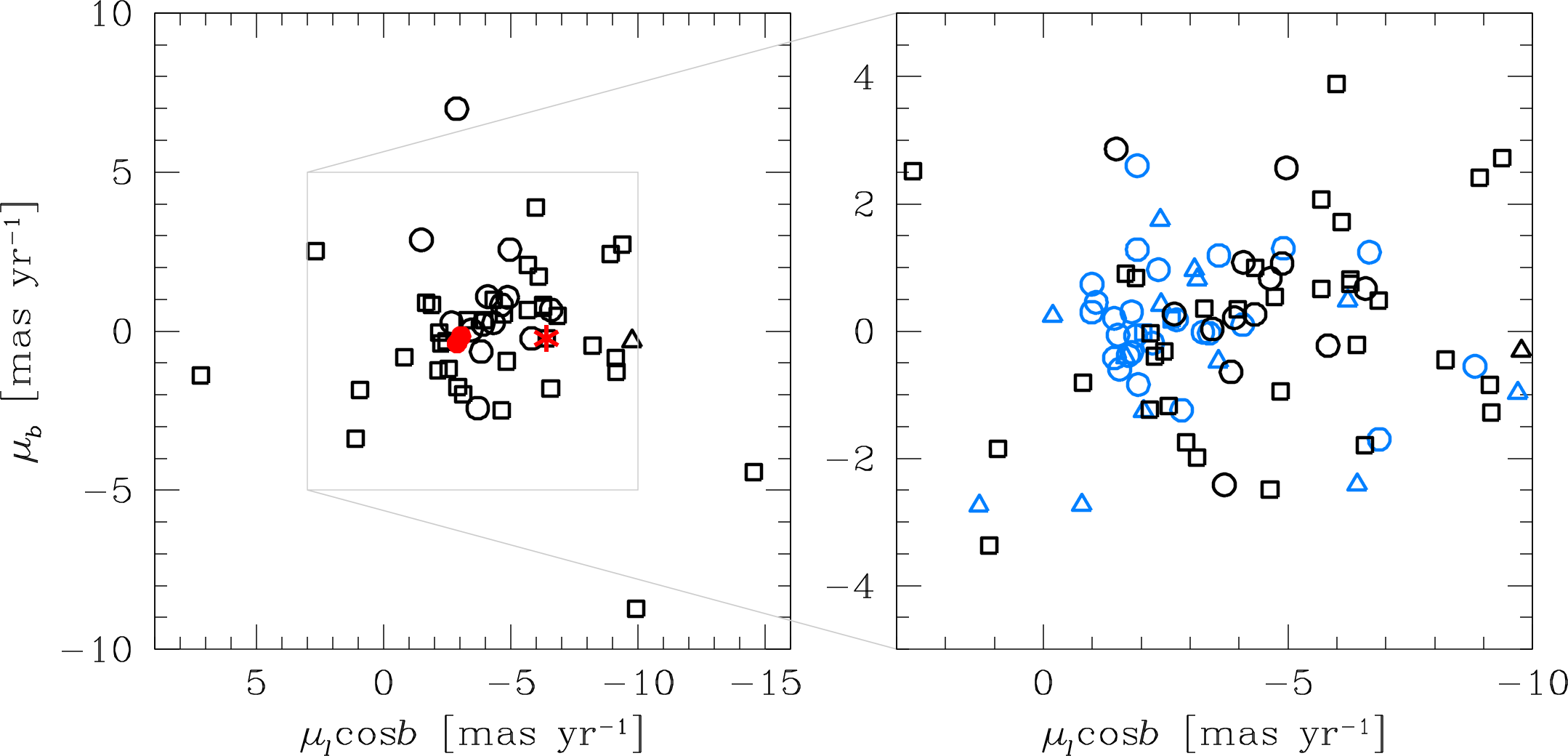}
  \caption{VPDs of the candidate MSs (black points). The red asterisk
    marks the PM of Sgr A*, while the two red points indicate the PMs
    of the Arches and the Quintuplet clusters. As in
    Fig.~\ref{fig:figure1415}, circles, squares and triangles
    represent MSs in the Primary, Secondary and Other samples,
    respectively. The right panel is a zoomed-in view of the left
    panel, in which we also plot in blue the confirmed MSs in our
    field, as a reference.}
  \label{fig:figure26}
\end{figure*}

In the left panel of Fig.~\ref{fig:figure26}, we compare the absolute
PMs of the candidate MSs with those of Sgr A* and the Arches and
Quintuplet clusters. The right panel of Fig.~\ref{fig:figure26} shows
a zoom-in around the location of most candidate MSs in the VPD. Blue
circles represent the confirmed MSs analyzed in
Sect.~\ref{confirmed}. The PM distribution of candidate MSs is broader
and less centered around the clusters' positions than that of the
confirmed MSs, suggesting that stars in these two groups have a
different origin. However, as for the confirmed MSs, we lack of LOS
velocities and distances and we cannot exclude the possibility they
could have been born closer to the GC.

\begin{figure*}
  \centering
  \includegraphics[width=\textwidth]{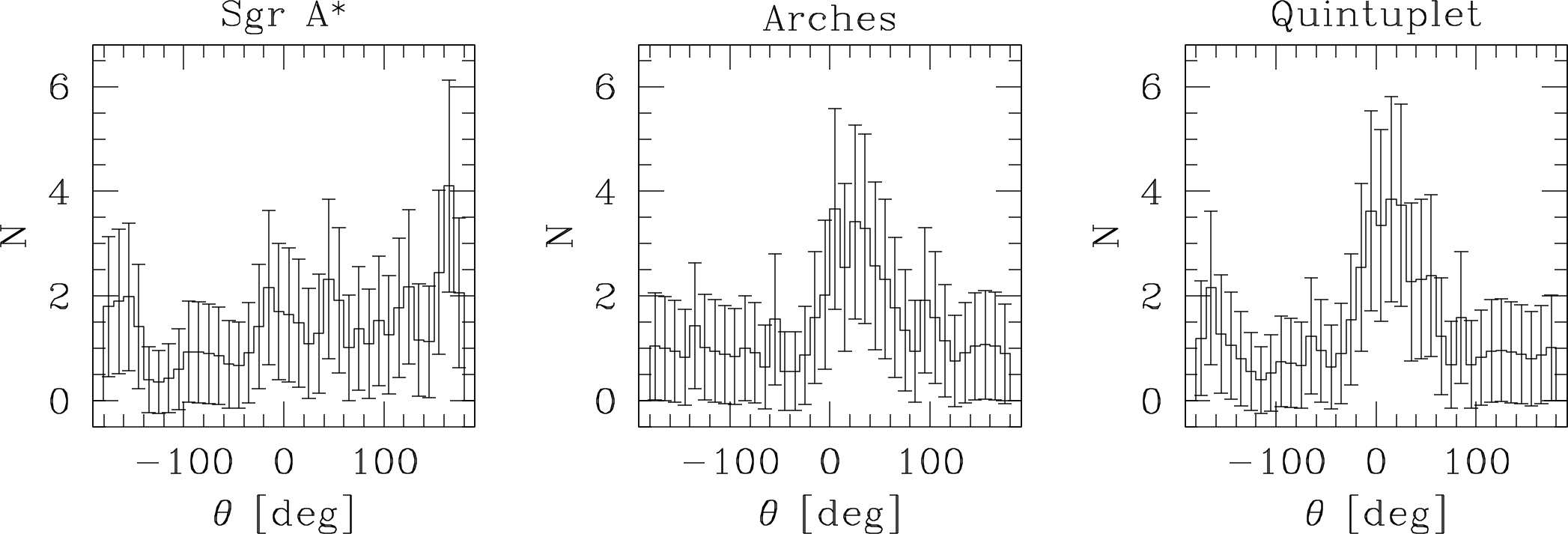}
  \caption{The histograms of the PM position angle $\theta$ for the
    candidates MSs (error bars are defined as $\sqrt{N}$). The
    histograms for $\theta$ computed with respect to Sgr A*, the
    Arches and the Quintuplet are show in the left, middle and right
    panels, respectively.}
  \label{fig:figure27}
\end{figure*}

Similarly to the analyses described in Sect.~\ref{confirmed_run}, we
computed the PM position angle $\theta$ for candidate MSs. We
identified again a few potential escapers (Table~\ref{tab:candidate})
but their flight time is short ($<0.6$ Myr), as in the case of the
confirmed MSs, and some objects passed all criteria for more than one
origin. Figure~\ref{fig:figure27} presents the histogram of the PM
position angle $\theta$ for candidate MSs. There is not a clear peak
within $|\theta| < 10$ deg (escaping objects) in any histogram. Both
the Arches and Quintuplet histograms present a significant
($>$3$\sigma$) peak at $\theta \sim 27$ deg and $\sim 17$ deg,
respectively, but these peaks are likely due to the relative position
and PM of the candidate MSs with respect to the clusters. All these
pieces of evidence favor the idea that these potential escapers might
in fact just be the result of perspective effects.

Finally, the majority of candidate MSs passed the tidal-tail test.

\begin{table*}
  \caption{Overview of the candidate MSs selected as likely
    escapers.}
  \centering
  \label{tab:candidate}
  \begin{threeparttable}
    \setlength\tabcolsep{3 pt}
    \begin{tabular}{cccccccc}
      \hline
      \hline
      ID & $\Delta$PM & $\theta$ & Flight time & Possible origin & List \\
         & \masyr & deg & Myr & & & \\
      \hline 
      427662  & $1.83 \pm 1.08$ & $-4.8 \pm 0.9$  & $0.25 \pm 0.08$ & Sgr A* & Secondary \\
      1101618 & $4.13 \pm 0.83$ & $-10.4 \pm 0.3$ & $0.12 \pm 0.02$ & Sgr A* & Secondary \\
      1187124 & $0.59 \pm 0.78$ & $-5.1 \pm 1.6$  & $0.62 \pm 0.28$ & Sgr A* & Primary \\
      1210224 & $2.97 \pm 0.81$ & $-4.0 \pm 0.3$  & $0.16 \pm 0.03$ & Sgr A* & Primary \\
      1274214 & $7.52 \pm 1.13$ & $7.0 \pm 0.2$   & $0.07 \pm 0.01$ & Sgr A* & Secondary \\
      \hline 
      12280   & $6.73 \pm 0.72$ & $3.8 \pm 0.1$   & $0.13 \pm 0.01$ & Arches & Other \\
      236988  & $6.22 \pm 0.34$ & $2.5 \pm 0.1$   & $0.21 \pm 0.01$ & Arches & Secondary \\
      427662  & $5.19 \pm 1.08$ & $9.3 \pm 0.3$   & $0.24 \pm 0.02$ & Arches & Secondary \\
      656883  & $3.36 \pm 0.88$ & $-0.9 \pm 0.3$  & $0.34 \pm 0.05$ & Arches & Secondary \\
      1117192 & $3.38 \pm 1.06$ & $0.3 \pm 0.4$   & $0.12 \pm 0.03$ & Arches & Secondary \\
      1354402 & $1.60 \pm 0.79$ & $ -3.9 \pm 0.4$  & $0.18 \pm 0.04$ & Arches & Secondary \\
      \hline
      427662  & $5.34 \pm 1.08$ & $-2.0 \pm 0.3$  & $0.24 \pm 0.03$ & Quintuplet & Secondary \\
      1026874 & $6.26 \pm 1.10$ & $7.7 \pm 0.3$   & $0.06 \pm 0.01$ & Quintuplet & Secondary \\
      1187124 & $2.93 \pm 0.78$ & $6.4 \pm 0.4$   & $0.10 \pm 0.01$ & Quintuplet & Primary \\
      \hline
    \end{tabular}  
    \begin{tablenotes}
    \item \textbf{Notes.} (i) Star \# 427662 passed the criteria for
      Sgr A*, the Arches and the Quintuplet. Star \# 1187124 passed
      the criteria for both Sgr A* and the Quintuplet.
    \end{tablenotes}
  \end{threeparttable}
\end{table*}

\subsection{Non-massive stars}\label{nonims}

Five stars in our lists have a spectral type that suggests they cannot
be classified as massive stars. The {$m_{\rm F190N}$ versus
  ($m_{\rm F139M}-m_{\rm F190N}$) CMD and the VPDs show in
  Fig.~\ref{fig:figure28} suggest these objects to be Bulge/Bar stars.

\begin{figure*}
  \centering
  \includegraphics[width=0.9\textwidth]{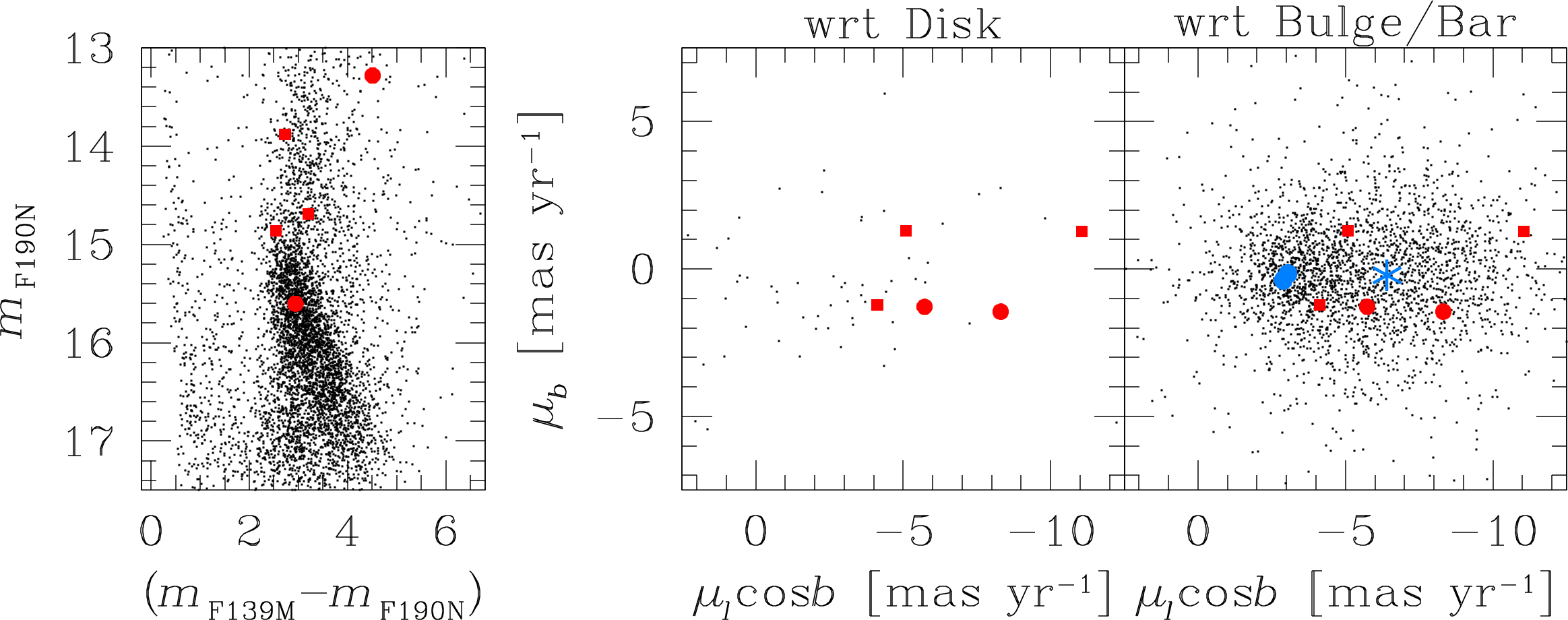}
  \caption{(Left panel): $m_{\rm F190N}$ versus
    ($m_{\rm F139M}-m_{\rm F190N}$) CMD for the stars in the WFC3/IR
    field. Red points represent the non-massive objects found in our
    list; all other stars are shown in black. (Middle and right
    panels): VPDs of the PMs for a sample of Disk (middle panel) and
    Bulge/Bar (right panel) stars (black points). Red points have the
    same meaning as in the CMD. Blue dots show the absolute PMs of the
    Arches and the Quintuplet, while the blue star indicates the
    motion of Sgr A*.}
  \label{fig:figure28}
\end{figure*}

\section{Fast-moving stars}\label{fast}

MSs represent the tip of the iceberg of the GC population of stars
born around Sgr A* or in a cluster and later ejected as a result of
dynamical effects. Some stars do not produce strong emission lines and
cannot be detected in emission-line surveys, but we can still find
some potential candidates thanks to PMs. It is unfeasible to search
for all potential escapers among the myriad of stars in our FoV, but
we can at least study objects with a relatively-high PM.

We focused only on well-measured stars in the WFC3/IR field with a PM
error lower than 2 \masyr. Since we are interested in fast-moving
objects, we also included in our sample stars that have a PM larger
than 70 \masyr (see discussion in Sect.~\ref{pm}). We removed all
stars that, according to their location in a CMD, are likely part of
the foreground Disk population. We then analyzed two samples of stars,
one bright (stars brighter than $m_{\rm F139M} \sim 18.8$, i.e., with
a magnitude similar to that of the confirmed MSs) and one faint (all
other objects). This choice is motivated by the fact that bright and
faint stars have different PM errors and the interpretation of the
results requires different considerations.

For each star, we computed its relative PM with respect to Sgr A*, the
Arches or the Quintuplet. Stars in the GC have a velocity dispersion
of more than 100 \kms ,which corresponds to about 3 \masyr at the
distance of Sgr A*. Therefore, we kept only sources with a PM larger
than 10 \masyr, i.e., three times the typical velocity dispersion in
the GC. Finally, we measured their PM position angles $\theta$.

Figure~\ref{fig:figure29} presents the results. In the left panels, we
show the histograms of the relative PMs with respect to the Quintuplet
(top panels), the Arches (middle panel) and Sgr A* (bottom panel). The
bright sample is in blue, while the faint sample is shown in red. Both
groups have similar histograms. The histograms of the PM position
angle $\theta$ of fast-moving objects are displayed in the right
panels.

\begin{figure}
  \centering
  \includegraphics[width=\columnwidth]{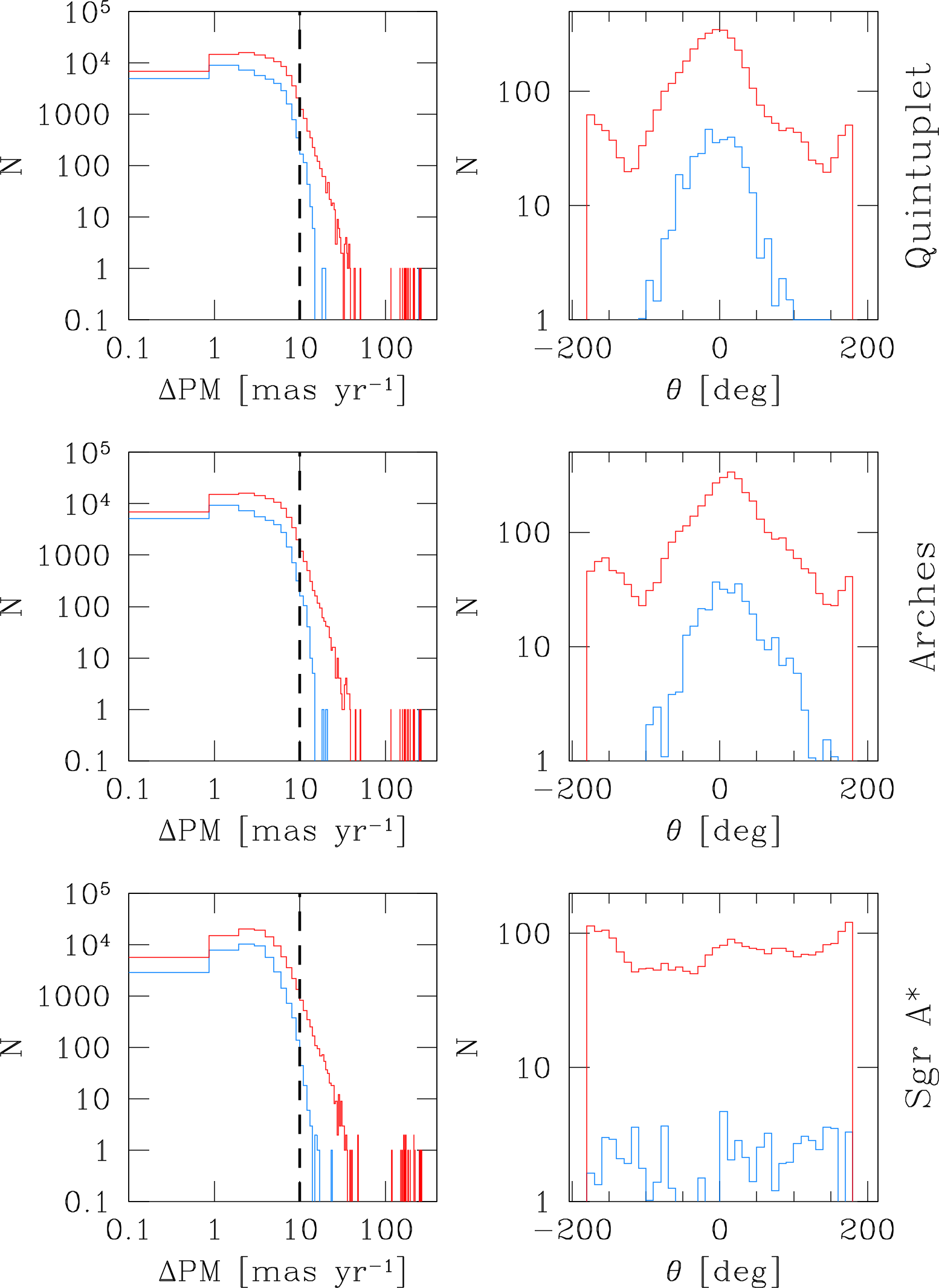}
  \caption{The histograms of the relative PMs with respect to the
    Quintuplet (top), the Arches (middle) and Sgr A* (bottom) are
    shown in the left panels. Blue and red contours represent the
    histograms for stars brighter and fainter than
    $m_{\rm F139M} \sim 18.8$, respectively. The black, dashed
    vertical lines sets the threshold for the selection of fast-moving
    objects. In the right panels, we show the histograms of the PM
    position angle $\theta$ for the fast-moving stars.}
  \label{fig:figure29}
\end{figure}

The shape of the histograms of the PM position angles shown in
Fig.~\ref{fig:figure29} can be again explained by the peculiar
location of Sgr A*, the Arches and the Quintuplet in our field. Stars
are located all around (both North and South, East and West, of) Sgr
A*, in front or behind it. As such, the distribution of $\theta$ is
expected to be flat.

The histograms of the bright stars in the two massive clusters show
hints of a peak at $\theta \sim 0$ deg, with a sharp drop out around
$\pm$90 deg. This can be explained in terms of selection effects: the
PM distribution of Bulge/Bar stars is broader along the $l \cos b$
direction than the $b$ direction, so that any selection based solely
on PM size disproportionately favors stars with larger $\mu_l \cos
b$. Faint stars, on the other hand, are able to populate all position
angles: this is most likely due to the fact that their much larger PM
errors make the PM distribution broader, and stars moving mainly along
the $b$ direction can still survive the selection based on PM
size. For this reason, the histograms of the faint stars show rather
flat distributions. The peak at $\theta \sim 0$ deg can be interpreted
as discussed in Sect.~\ref{MS}.

If we restrict our analysis to stars with a PM larger than 26
\masyr($\sim$1000 \kms at 8.178 kpc), we find three stars in the
WFC3/IR region escaping from Sgr A* ($|\Delta\theta| \lesssim 10$
deg). A visual inspection of the WFC3/IR stacked images reveals the
presence of additional very-faint, fast-moving objects not detected by
our reduction software. However, the WFC3/IR cosmetic and angular
resolution are worse than that of ACS/WFC and, with only two epochs of
data, it is not straightforward to infer if they are genuine HVSs or
image artifacts. Among bright stars, which PM measurements are more
robust, we do not find any HVS candidate.

\citet{2015ARA&A..53...15B} estimates that, assuming a continuous and
isotropic ejection of HVSs, the density of HVSs should scale as
$r^{-2}$ kpc$^{-3}$, with $r$ the distance from Sgr A*. Our
observations cover part of an annulus in the plane of the sky centered
on Sgr A* with inner and outer radii of 3 ($\sim$7 pc at 8.178 kpc)
and 17 arcmin ($\sim$40 pc), respectively. According to the relation
of \citet{2015ARA&A..53...15B}, we should expect $\sim$1 HVS within a
spherical region of radius 40 pc around Sgr A*. However, we are not
mapping the entire volume of this sphere centered on Sgr A* but only
part of a spherical shell, and HVSs could not be ejected continuously
and isotropically. For these reasons, the number of HVSs we should
find in our field is likely less than 1. If we focus on bright stars,
our null-detection would be in agreement with the theoretical
expectations. However, this result is biassed by the small region
analyzed in our work and the lack of LOS data (we cannot detect HVSs
with a motion predominantly along the LOS and not in the plane of the
sky). A larger FoV and complementary LOS-velocity data are needed to
reach a definitive conclusion about HVSs in the GC.

We do not find significant fast-moving Bulge/Bar stars in the ACS/WFC
field.

\section{Conclusions}

The location of the origin of MSs in the GC region is still a matter
of debate. Many pieces of information have been collected to
investigate the nature of these massive objects, but we still need
some key elements to fully reconstruct their history. In an effort to
further shed light on these MSs, we computed high-precision PMs of
stars near the GC with \hst data, and analyzed the kinematics of
confirmed and candidate MSs in the field.  We make our
astro-photometric catalogs publicly available. The description of the
catalogs is provided in Appendix~\ref{appendix:catalog}. The lack of
LOS velocities and distances of our targets is now the main source of
uncertainty of our work and makes some of the conclusions mainly
speculations. However, PMs allow us to provide constraints for some
proposed scenarios even without a complete analysis of the Bulge
kinematics.

The location of the origin of most confirmed, isolated MSs is still
uncertain. We estimate an upper limit for their distance from Sgr A*
of 3--4 kpc by comparing the PMs of the MSs with those of spiral-arm
objects. Photometric and kinematic properties of these isolated MSs
are different from what we would expect for a genuine population in
the GC. However, their nature (young, massive stars), the high and
variable extinction, and peculiar motions (like those of the Arches
and the Quintuplet) could explain the discrepancies we observe in CMDs
and VPDs. As such, most of these isolated MSs could be in the GC, born
a few Myrs ago \textit{in-situ} from a single molecular cloud. A few
objects might be former members of the Central Cluster, the Arches or
the Quintuplet, either ejected at velocities lower than 1\,000 \kms or
part of a tidal-tail structure. Thanks to our PMs, we add strength to
the argument that a selection of previously classified MSs are in fact
likely Arches' members \citep[in agreement
with][]{2018A&A...617A..65C}. One object (star \# 1398622 in our
ACS/WFC catalog) in the Secondary list of \citet{2011MNRAS.417..114D}
is one of the most interesting targets because it is located outside
the nominal tidal radius of the Arches and its kinematics is
consistent with being an Arches' (former) member.

We also analyzed a sample of MSs that still require a clear
spectroscopic characterization. We excluded Disk interlopers from our
list by means of Gaia-DR2 parallaxes, optical-NIR CMDs and PMs. Most
of the remaining candidate MSs are consistent with a Bulge/Bar field
population according to their locations in CMDs and VPDs. The PM
position angles of candidate MSs support an \textit{in-situ}
formation.

We searched our data set for fast-moving objects that could have been
radially ejected from Sgr A*, the Arches or the Quintuplet. The
histograms of the PM position angles show no clear excess of
high-velocity escapers. The peaks in these histograms can be simply
interpreted by taking into account for the location of our FoV with
respect to Sgr A* or the clusters, and the overall kinematics of the
stars toward the GC. Large FoVs and LOS-velocity data are needed to
complete the census of fast-moving objects in the GC.

Additional follow-ups to obtain LOS velocities are required to give a
full 3D kinematic picture of this region. Furthermore, an additional
third epoch of data, possibly with the \textit{James Webb Space
  Telescope}, would further improve the astrometric precision of and
overall quality of our catalogs, and would allow us to further explore
the wealth of information in this GC region.

\section*{Acknowledgments}

ML and AB acknowledge support from STScI grants GO 12915 and 13771.
DJL acknowledges support from the Spanish Government Ministerio de
Ciencia, Innovaci\'on y Universidades through grants
PGC-2018-091\,3741-B-C22 and from the Canarian Agency for Research,
Innovation and Information Society (ACIISI), of the Canary Islands
Government, and the European Regional Development Fund (ERDF), under
grant with reference ProID2017010115. LRP acknowledges support from
the Generalitat Valenciana through the grant PROMETEO/2019/041. LRB
acknowledges partial support by MIUR under PRIN programme no.
2017Z2HSMF. The authors thank the anonymous referee for the useful
comments and suggestions. This work has made use of data from the
European Space Agency (ESA) mission {\it Gaia}
(\url{https://www.cosmos.esa.int/gaia}), processed by the {\it Gaia}
Data Processing and Analysis Consortium (DPAC,
\url{https://www.cosmos.esa.int/web/gaia/dpac/consortium}). Funding
for the DPAC has been provided by national institutions, in particular
the institutions participating in the {\it Gaia} Multilateral
Agreement. This research made use of
Astropy,\footnote{\href{http://www.astropy.org}{http://www.astropy.org}}
a community-developed core Python package for Astronomy
\citep{2013A&A...558A..33A,2018AJ....156..123A}. This research has
made use of the SIMBAD database, operated at CDS, Strasbourg, France.

\section*{Data availability}

The catalogs described in this article are available in the article
and in its online supplementary material.

\appendix

\section{Comparison with Gaia DR2 PMs}\label{appendix:gaia}

Figure~\ref{fig:figure30} shows a comparison between the \hst-based
and the Gaia-DR2 PMs. The red lines are the plane bisectors, the blue
lines (not always visible) are the weighted straight-line fits to the
points. Stars in common between \hst and Gaia are bright Disk
objects. These plots show that there is good agreement between the two
sets of PMs. Figure~\ref{fig:figure31} compares the PM errors of stars
in common between the Gaia-DR2 and our \hst catalogs. At the faint-end
of Gaia ($G \gtrsim 18$), our \hst PM errors are a factor of 2 smaller
than in the Gaia catalog.

Figure~\ref{fig:figure32} presents a comparison between Gaia-DR2 and
\hst PMs as a function of the X and Y positions in the FoV. The
largest deviations (all within 3$\sigma$) are visible at the edges of
the FoV. In these regions, the local network of reference stars used
to transform \hst positions on to the master frame are not uniformly
distributed, thus possibly introducing some systematic error that can
propagate to PM measurements.

The comparisons between the WFC3/IR-based and the Gaia-DR2 PMs as a
function of the $(G-m_{\rm F139M})$ color and $m_{\rm F139M}$
magnitude are shown in the left panels of Fig.~\ref{fig:figure33}.
Similar comparisons between ACS/WFC and Gaia PMs as a function of
$(G-m_{\rm F850LP})$ and $m_{\rm F850LP}$ are shown in the right
panels. Again, our PMs are in agreement with those of Gaia at the
3$\sigma$ level. A systematic trend as a function of color for stars
redder than the bulk of the Disk sequence is present. However, most
stars in this color range have $G > 18.5$ (see CMDs in
Fig.~\ref{fig:figure34}), i.e., their Gaia PM errors are large. Gaia
PMs in this faint regime are not generally used in our kinematic
analyses.

\begin{figure}
  \centering
  \includegraphics[width=\columnwidth]{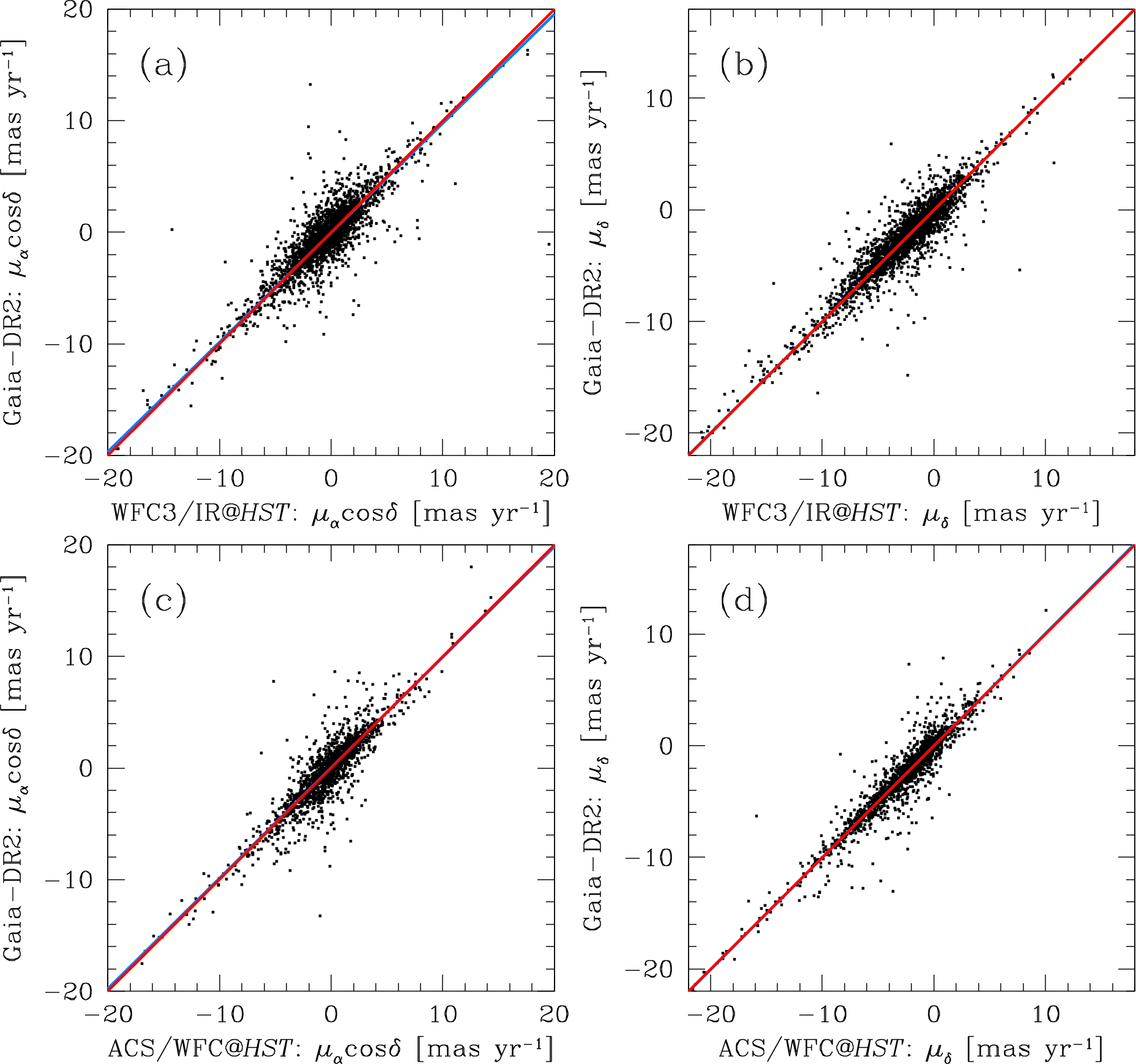}
  \caption{Comparison between Gaia DR2 PMs and WFC3/IR (panels a and
    b) or the ACS/WFC (panels c and d) PMs. In each panel, the red
    line is the plane bisector, while the blue lines (when not hidden
    below the red lines) represent the weighted straight-line fit to
    the points.}
  \label{fig:figure30}
\end{figure}

\begin{figure}
  \centering
  \includegraphics[width=\columnwidth]{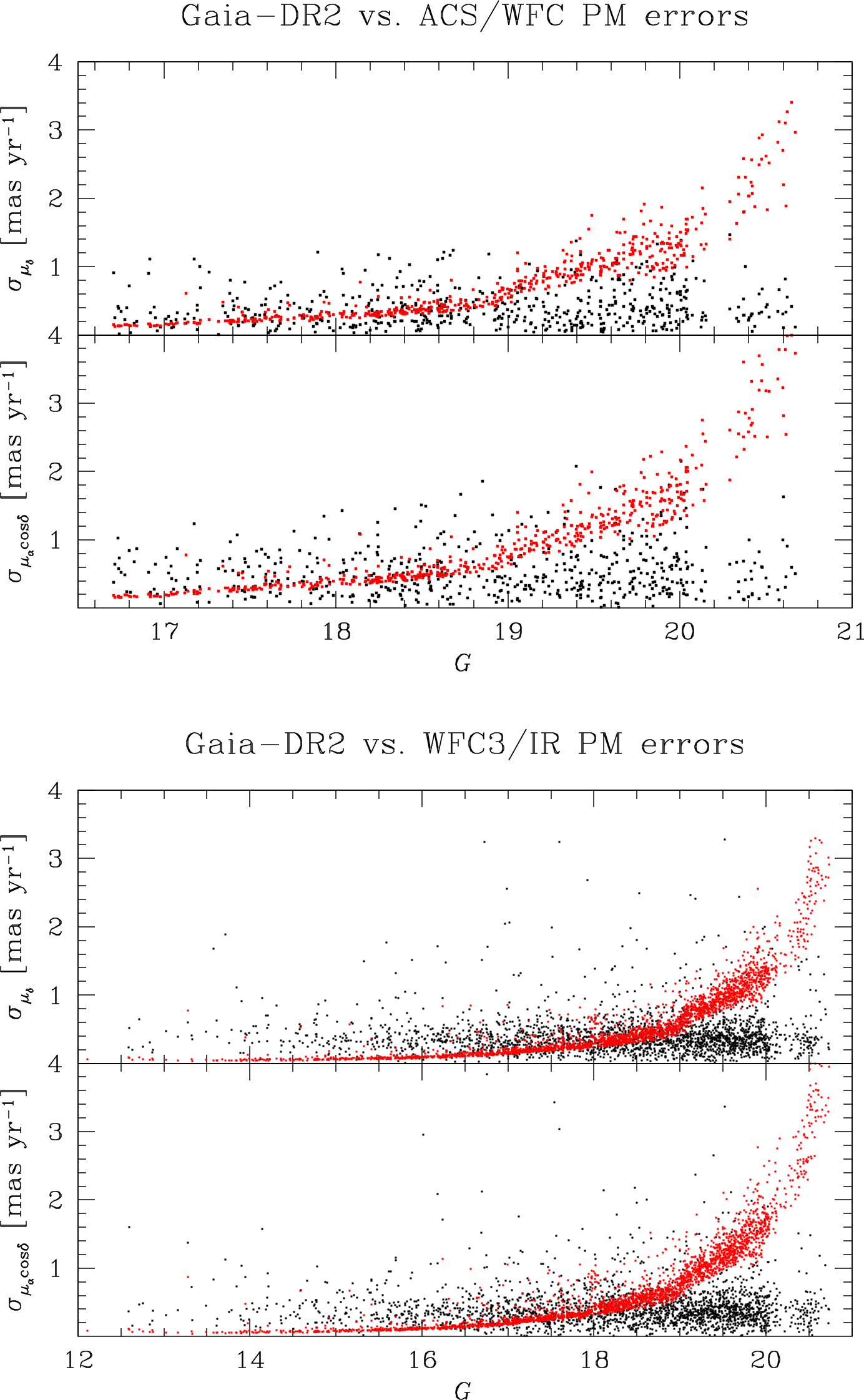}
  \caption{PM errors in the Gaia DR2 (red points) and our \hst
    catalogs (black points) as a function of the $G$
    magnitude. ACS/WFC PM errors are in the top panels, while WFC3/IR
    PM errors are in the bottom panels.}
  \label{fig:figure31}
\end{figure}

\begin{figure*}
  \centering
  \includegraphics[width=0.56\textwidth]{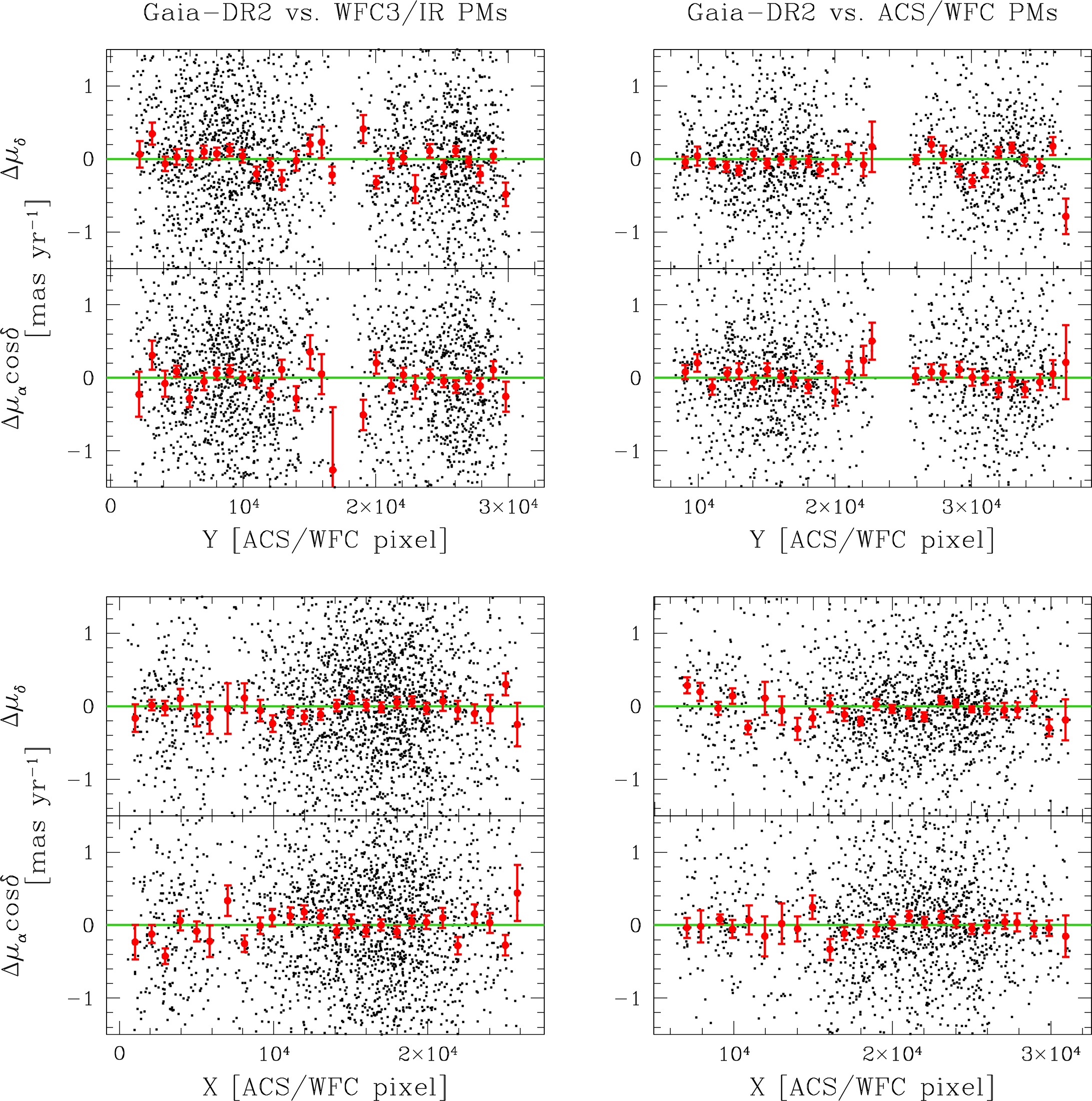}
  \caption{PM difference between Gaia and WFC3/IR (left) or ACS/WFC
    (right) PMs as a function of the Y (top panels) and X (bottom
    panels) positions in the FoV. Black points are all stars in
    common, red dots (with error bars) are the median values of the PM
    difference over 1000-pixel-wide bins. The green line is set at 0
    \masyr.}
  \label{fig:figure32}
\end{figure*}

\begin{figure*}
  \centering
  \includegraphics[width=0.56\textwidth]{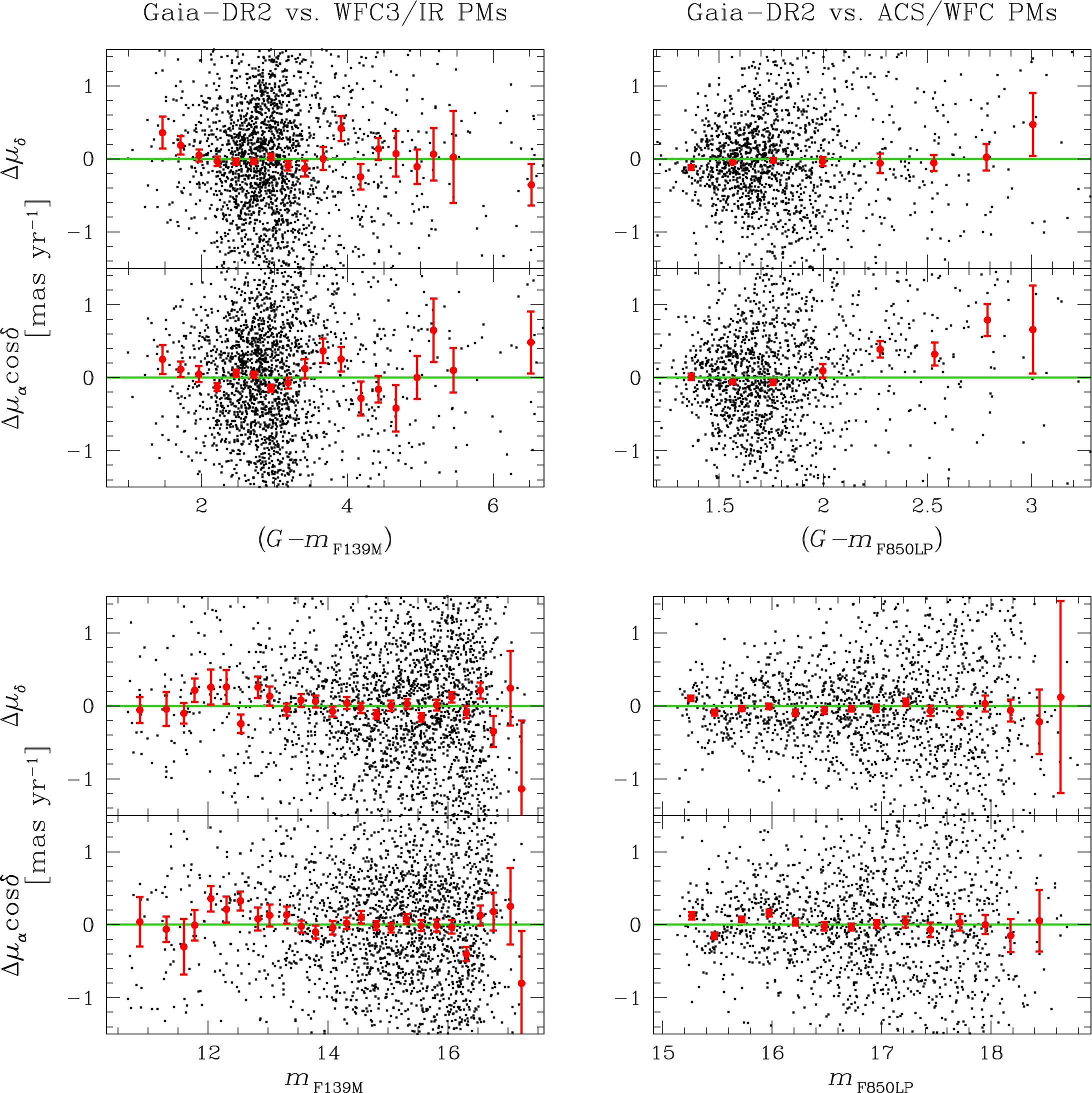}
  \caption{(Left panels): PM difference between Gaia and WFC3/IR PMs
    as a function of $(G-m_{\rm F139M})$ (top panels) and
    $m_{\rm F139M}$ (bottom panels). As in Fig.~\ref{fig:figure32},
    the green line is set at 0 \masyr. Black points represent
    individual stars, red points (with error bars) are the median
    values of the PM differences over 0.25-mag-wide bins. (Right
    panels): PM difference between Gaia and ACS/WFC PMs as a function
    of $(G-m_{\rm F850LP})$ (top panels) and $m_{\rm F850LP}$ (bottom
    panels). Color coding and bin sizes are the same as in the left
    panels.}
  \label{fig:figure33}
\end{figure*}

\begin{figure*}
  \centering
  \includegraphics[width=0.56\textwidth]{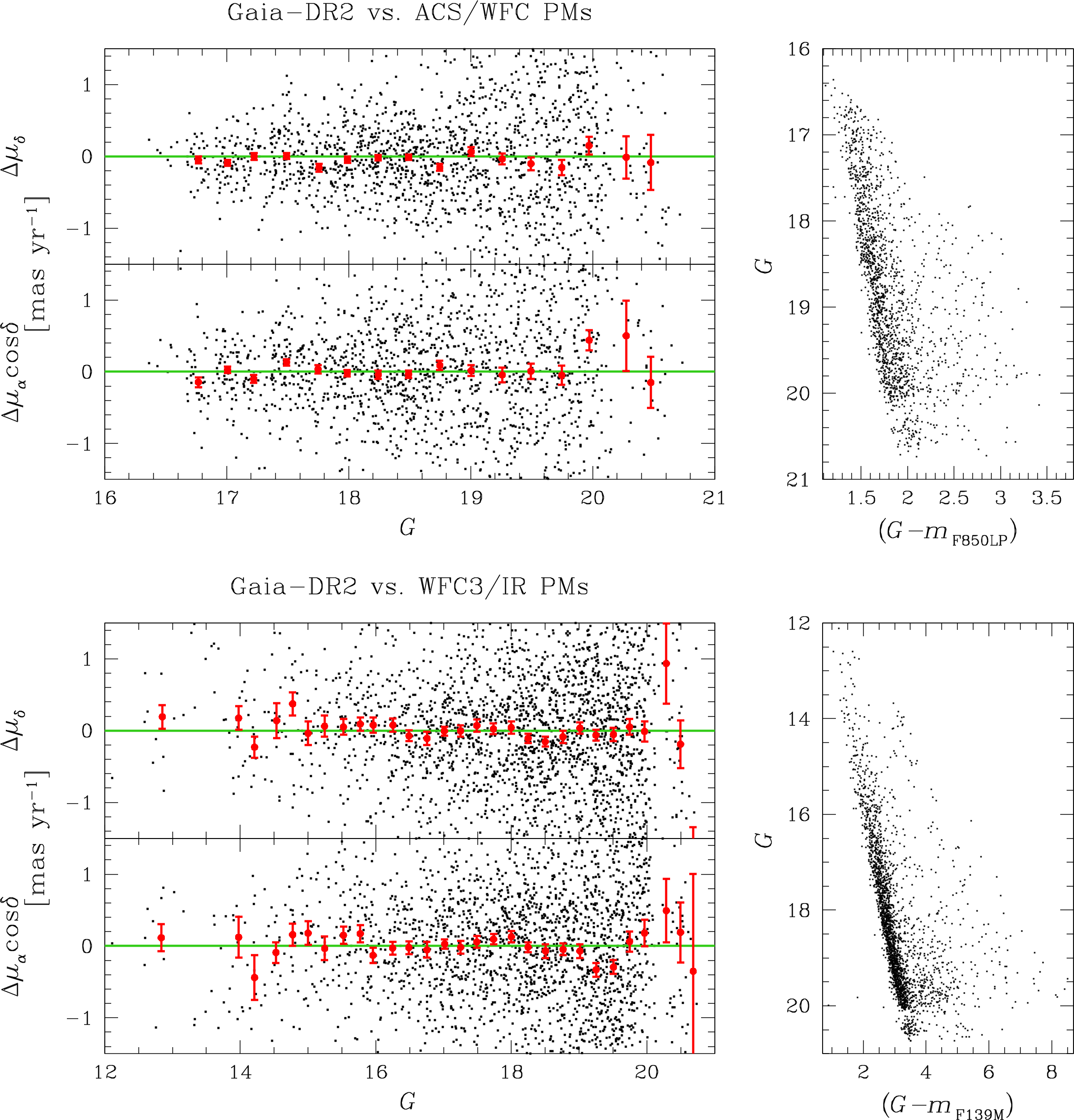}
  \caption{The Figure shows that most of the color-related PM trends
    visible in Fig.~\ref{fig:figure33} are mainly due the Gaia-DR2 PMs
    of faint stars. In the left panels, we show the PM difference
    between \hst (top panels for the ACS/WFC, bottom panels for the
    WFC3/IR) and Gaia data as a function of the $G$ magnitude. Largest
    scatters are visible at $G > 18.5$. These stars are the reddest
    objects in the CMDs on the right. This finding supports the idea
    that the color trends in Fig.~\ref{fig:figure33} are mainly caused
    by Gaia-DR2 PMs.}
  \label{fig:figure34}
\end{figure*}

\section{Comparison between WFC3- and ACS-based PMs}\label{appendix:hst}

The comparison between the PMs obtained with the ACS/WFC and the
WFC3/IR data for all stars in common between the two data sets (bottom
panels of Fig.~\ref{fig:figure35}) shows that there is a systematic
trend. Most of the stars responsible of the deviation from the 1:1
relation have a bright neighbor in the WFC3/IR data. These stars can
easily be removed from the sample by using two diagnostic parameters
described in Sect.~\ref{datared}: the \texttt{RADXS} parameter
($|$\texttt{RADXS}$|$$<$0.04) and the fraction of neighbor flux within
the fitting radius before neighbor subtraction
($o<$1). After removal, the comparison between the two sets of PMs is
improved (top panels).

\begin{figure*}
  \centering
  \includegraphics[width=0.56\textwidth]{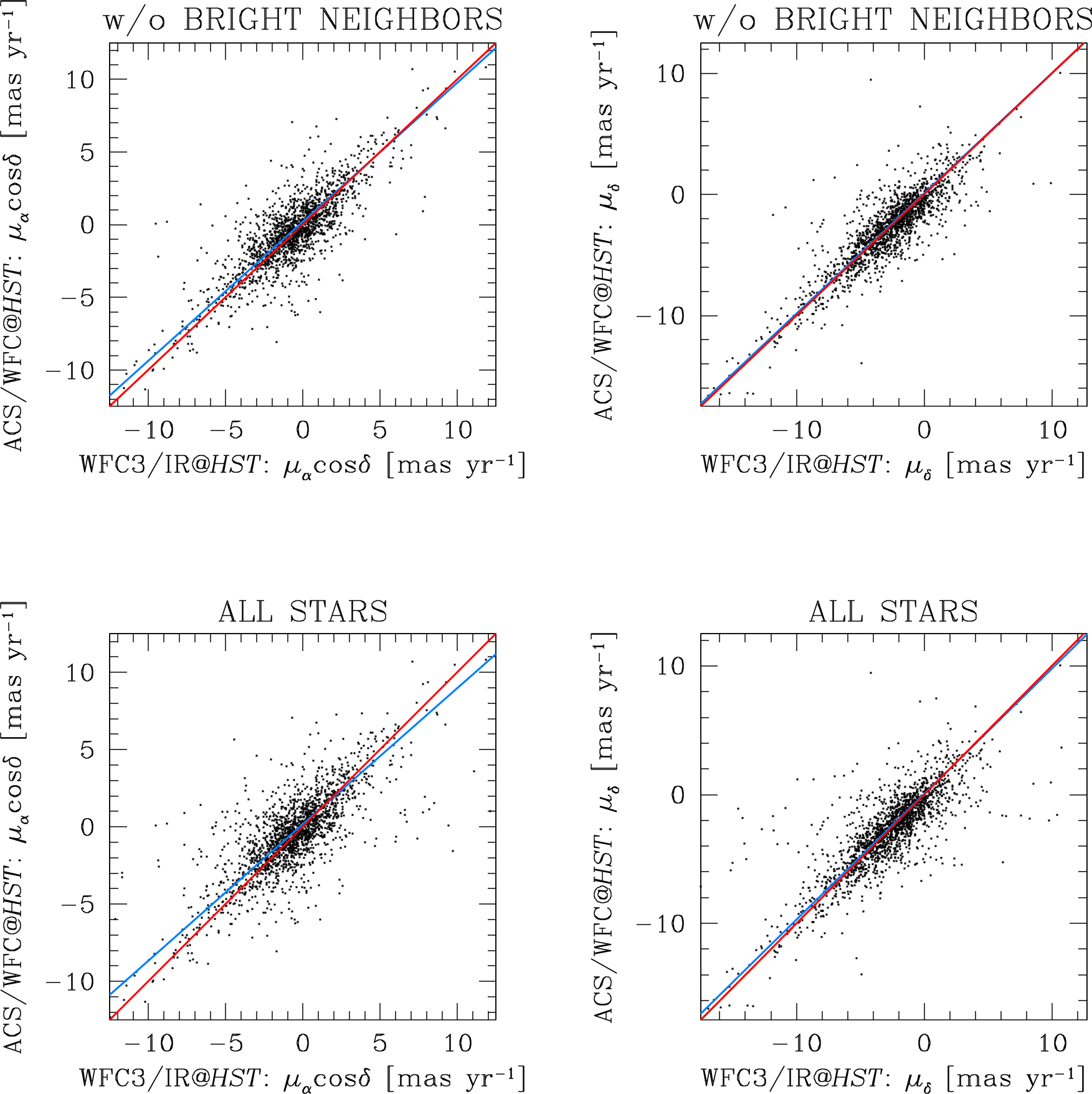}
  \caption{Similar to Fig.~\ref{fig:figure30}, but for the WFC3/IR-
    and the ACS/WFC-based PMs. The comparison obtained by using only
    stars without bright neighbors is presented in the top
    panels. Bottom panels show all stars in common.}
  \label{fig:figure35}
\end{figure*}

\section{Description of the astro-photometric catalogs}\label{appendix:catalog}

We release a photometric catalog for each instrument/epoch, and a PM
catalog for each instrument. The ID entries in the astrometric
catalogs are the internal IDs of the reduction process.

Table~\ref{tab:photcat} shows the first 10 lines of the GO-12915
WFC3/IR data (all other photometric catalogs contain the same column
information). VEGA magnitudes can be converted into instrumental
magnitudes (at a 1-s exposure time) by subtracting the zero-points
listed in Table~\ref{tab:photzp}.

The first 10 lines of the WFC3/IR PM catalog are shown in
Table~\ref{tab:pmcat}. The X and Y positions (pixel scale 50 mas
pixel$^{-1}$) are the average between the first- and second-epoch
positions obtained in the PM-computation process. PM errors equal to
99.99 \masyr refer to PMs obtained with only measurement in at least
one epoch. The average temporal baseline is defined as the difference
between the average time of the first- and second-epoch images used to
measure PMs. The values $\rm n_1$ and $\rm n_2$ are the numbers of
images used to compute the averaged first- and second-epoch positions,
respectively.

\begin{table*}
  \caption{First 10 lines of the WFC3/IR F139M photometric catalog
    based on the GO-12915 data.}
  \centering
  \label{tab:photcat}
  \begin{threeparttable}
    \begin{tabular}{ccccccccc}
      \hline
      \hline
      VEGA mag & rms mag & \texttt{QFIT} & $o$ & \texttt{RADXS} & $\rm n_f$ & $\rm n_u$ & Local sky & Local-sky rms \\
      (1) & (2) & (3) & (4) & (5) & (6) & (7) & (8) & (9) \\
      \hline
      16.6034 & 22.6732 & 0.999 & 0.0030 & $-0.0132$ & 1 & 1 &  2.15 & 0.72 \\
      13.9209 &  1.9165 & 0.999 & 0.0003 & $ 0.0052$ & 1 & 1 &  0.00 & 0.00 \\
      16.8250 & 27.8072 & 0.999 & 0.0005 & $ 0.0177$ & 1 & 1 &  2.02 & 0.28 \\
      15.9924 & 12.9156 & 0.998 & 0.0004 & $ 0.0175$ & 1 & 1 &  3.23 & 0.99 \\
      13.7264 &  1.6022 & 0.999 & 0.0000 & $ 0.0126$ & 1 & 1 &  9.34 & 3.65 \\
      17.1066 &  0.0019 & 0.998 & 0.0056 & $-0.0101$ & 2 & 2 &  2.14 & 0.92 \\
      16.6040 & 22.6854 & 1.000 & 0.0023 & $ 0.0056$ & 1 & 1 &  2.04 & 0.57 \\
      16.6611 &  0.0210 & 0.999 & 0.0014 & $-0.0070$ & 2 & 2 &  1.97 & 0.40 \\
      12.9643 &  0.7941 & 0.999 & 0.0000 & $-0.0189$ & 1 & 1 & 18.41 & 8.27 \\
      16.3636 & 18.1810 & 0.999 & 0.0019 & $ 0.0146$ & 1 & 1 &  2.30 & 0.50 \\
      $[...]$ & $[...]$ & $[...]$ & $[...]$ & $[...]$ & $[...]$ & $[...]$ & $[...]$ & $[...]$ \\
      \hline
    \end{tabular}  
    \begin{tablenotes}
    \item \textbf{Columns:} (1) VEGA magnitude; (2) magnitude rms; (3)
      \texttt{QFIT}; (4) fraction of neighbor flux within the fitting
      radius before neighbor subtraction; (5) excess/deficiency of
      flux just outside of the fitting radius with respect to that
      expected from the PSF; (6) number of single exposures a star was
      found; (7) number of measurements used to compute the
      photometric quantities of a star; (8) local sky (in counts); (9)
      local-sky rms (in counts). 
    \item \textbf{Notes:} (i) Stars are ordered as in the
      corresponding PM catalog. (ii) Stars measured in only one image
      have a magnitude rms equal to 9.99 mag. (iii) Stars with rms
      mag, \texttt{QFIT}, $o$, \texttt{RADXS}, $\rm n_f$ and $\rm n_u$
      equal to 0 are saturated in the majoirity of the images in which
      they were found.
    \end{tablenotes}
  \end{threeparttable}
\end{table*}

\begin{table}
  \caption{Adopted zero-points used to calibrate our instrumental
    photometry into the VEGA-mag flight system.}
  \centering
  \label{tab:photzp}
  \begin{threeparttable}
    \begin{tabular}{cccc}
      \hline
      \hline
      GO & Instrument/Camera & Filter & Zero-point \\
      \hline
      12915 & WFC3/IR & F139M  & 23.3039 \\
      12915 & ACS/WFC & F850LP & 24.2151 \\
      13771 & WFC3/IR & F139M  & 23.3089 \\
      13771 & ACS/WFC & F850LP & 24.2161 \\
      \hline
    \end{tabular}
  \end{threeparttable}
\end{table}

\begin{table*}
  \caption{First 10 lines of the WFC3/IR PM catalog.}
  \centering
  \label{tab:pmcat}
  \begin{threeparttable}
    \begin{tabular}{cccccccccccc}
      \hline
      \hline
      R.A. & Dec. & X & Y & $\mu_\alpha \cos\delta$ & $\sigma_{\mu_\alpha \cos\delta}$ & $\mu_\delta$ & $\sigma_{\mu_\delta}$ & $\Delta$time & $\rm n_1$ & $\rm n_2$ & ID \\
      deg & deg & pixel & pixel & \masyr  & \masyr  & \masyr  & \masyr & yr & & & \\
      (1) & (2) & (3) & (4) & (5) & (6) & (7) & (8) & (9) & (10) & (11) & (12) \\
      \hline
      266.2905382 & $-29.2501896$ & 19418.5218 & 1236.5295 & $-1.87299$ & 99.99000 & $-3.61488$ & 99.99000 & 2.8222 & 1 & 1 & 3 \\
      266.2895192 & $-29.2501574$ & 19482.5327 & 1238.8051 & $-2.56358$ & 99.99000 & $-4.89845$ & 99.99000 & 2.8222 & 1 & 1 & 4 \\
      266.2935075 & $-29.2501091$ & 19231.9894 & 1242.4437 & $-1.85439$ & 99.99000 & $-4.21034$ & 99.99000 & 2.8222 & 1 & 1 & 5 \\
      266.2939944 & $-29.2499806$ & 19201.4104 & 1251.7144 & $-7.62024$ & 99.99000 & $-2.58183$ & 99.99000 & 2.8222 & 1 & 1 & 6 \\
      266.2938486 & $-29.2495592$ & 19210.5888 & 1282.0495 & $-2.51380$ & 99.99000 & $-9.97158$ & 99.99000 & 2.8222 & 1 & 1 & 7 \\
      266.2911804 & $-29.2493321$ & 19378.2178 & 1298.2947 & $-2.77795$ &  1.39870 & $-3.34788$ &  1.05766 & 2.8222 & 2 & 2 & 9 \\
      266.2955196 & $-29.2489240$ & 19105.6429 & 1327.8556 & $-2.34727$ & 99.99000 & $-2.63357$ & 99.99000 & 2.8222 & 1 & 1 & 10 \\
      266.2902060 & $-29.2488892$ & 19439.4495 & 1330.1431 & $-2.93934$ &  1.53431 & $-2.57190$ &  0.88552 & 2.8222 & 2 & 2 & 11 \\
      266.2870774 & $-29.2488844$ & 19635.9957 & 1330.3597 & $-8.50908$ & 99.99000 & $-3.35906$ & 99.99000 & 2.8222 & 1 & 1 & 12 \\
      266.2951208 & $-29.2488618$ & 19130.6988 & 1332.3214 & $-5.23683$ & 99.99000 & $-6.11044$ & 99.99000 & 2.8222 & 1 & 1 & 13 \\
      $[...]$ & $[...]$ & $[...]$ & $[...]$ & $[...]$ & $[...]$ & $[...]$ & $[...]$ & $[...]$ & $[...]$ & $[...]$ & $[...]$ \\
      \hline
    \end{tabular}  
    \begin{tablenotes}
    \item \textbf{Columns:} (1) Right ascension; (2) Declination; (3)
      X position; (4) Y position; (5) PM along $\alpha\cos\delta$; (6)
      PM error along $\alpha\cos\delta$; (7) PM along $\delta$; (8) PM
      error along $\delta$; (9) Average temporal baseline; (10) Number
      of measurements in epoch 1; (11) Number of measurements in epoch
      2; (12) ID.  
    \item \textbf{Notes:} (i) Stars measured in only one image have a
      PM error of 99.99 \masyr in each coordinate.
    \end{tablenotes}
  \end{threeparttable}
\end{table*}

\section{List of confirmed, candidate and non MSs}\label{appendix:mslist}

Table~\ref{tab:mslist} presents the list of confirmed MSs, candidate
MSs and non-massive objects analyzed in Sect.~\ref{MS}.

\begin{table*}
  \caption{First 10 lines of the list of confirmed MSs, candidate MSs
    and non-massive objects analyzed in Sect.~\ref{MS}.}
  \centering
  \label{tab:mslist}
  \begin{threeparttable}
    \begin{tabular}{ccccc}
      \hline
      \hline
      R.A. & Dec. & ID & List & Alternative ID \\
      deg & deg & & & \\
      (1) & (2) & (3) & (4) & (5) \\
      \hline
      266.2907358 & $-29.2369064$ & 189   & Primary & 151 \\
      266.2873217 & $-29.2049508$ & 1616  & Primary & 147 \\
      266.2793346 & $-29.2001464$ & 1869  & Primary & 53  \\
      266.3411210 & $-29.1998483$ & 1886  & Primary & 49  \\
      266.2701831 & $-29.1962723$ & 2061  & Other   & 2MASS J17450483-2911464 \\
      266.2619529 & $-29.1499028$ & 5289  & Primary & 50  \\
      266.3196554 & $-28.9736557$ & 13644 & Primary & 134 \\
      266.3703120 & $-28.9573639$ & 14221 & Other   & XID \#947 \\
      266.3811732 & $-28.9546972$ & 14332 & Primary & 36  \\
      266.4005566 & $-28.9440693$ & 14733 & Primary & 111 \\
      $[...]$ & $[...]$ & $[...]$ & $[...]$ & $[...]$ \\
      \hline
    \end{tabular}  
    \begin{tablenotes}
    \item \textbf{Columns:} (1) Right ascension; (2) Declination; (3)
      ID; (4) Original list (see Sect.~\ref{MS}); (5) ID in the
      original list in column (4). 
    \item \textbf{Notes:} (i) ``Primary'' and ``Secondary'' are the
      lists of confirmed and potential, respectively, Paschen $\alpha$
      emitters of \citet{2011MNRAS.417..114D}. ``Other'' is the
      collection of objects from other sources (see
      Sect.~\ref{MS}). (ii) Objects with alternative IDs in the
      ``Other'' list starting with ``C'' are from
      \citet{2018A&A...617A..65C}. (iii) We refer to the papers listed
      in Sect.~\ref{MS} for the spectral-type characterization of
      these stars.
    \end{tablenotes}
  \end{threeparttable}
\end{table*}

\bibliographystyle{mnras}

\bsp	
\label{lastpage}

\end{document}